\documentclass[%
 aip,
 amsmath,amssymb,
 reprint,%
]{revtex4-1}

\usepackage{graphicx}
\usepackage{dcolumn}
\usepackage{bm}

\usepackage[utf8]{inputenc}
\usepackage[T1]{fontenc}
\usepackage{mathptmx}
\usepackage{etoolbox}
\usepackage{bigints}
\usepackage[dvipsnames]{xcolor}
\usepackage{comment} 

\makeatletter
\def\@email#1#2{%
 \endgroup
 \patchcmd{\titleblock@produce}
  {\frontmatter@RRAPformat}
  {\frontmatter@RRAPformat{\produce@RRAP{*#1\href{mailto:#2}{#2}}}\frontmatter@RRAPformat}
  {}{}
}%
\makeatother
\begin{document}

\preprint{AIP/123-QED}

\title[Nitrogen hydrates]{Dissociation line and driving force for nucleation of the nitrogen hydrate from computer simulation}

\author{Jes\'us Algaba}
\affiliation{Laboratorio de Simulaci\'on Molecular y Qu\'imica Computacional, CIQSO-Centro de Investigaci\'on en Qu\'imica Sostenible and Departamento de Ciencias Integradas, Universidad de Huelva, 21006 Huelva Spain}

\author{Miguel J. Torrej\'on}
\affiliation{Laboratorio de Simulaci\'on Molecular y Qu\'imica Computacional, CIQSO-Centro de Investigaci\'on en Qu\'imica Sostenible and Departamento de Ciencias Integradas, Universidad de Huelva, 21006 Huelva Spain}

\author{Felipe J. Blas}
\affiliation{Laboratorio de Simulaci\'on Molecular y Qu\'imica Computacional, CIQSO-Centro de Investigaci\'on en Qu\'imica Sostenible and Departamento de Ciencias Integradas, Universidad de Huelva, 21006 Huelva Spain}
\email{felipe@uhu.es}

\begin{abstract}

In this work, we determine the dissociation line of the nitrogen (N$_{2}$) hydrate by computer simulation using the TIP4P/Ice model for water and the TraPPE force field for N$_{2}$. We use the solubility method proposed recently by some of us to evaluate the dissociation temperature of the hydrate at different pressures, from $500$ to $1500\,\text{bar}$. Particularly, we calculate the solubility of N$_2$ in the aqueous solution when is in contact with a N$_{2}$-rich liquid phase and when in contact with the hydrate phase via planar interfaces as functions of temperature. Since the solubility of N$_{2}$ decreases with temperature in the first case and increases with temperature in the second case, both curves intersect at a certain temperature that determines the dissociation temperature at a given pressure. We find a good agreement between the predictions obtained in this work and the experimental data taken from the literature in the range of pressures considered in this work. From the knowledge of the solubility curves of N$_{2}$ in the aqueous solution, we also determine the driving force for nucleation of the hydrate, as a function of temperature, at different pressures. In particular, we use two different thermodynamic routes to evaluate the change of the chemical potential for the formation of the hydrate. Although the driving force for nucleation slightly decreases (in absolute value) when the pressure is increased, our results indicate that the effect of pressure can be considered negligible in the range of pressures studied in this work. To the best of our knowledge, this is the first time the driving force for nucleation of a hydrate that exhibits crystallographic structure sII, along its dissociation line, is studied from computer simulation.

\end{abstract}

\maketitle

%

\section{Introduction}

Hydrates are non-stoichiometric inclusion solid compounds in which guest molecules, such as methane (CH$_{4}$), carbon dioxide (CO$_{2}$), hydrogen (H$_{2}$), nitrogen (N$_{2}$), and tetrahydrofuran (THF), among many other species, are enclathrated in the voids left by a periodic network of water molecules or
host.~\cite{Sloan2008a,Ripmeester2022a} Water molecules interact not only via dispersive attractive and repulsive interactions, but also through specific, short-range, and highly directional hydrogen-bonding interactions, that cause the network arrangement of the system. The guests generally do not compete with the specific interactions that keep together the coordinated lattice formed by the host. Hydrates crystallize in the
well-known structures sI, sII, and sH. Hydrates of small molecules, such as CH$_{4}$ and CO$_{2}$, form sI hydrates, while some small molecules, including N$_{2}$ and H$_{2}$, and in general larger substances, like THF, usually form sII hydrate structures.~\cite{Sloan2008a,Ripmeester2022a}

The formation of sII structures of hydrates with small molecules like N$_{2}$ and H$_{2}$ is unusual. However, the non-stoichiometric nature of hydrates offers the possibility of multiple cage occupancy.~\cite{Sloan2008a,Ripmeester2022a} In 1997, Kuhs \emph{et al.}~\cite{Kuhs1997a} carried out a series of high-resolution neutron diffraction experiments on N$_{2}$ hydrates that demonstrated that small cages exhibit single occupancy but large cages are doubly occupied at relatively high pressures. The experimental results were also corroborated by molecular dynamics computer simulations~\cite{vanKlaveren2001a,vanKlaveren2001b,VanKlaveren2002a} and additional experiments including neutron powder diffraction~\cite{Chazallon2002a} and Raman scattering measurements.~\cite{Sasaki2003a} More recently, Tsimpanogiannis \emph{et al.} have also confirmed this idea via Monte Carlo simulations of the N$_{2}$ hydrate. It is also important to mention the pioneering works of Alavi \emph{et al.}~\cite{Alavi2005a,Alavi2006a}  on the study of the stability of H$_{2}$ and H$_{2}$ + tetrahydrofurane hydrates in the context of multiple occupancy from molecular dynamics simulations. Under this perspective, it is possible to understand, from a molecular perspective, why small molecules like N$_{2}$ and H$_{2}$ form, at least at low and moderate pressures, sII hydrate structure. The explanation of the preference to form sII hydrates instead of sI one is because N$_{2}$ and H$_{2}$ molecules better stabilize small hydrate cages which are in greater number in sII crystallographic structure~\cite{Barnes2013a} (see below a more detailed description of the type and number of cages in each structure).

Fundamental and applied research on hydrates has been motivated by several reasons.~\cite{Sloan2008a,Ripmeester2022a,Ripmeester2016a,Ratcliffe2022a} One of the most important applications of hydrates in the current environmental, energetic, and economic context is the use of hydrates of N$_{2}$, H$_{2}$, their mixtures, and in combination with THF, as strategic materials for gas transportation and storage.~\cite{Tsimpanogiannis2017a,Brumby2019a,Yi2019a,Michalis2022a} This represents an alternative to the currently used metal hydrides, whose development has not been implemented yet due to factors such as the incomplete characterization of the thermodynamics and kinetics of these new storage media. The use of hydrates in this application would represent a reduction in raw material costs, with similar volumetric storage performance. To this end, it is necessary an accurate and precise knowledge of the thermodynamics, with particular emphasis on phase equilibria, but also on the kinetics of the formation and growth of these hydrates.

It is clear from the previous discussion that the N$_{2}$ hydrate is a strategic material from practical and industrial points of view. However, the study of the N$_{2}$ hydrate it is also relevant from a theoretical perspective due to its crystallographic structure. The N$_{2}$ hydrate exhibits sII structure formed from $136$ water molecules distributed in $16$ D (pentagonal dodecahedron or $5^{12}$) cages and $8$ H (hexakaidecahedron or $5^{12}6^{4}$) cages. Assuming  single full occupancy, the sII unit cell has 24 additional N$_{2}$ molecules. Interestingly, although sI and sII hydrate structures are very different, their stoichiometry is very similar. Hydrates with sI structure have 1 guest molecule (CO$_{2}$ or CH$_{4}$ for example) per 5.75 water molecules in the  unit cell (or $8$ guest molecules per $46$ water molecules, i.e., $46/8=5.75$) and hydrates with sII structure has 1 guest molecule (N$_{2}$) per 5.67 water molecules in the  unit cell, approximately (or $24$ N$_{2}$ molecules per $136$ water molecules, i.e, $136/24\approx 5.67$). Note that in the sII structure, the $66.7\%$ of the cages are D cages, whereas only the $25\%$ are D cages in the sI structure. Due to this and the reasons previously explained, the N$_{2}$ hydrate is one of the few hydrates forming sII crystallographic structure with a small guest as N$_{2}$ that is able to exhibit multiple occupancy in their cages. In this work, we concentrate on the N$_{2}$ hydrate from a molecular perspective and perform molecular simulations to study the dissociation line of the hydrate and one of the key parameters that controls its nucleation, $\Delta\mu_{\text{N}}$, the driving force for nucleation.~\cite{Kashchiev2000a,Kashchiev2002a,Kashchiev2002b,Kashchiev2003a}

The dissociation line of the N$_{2}$ hydrate has been previously determined experimentally by several authors.~\cite{vanCleeff1960a,Marshall1964a,Jhaveri1965a,Sugahara2002a,Mohammadi2003a} In addition, it is also possible to describe theoretically the phase equilibria of the hydrate using the van der Waals and Platteeuw (vdW\&P) formalism,~\cite{vanderWaals1950a,vanderWaals1951a} in combination with an equation of state.~\cite{Dufal2012a} From the point of view of molecular simulation, several authors have studied some properties of the N$_{2}$ hydrate, including  thermodynamic stability of the hydrate under the perspective of single/double cage occupancy using Monte Carlo~\cite{Tanaka1993a,Tanaka1993b,Tanaka1994a,Tanaka2004a,Tsimpanogiannis2012a} and molecular dynamics simulations.~\cite{vanKlaveren2001a,vanKlaveren2001b,VanKlaveren2002a} However, none of the previous works are devoted to determining the three-phase line of the N$_{2}$ hydrate. Particularly interesting for the present study is the work of Yi \emph{et al.}\cite{Yi2019a} In this paper, the authors use the well-known TIP4P/2005 model for water~\cite{Abascal2005a} and the 2CLJQ model for N$_{2}$ of Vrabec \emph{et al.}~\cite{Vrabec2001a} to determine the dissociation line of the N$_{2}$ hydrate and to investigate the pressure influence on hydrate growth. Particularly, they have checked that the amount of cages occupied by two N$_{2}$ molecules (double occupancy of the large cages of the hydrate) is increased when the pressure is very high, above $1000\,\text{bar}$. Although the TIP4P/2005 model is one of the most accurate and widely used rigid non-polarizable water model in the literature, it is not particularly suited to deal with solid phases. In these cases, the also well-known TIP4P/Ice water model is preferable since predicts very accurately the melting point of water in a wide range of pressures.~\cite{Abascal2005b} This is also corroborated by a nice paper of Conde and Vega in which the authors show that it is necessary a good model of water for describing accurately the melting temperature of ice Ih in order to get a good prediction of dissociation temperatures of hydrates.~\cite{Conde2013a}

The usual methodology to calculate dissociation lines of hydrates from molecular simulations, based on the original ideas of Ladd and Woodcock proposed in the 1970s, is the well-known direct coexistence technique.~\cite{Ladd1977a} Conde and Vega published the first paper in which the technique was used to determine the three-phase coexistence line of the CH$_{4}$ methane.~\cite{Conde2010a,Conde2013a} From that paper, several authors have determined the dissociation line of several hydrates using the same approach,~\cite{Miguez2015a,Perez-Rodriguez2017a,Michalis2015a,Costandy2015a,Fernandez-Fernandez2019a,Fernandez-Fernandez2021a,Blazquez2023b} including Yu \emph{et al.}~\cite{Yi2019a} that have determined the dissociation line of the N$_{2}$. However, in this work we propose to use a new and alternative technique, the so-called solubility method, introduced by some of us to determine the dissociation temperature, at $400\,\text{bar}$, of the CH$_{4}$ and CO$_{2}$ hydrates.~\cite{Grabowska2022a,Algaba2023a} Is it possible to use the solubility method for hydrates that exhibit sII crystallographic structure? The method is based on the calculation of the solubility of the guest, in this case, N$_{2}$, in the aqueous solution in equilibrium via planar interfaces with the N$_{2}$-rich liquid phase and with the hydrate. In the first case, the molar fraction of N$_{2}$ (when it is in contact with the N$_{2}$ phase) is a decreasing function of the temperature. In the second case, the molar fraction of N$_{2}$ (when it is in contact with the hydrate phase) is an increasing function of the temperature. Consequently, both curves intersect, at a given pressure, and this allows to determine the $T_{3}$ of the hydrate. A schematic representation of the solubility method has been represented in Fig. 1. As can be seen, the solubility curves of N$_2$ in the aqueous phase, as functions of temperature, are calculated when it is in contact with a N$_2$-rich liquid phase and a hydrate phase via planar interfaces. The temperature at which both curves cross is the point at which the aqueous, hydrate, and N$_2$-rich liquid phases are in equilibrium, i.e. the so-called $T_3$ at the corresponding pressure.

\begin{figure}
\includegraphics[width=\columnwidth]{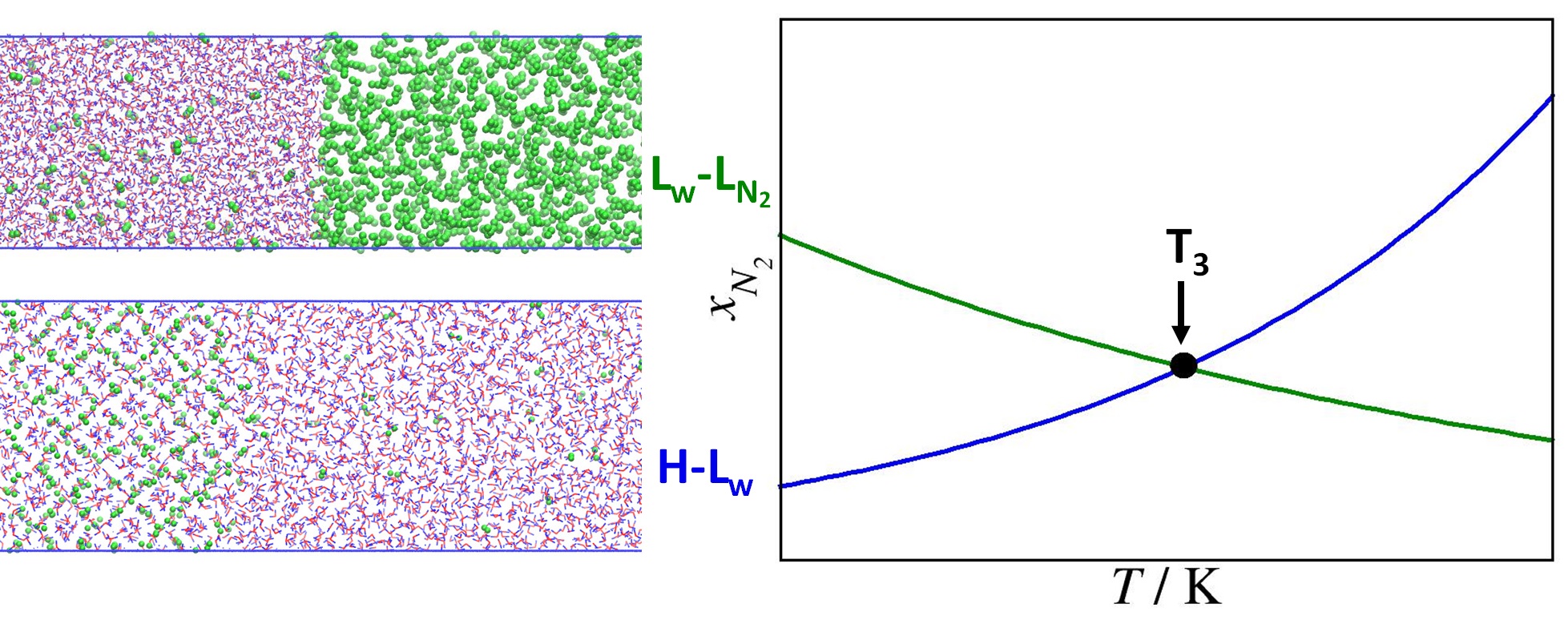}\\
\caption{Schematic representation of the solubility method. The top left picture represents the simulation box used to study the solubility of N$_2$ in the aqueous phase (L$_{\text{w}}$) when in contact with a N$_2$-rich (L$_{\text{N}_{2}}$) liquid phase. The bottom left picture represents the simulation box used to study the solubility of N$_2$ in the aqueous phase (L$_{\text{w}}$) when in contact with a hydrate phase (H). Both pictures have been obtained from VMD. Nitrogen atoms are represented in purple, and hydrogens and oxygen from water molecules in blue and red respectively. Finally, the right picture is a schematic representation of the solubility results obtained from H-L$_{\text{w}}$ (blue curve) and L$_{\text{w}}$--L$_{\text{N}_{2}}$ (purple curve) equilibria. The filled black circle represents the temperature at which both curves cross ($T_3$).}
\label{scheme}
\end{figure}

The solubility method is a very interesting and clean way to calculate the dissociation temperatures of hydrates. However, the determination of the two solubility curves is also key in the context of nucleation of the N$_{2}$ hydrate. The first solubility curve (L$_{\text{w}}$--L$_{\text{N}_{2}}$) corresponds to the thermodynamic states at which hydrates are formed at experimental conditions when both phases are in equilibrium via a planar interface. The second solubility curve (H--L$_{\text{N}_{2}}$) is also key to determining the driving force for nucleation of hydrates using a new and alternative route presented by some of us in a previous work.~\cite{Algaba2023a} From this information, in this work, we also propose to evaluate the driving force for nucleation of the N$_{2}$ hydrates considering the formation of the hydrate as a chemical reaction in which water and N$_{2}$ molecules ``react'' in the aqueous solution phase.~\cite{Kashchiev2000a,Kashchiev2002a,Kashchiev2002b,Kashchiev2003a} This strategy has been already used by Grabowska \emph{et al.} and Algaba \emph{et al.} to evaluate driving forces for nucleation of the CH$_{4}$ and CO$_{2}$ hydrates.~\cite{Grabowska2022a,Algaba2023a} Is it possible to follow a similar strategy and calculate the driving force for nucleation of the N$_{2}$ hydrate that exhibit sII structure? According to these studies, the change in chemical potential of this reaction can be identified as $\Delta\mu_{\text{N}}$, the driving force for nucleation. We propose to use two different thermodynamic routes,~\cite{Grabowska2022a,Algaba2023a} at three different pressures, to estimate the driving force for nucleation of a hydrate that exhibits sII crystallographic structure. To the best of our knowledge, it is the first time the driving force for nucleation of the N$_{2}$ hydrate is calculated from computer simulation.

To answer the questions of the two previous paragraphs on the possibility of determining the dissociation line from the solubility method and the driving force for nucleation using route 1 and dissociation route of the N$_{2}$ hydrate, a clathrate that exhibits the sII crystallographic structure, we follow here the simplest strategy taken into account the complexity of the N$_{2}$ hydrate. We assume that both cages of the hydrate, the D and H cages, are fully occupied. In addition to that, we only consider single-cage occupancy. According to the literature, this is a good approximation at low pressures and can be considered a first approach to determine the dissociation line and driving force for nucleation of the N$_{2}$. A more detailed study of how occupancy affects these properties is out of the scope of this preliminary study and will be the subject of a future work.

The organization of this paper is as follows: In Sec. II, we describe the models, simulation details, and methodology used in this work. The results obtained, as well as their discussion, are described in Sec. III. Finally, conclusions are presented in Sec. IV.

\section{Models, simulation details, and methodology}

In this work, all molecular dynamics simulations are carried out using the GROMACS molecular dynamics package (2016.5 version).~\cite{VanDerSpoel2005a} N$_{2}$ molecules are modeled through the TraPPE (Transferable Potentials for Phase Equilibria) force field~\cite{Potoff2001a} and water molecules are described using the well-known TIP4P/Ice model.~\cite{Abascal2005b} Unlike dispersive interactions between the oxygen atoms of water and the nitrogen atoms of N$_{2}$ are calculated using a modified Berthelot combining rule as,

\begin{equation}
    \epsilon_{\text{ON}}=\xi_\text{{ON}}(\epsilon_{\text{OO}}\,\epsilon_\text{{NN}})^{1/2}
    \label{unlike}
\end{equation}

\noindent where $\epsilon_{\text{ON}}$ is the well depth associated with the LJ potential for the unlike interactions between the oxygen atoms of water, O, and the nitrogen atoms of N$_2$, N. $\epsilon_\text{{OO}}$ and $\epsilon_\text{{NN}}$ are the well depth for the like interactions between the O and N atoms respectively, and $\xi_\text{{ON}}$ is the factor that modifies the Berthelot combining rule. Some of the simulations are performed using the standard Berthelot combining rule with $\xi_\text{{ON}}=1.0$. We also perform simulations using Eq.~\eqref{unlike} with an optimized value of $\xi_\text{{ON}}$. As we will see later in Section III, this allows to accurately describe the three-phase coexistence line of the N$_2$ hydrate along a wide range of pressures. To obtain the optimal value of the unlike interaction parameters between the oxygen and nitrogen atoms of water and nitrogen molecules, respectively, we try different values of the combining parameter $\xi_{\text{ON}}$, including $1.00$, $1.10$, and $1.15$. With these values, we determine the solubility of N$_{2}$ in the aqueous phase when in contact with the N$_{2}$-rich liquid phase and the hydrate. From this information, we obtain the values of the dissociation temperature, $T_{3}$, at $500\,\text{bar}$. We then compare the predictions from the model and the experimental data from the literature, finding that $\xi_{\text{ON}}=1.15$ provides the best description of the $T_{3}$ at this pressure. Then, we use this parameter to predict the $T_{3}$ of the hydrate at different pressures. It is interesting to note that we observe that an increase of $0.05$ units in the combining parameter $\xi_{\text{ON}}$ produces an increment of the $T_{3}$ of $5\,\text{K}$ approximately.

We use a Verlet leapfrog\cite{Cuendet2007a} algorithm with a time step of $2\,\text{fs}$ to solve the Newton's motion equations. We also use the Nosé-Hoover thermostat\cite{Nose1984a} and the Parrinello-Rahman barostat,\cite{Parrinello1981a} with a time constant of $2\,\text{ps}$, to ensure that simulations are performed at constant temperate and pressure. In all cases, we use a cut-off of $1.0\,\text{nm}$ for the Coulombic and dispersive interactions. The Fourier term of the Ewald sums is evaluated using the particle mesh Ewald (PME) method.~\cite{Essmann1995a} The width of the mesh is $0.1\,\text{nm}$ with a relative tolerance of $10^{-5}$.

We simulate liquid-liquid (L$_{\text{w}}$--L$_{\text{N}_{2}}$) and hydrate-liquid (H--L$_{\text{w}}$) two-phase equilibria to determine the three-phase coexistence line of the N$_2$ hydrate according to the solubility method.~\cite{Grabowska2022a, Algaba2023a} We also perform bulk simulations to calculate the driving force for nucleation of this hydrate at different conditions of pressure and supercooling. For systems that exhibit L$_{\text{w}}$-L$_{\text{N}_{2}}$ equilibria, the simulations are carried out in the $NP_{z}\mathcal{A}T$ ensemble, in which only $L_z$, the side of the simulation box perpendicular to the L$_{\text{w}}$--L$_{\text{N}_{2}}$ planar interface, is varied. In the case of systems that exhibit H--L$_{\text{w}}$ equilibria, we perform simulations in the anisotropic $NPT$ ensemble in which each side of the simulation box is allowed to fluctuate independently to keep the pressure constant. This ensures that the solid hydrate structure is equilibrated without stress. Finally, bulk systems of N$_2$, water, and hydrate are performed in the isotropic $NPT$ ensemble, i.e., the sides of the simulation box fluctuate isotropically to keep the pressure constant.

\section{Results}

\subsection{L$_{\text{w}}$--L$_{\text{N}_{2}}$ equilibria: Solubility of N$_2$ in liquid water and interfacial tension}

In this section, we study the phase equilibria of the water + N$_{2}$ binary mixture at conditions at which the system exhibits liquid-liquid immiscibility. We first concentrate on the determination of the solubility curve of N$_{2}$ in the aqueous solutions in equilibrium with the N$_{2}$-rich liquid phase, which is key to knowing the conditions at which nucleation takes place and to determining the dissociation line of the N$_{2}$ hydrate at a given pressure. We also consider in the following section the temperature dependence of the liquid-liquid interfacial tension at several pressures.

\subsubsection{Solubility of N$_2$ in liquid water}

We determine the solubility of N$_2$ in the aqueous solution, as a function of temperature, when is in contact with N$_{2}$-rich liquid phase via a planar interface. In particular, we consider three different pressures, $500$, $1000$, and $1500\,\text{bar}$. To this end, we prepare an initial simulation box consisting of two fluid phases, a N$_2$-rich liquid phase formed of 1223 guest molecules, and a water phase with 2800 water molecules. Both simulation boxes are put in contact via a planar interface according to the direct coexistence simulation technique. Arbitrary, we choose the $z$-axis perpendicular to the planar interface and the $xy$ plane parallel to the interface. The total dimensions of the initial simulation box are $L_{x}=L_{y}=3.8\,\text{nm}$ and $L_{z}\simeq15\,\text{nm}$.

Since we are dealing with an inhomogeneous system in which a planar interface exists perpendicular to the $z$-axis, simulations are performed in the $NP_{z}\mathcal{A}T$ ensemble. According to the rule phase, this ensures that the system evolves to the equilibrium state at the temperature and pressure fixed during the simulations. Note that in the $NP_{z}\mathcal{A}T$ ensemble, only the length of the simulation box along the $z$-axis is allowed to vary, keeping the interfacial area $\mathcal{A}=L_{x}\times L_{y}$ constant along the simulation. In other words, the $z$ component of the pressure tensor (perpendicular to the interface) is constant and equal to the equilibrium pressure. Due to the low solubility of N$_2$ in water, long simulation times are required in order to calculate the solubility of this compound in the aqueous solution. In particular, the simulations run during $800\,\text{ns}$. We consider an equilibration period of $200\,\text{ns}$, followed by a production period of $600\,\text{ns}$ during which the corresponding magnitudes are calculated and accumulated.

\begin{figure}
\includegraphics[width=0.9\columnwidth]{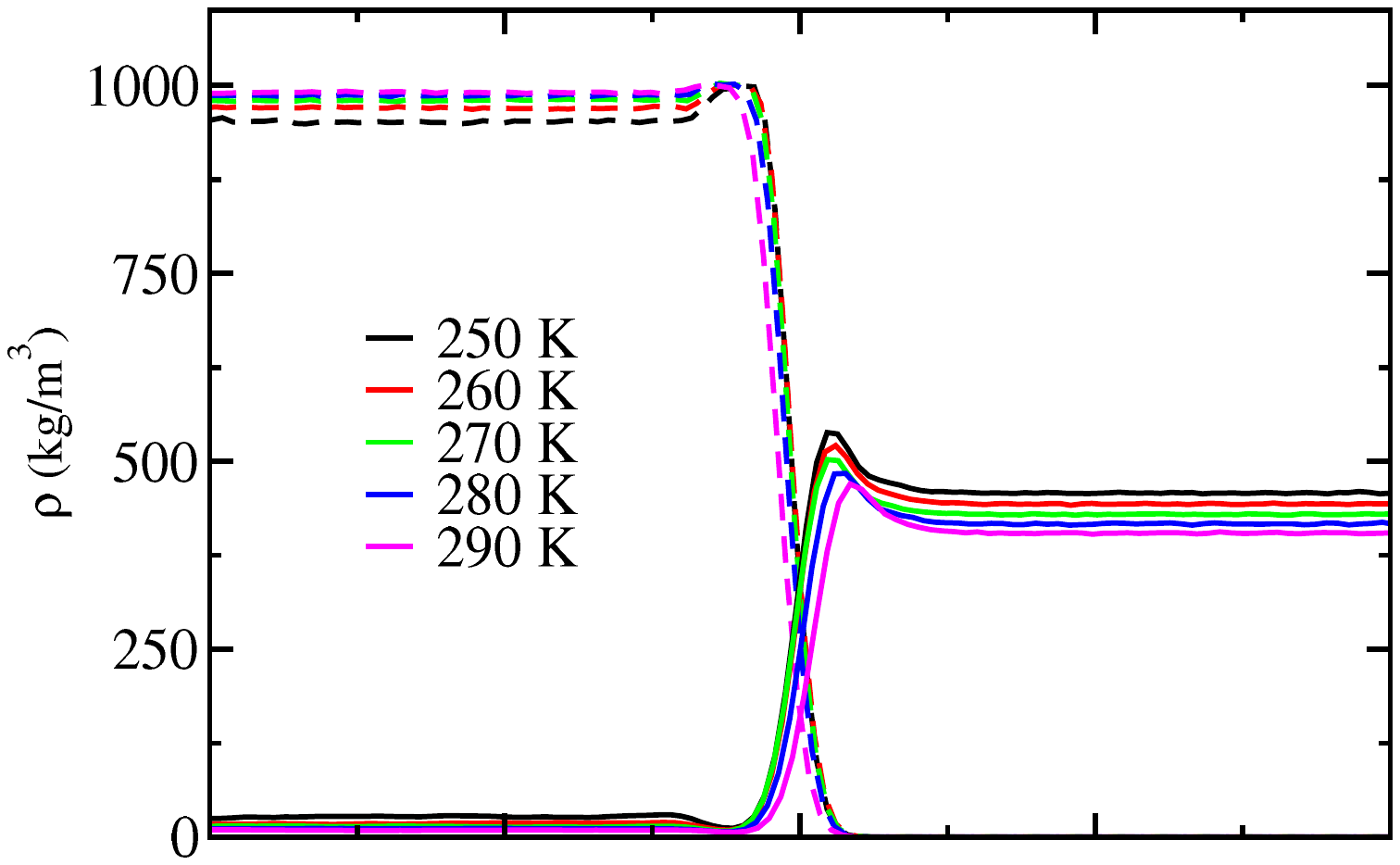}

\vspace{-0.9cm}
\includegraphics[width=0.9\columnwidth]{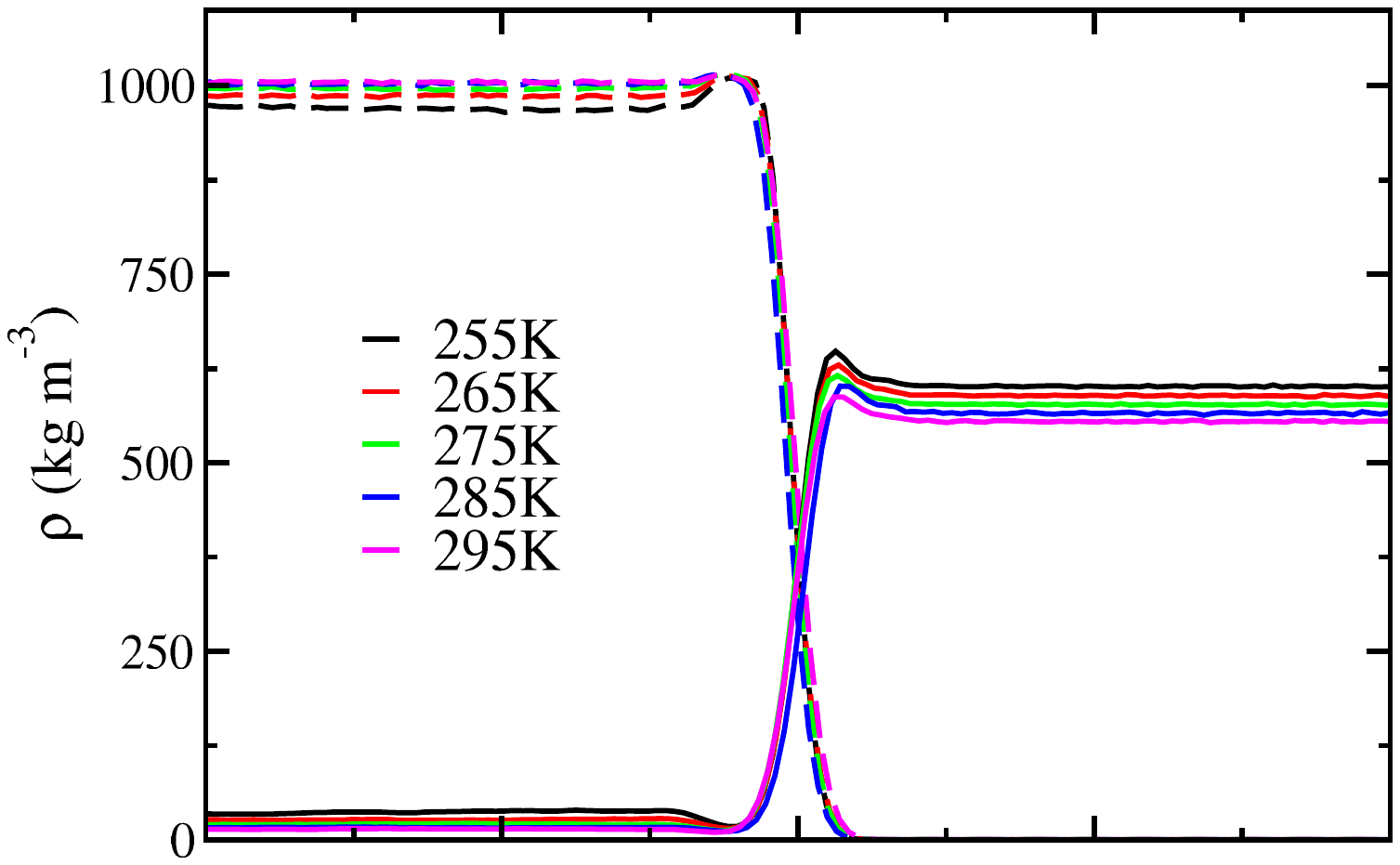}

\vspace{-0.9cm}
\includegraphics[width=0.9\columnwidth]{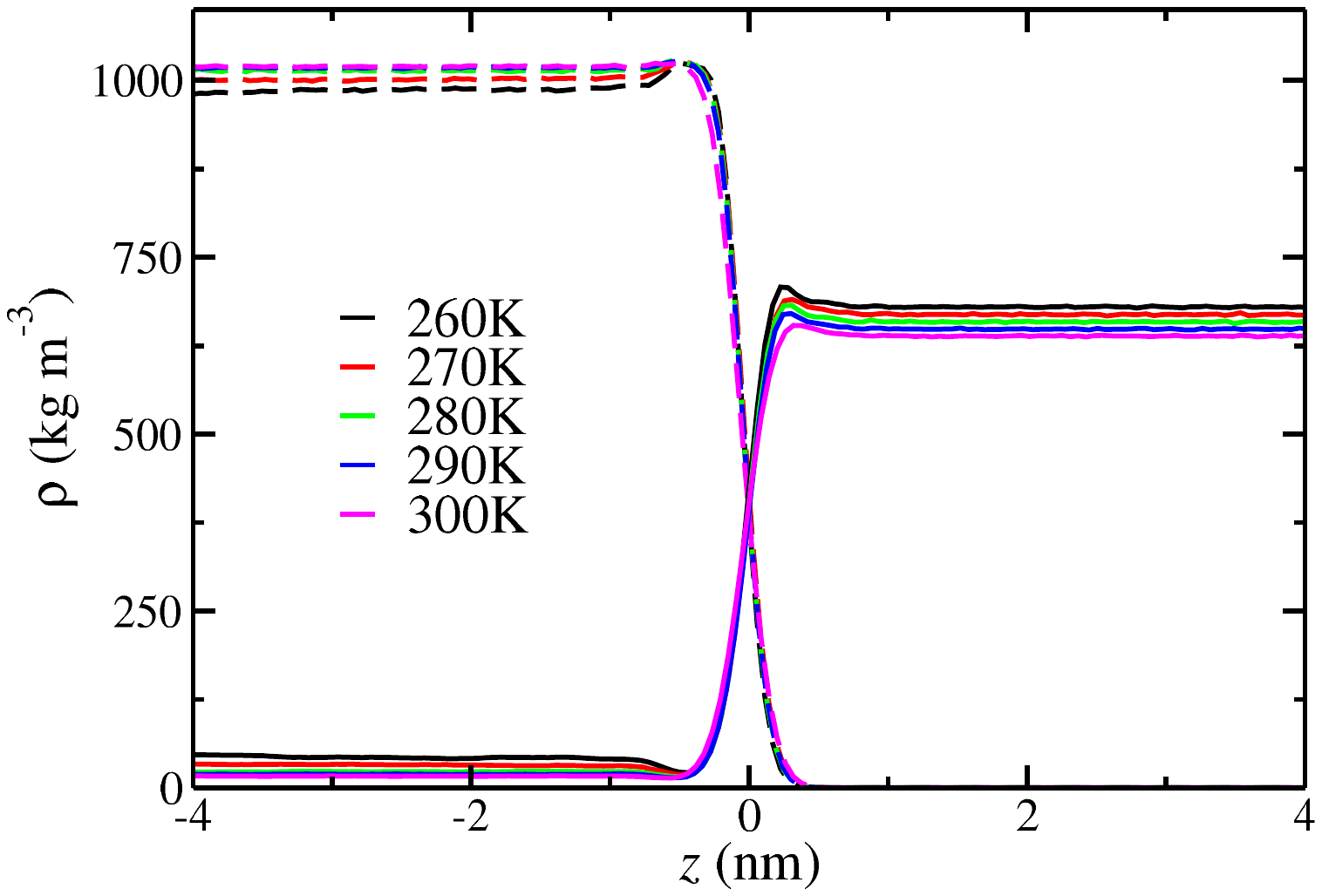}
\caption{Simulated equilibrium density profiles, $\rho(z)$, across the L$_{\text{w}}$--L$_{\text{N}_{2}}$ interface of N$_{2}$ (continuous curves) and water (dashed curves) as obtained from MD $NP_{z}\mathcal{A}T$ simulations, using $\xi_{\text{ON}}=1.15$, at $500$ (a), $1000$ (b), and $1500\,\text{bar}$ (c) and several temperatures (see legends).}
\label{figure1}
\end{figure}

In order to calculate the solubility of N$_2$ in the aqueous solution, we determine the density profiles of N$_{2}$ and water of the system at the three selected pressures and several temperatures. As we have previously mentioned, we use two different values of the $\xi_{\text{ON}}$ parameter, the deviation of $\epsilon_{\text{ON}}$ from the Berthelot rule. Particularly, we consider $\epsilon_{\text{ON}}=1.0$ and $1.15$. Figure~\ref{figure1} shows the density profiles at the studied conditions when $\xi_{\text{ON}}=1.15$. For better visualization and for avoiding repetition, only one of the two interfaces has been represented in Fig.~\ref{figure1}. The left side corresponds to the aqueous phase, which initially was a pure water phase, and the right side to the N$_2$-rich liquid phase. Note that due to the extremely low solubility of water in the N$_{2}$-rich liquid phase, this phase can be considered a pure N$_{2}$ liquid phase. Density profiles have been obtained by dividing the simulation box into 200 slabs perpendicular to the $z$ direction along which the system exhibits a planar L$_{\text{w}}$--L$_{\text{N}_{2}}$ interface. The position of the center of mass of the molecules is assigned to each slab and the density profiles are obtained from mass balance considerations.

Since the density profiles of both mixture components exhibit the same qualitative behavior, we simultaneously analyze the behavior of the density profiles at the three pressures. We first consider the behavior of the density profile of water along the interface (dashed curves). The density of water in the aqueous phase (left side of the profile) increases when the temperature is increased. In the N$_2$ liquid phase (right side of the profile), the density of water is practically zero in all cases. As we have already mentioned, this indicates that the solubility of water in the N$_{2}$ liquid phase can be considered negligible. In other words, the L$_{\text{w}}$--L$_{\text{N}_{2}}$ phase behavior of the water + N$_{2}$ mixture can be described as a saturated aqueous solution of N$_2$ in contact with a pure N$_2$ liquid phase at different temperatures and pressures. It is interesting to note that there exists preferential adsorption of water at the interface. This adsorption is observed in all cases but it decreases when the pressure is increased and/or when the temperature is increased. 

We now focus on the behavior of the density profiles of N$_2$ at the interface (continuous curves). As can be observed, the density of N$_2$ in the aqueous phase decreases when the temperature is increased. This is the expected behavior since the solubility of gases in water increases when temperature is decreased. Also, preferential desorption can be observed in the aqueous solution phase near the interface in all cases. This is in agreement with the preferential adsorption of water explained in the previous paragraph. As previously shown, preferential desorption decreases when pressure and/or temperature are increased. There is also preferential adsorption of N$_2$ near the interface but on the side corresponding to the N$_2$ liquid phase. As can be seen in Fig.~\ref{figure1}, the adsorption decreases when the pressure and/or the temperature are increased. Finally, it is interesting to point out that the density of the N$_2$ liquid phase decreases when the temperature is increased and/or when the pressure is decreased. Although the results presented in Fig.~\ref{figure1} correspond only to $\xi_{\text{ON}}=1.15$, we have also obtained the same qualitative results in the case in which $\xi_{\text{ON}}=1.0$.

\begin{figure}
\includegraphics[width=\columnwidth]{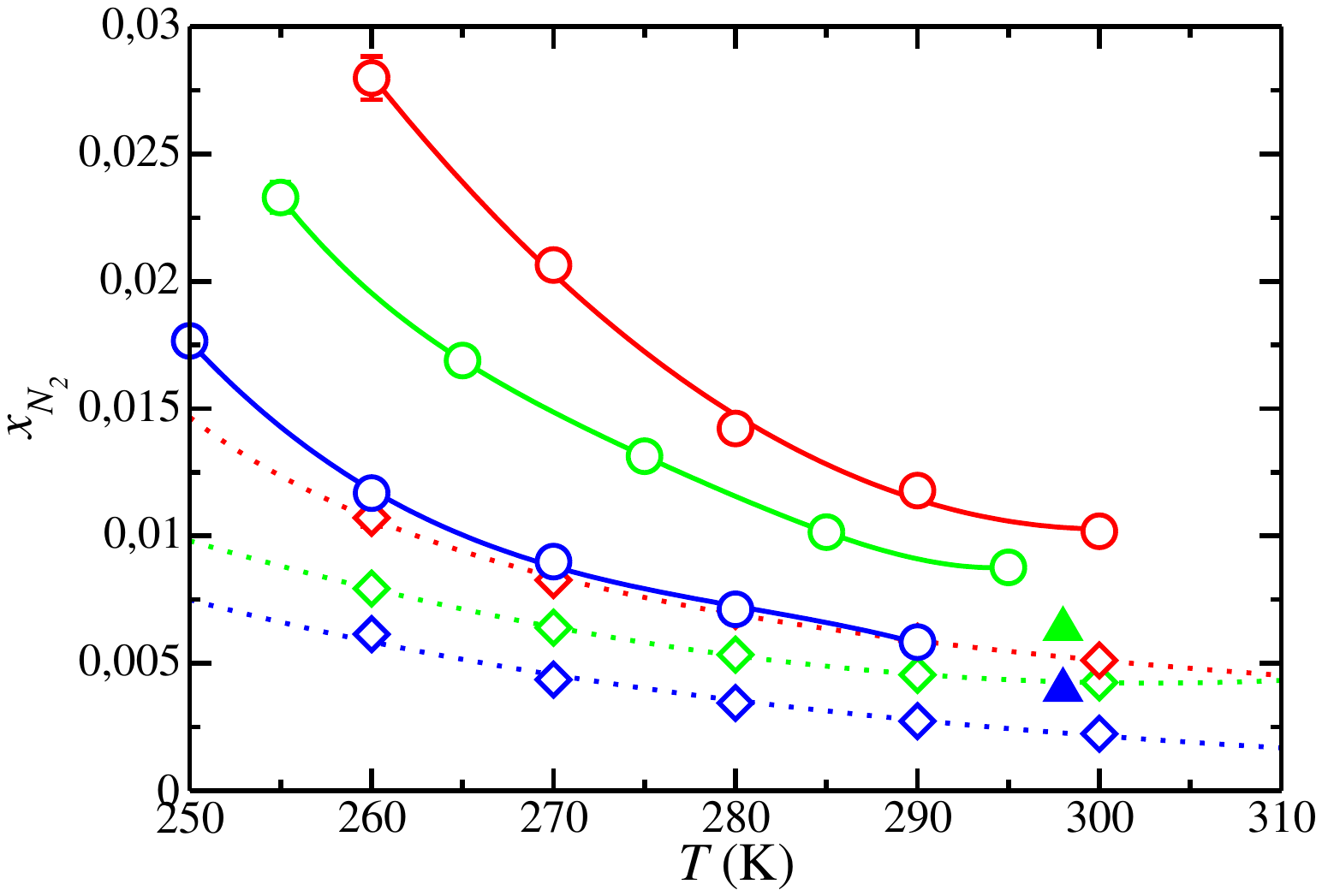}\\
\caption{Solubility of N$_2$ in the aqueous phase, as a function of temperature, at 
$500$ (blue), $1000$ (purple), and $1500\,\text{bar}$ (red) when the solution is in contact with the N$_{2}$ liquid phase via a planar interface. The symbols correspond to solubility values obtained from MD $NP_{z}\mathcal{A}T$ simulations using $\xi_{\text{ON}}=1.0$ (open diamonds and dashed curves) and $\xi_{\text{ON}}=1.15$ (open circles and continuous curves). In all cases, the curves are included as guides to the eyes. The filled blue (500 bar) and purple (1000 bar) up triangles represent the experimental data taken from the literature.~\cite{Wiebe1933}}
\label{figure2}
\end{figure}

From the analysis of the density profiles, it is possible to calculate the equilibrium molar fraction of N$_2$ in the aqueous solution phase when the initial pure water and pure N$_2$ phases are put in contact via a planar interface. The equilibrium density of water and N$_2$ in the aqueous solution phase can be easily obtained by averaging the density of each component over the aqueous region. At this point, it is important to remark that the region chosen has to be far enough from the interface in order to calculate accurately the bulk densities in the aqueous solution region. The results obtained in this work for both values of $\xi_{\text{ON}}$, 1.0 and 1.15, at 500, 1000, and 1500 bar, and at several temperatures have been represented in Fig.~\ref{figure2}.

According to Fig.~\ref{figure2}, the solubility of N$_2$ in water decreases when the temperature is increased, at a given pressure. Note that the same qualitative behavior is obtained from both values of $\xi_{\text{ON}}$. In addition to this, at a given temperature, the solubility of N$_2$ in water increases when the pressure is increased. Note that this is the expected behavior of the solubility of a gas in an aqueous phase. It is especially interesting to discuss the effect of $\xi_{\text{ON}}$ on the solubility of N$_2$ in the aqueous solution. As can be seen, the solubility of N$_2$ in water is small. An increment of the value of $\xi_{\text{ON}}$ provokes an increment of the solubility. The differences in solubility for different values of $\xi_{\text{ON}}$ become larger when the temperature is decreased and the pressure is increased. In particular, at the lowest common temperature of $260\,\text{K}$, the solubility of N$_{2}$ becomes 2, 2.5, and 3 times higher, at 500, 1000, and $1500\,\text{bar}$, respectively when $\xi_{\text{ON}}$ is increased from 1.0 to 1.15. On the other hand, at the highest common temperature of $290\,\text{K}$, the solubility becomes 1.9, 2, and 2.1 times higher, at 500, 1000, and $1500\,\text{bar}$, respectively. Notice that the effect of the temperature over the solubility is larger when the pressure and $\xi_{\text{ON}}$ are increased. Solubility results obtained in this work have been compared with experimental data taken from the literature$^{56}$ (see Fig.~3). As far as the authors know, at the conditions of pressure and temperature at which this study has been performed, there are only two experimental points at $500$ and $1000\,\text{bar}$ and $298\,\text{K}$ in both cases. As can be seen, simulation results obtained when $\xi_{\text{ON}}=1.0$ and $1.15$ are located between the experimental data, i.e., when $\xi_{\text{ON}}=1.0$, experimental data are slightly underestimated. Contrary, when $\xi_{\text{ON}}=1.15$, experimental data are slightly overestimated. In both cases, agreement between experimental data and simulation results is good.

\subsubsection{Interfacial tension}

It is possible to calculate the L$_{\text{w}}$--L$_{\text{N}_{2}}$ interfacial tension of the water + N$_{2}$ binary mixture from the same simulations performed to determine the density profiles and mutual solubilities. Contrary to the case of solid-fluid interfacial free energy,~\cite{Frenkel2002a,Allen2017a} the fluid-fluid interfacial tension can be easily calculated from the diagonal components of the pressure tensor as,~\cite{Rowlinson1982b,deMiguel2006a,deMiguel2006b}

\begin{equation}
\gamma=\frac{L_{z}}{2}\left[\left<P_{zz}\right>-\frac{\left<P_{xx}\right>+\left<P_{yy}\right>}{2}\right]
\label{EQ_tension}
\end{equation}

\begin{figure}
\includegraphics[width=\columnwidth]{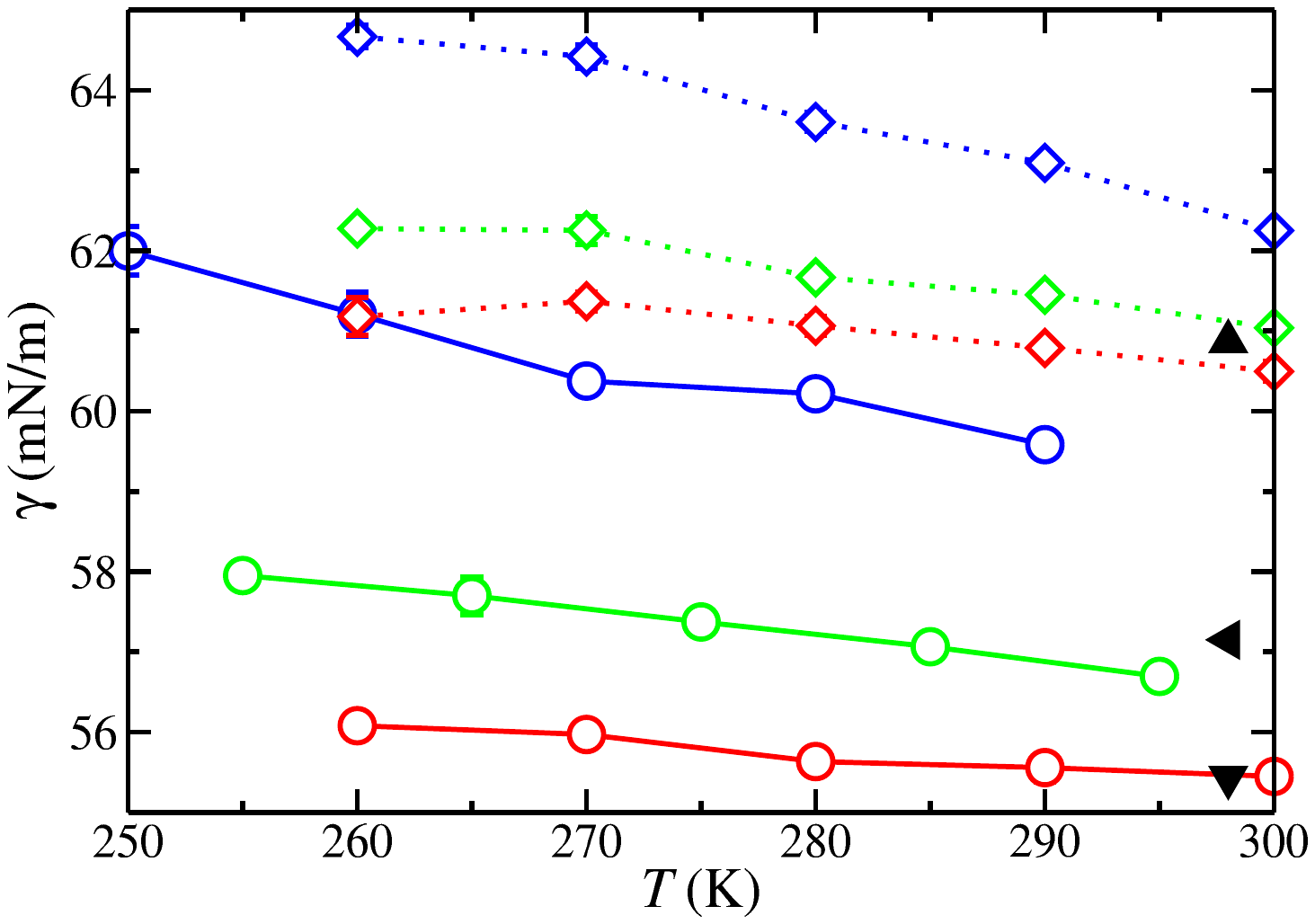}

\caption{L$_{\text{w}}$--L$_{\text{N}_{2}}$ interfacial tension, $\gamma$, as a function of the temperature between the aqueous solution and the N$_{2}$ liquid phase at $500$ (blue), $1000$ (purple), and $1500\,\text{bar}$ (red) obtained from MD $NP_{z}\mathcal{A}T$ simulations using $\xi_{\text{ON}}=1.0$ (open diamonds) and $\xi_{\text{ON}}=1.15$ (open circles). The filled black triangles up ($500\,\text{bar}$), triangles left ($1000\,\text{bar}$), and triangles down ($1500\,\text{bar}$) represent the experimental data taken from the literature.\cite{Wiegand1994} In all cases, the curves are included as guides to the eyes.}
\label{figure3}
\end{figure}

\noindent
In Eq.~\eqref{EQ_tension}, the additional factor $1/2$ arises because of the existence of two interfaces in the system, and $L_z$ is the size of the simulation box along the $z$ direction. Results corresponding to the L$_{\text{w}}$--L$_{\text{N}_{2}}$ interfacial tension of the water + N$_{2}$ binary mixture, at several pressures and temperatures, are shown in Fig.~\ref{figure3}. L$_{\text{w}}$--L$_{\text{N}_{2}}$ interfacial tensions have been obtained by averaging the diagonal components of the pressure tensor over the last $600\,\text{ns}$ without taking into account the first $200\,\text{ns}$ of equilibration. The final $600\,\text{ns}$ have been then divided into $10$ independent blocks of $60\,\text{ns}$ each, and the statistical error has been obtained as the standard deviation of the average.~\cite{Flyvbjerg1989a}

Interfacial tension is an extremely sensible property that strongly depends on the molecular details of the model used. This can be clearly seen in Fig.~\ref{figure3}, where the results for the interfacial tension are clearly different for the two values of $\xi_{\text{ON}}$ considered in this work. An increment of $\xi_{\text{ON}}$ from 1.0 to 1.15 provokes a reduction of $\gamma$ of the $9\%$, approximately. This is true for all the thermodynamic conditions studied in this work. However, as can be seen, the L$_{\text{w}}$--L$_{\text{N}_{2}}$ interfacial tension of the mixture exhibits the same qualitative $\gamma$ for both values of $\xi_{\text{ON}}$, as a function of temperature. Particularly, at a fixed pressure, $\gamma$ decreases when the temperature is increased. However, this dependency on the temperature becomes smaller when the pressure is increased. At 1500 bar, $\gamma$ is practically constant with the temperature. Contrary, at a fixed temperature, $\gamma$ increases when the pressure is reduced.

We have also included experimental data taken from the literature of the interfacial tension of the water + N$_{2}$ binary mixture at the same pressures considered in this work.~\cite{Wiegand1994} Unfortunately, there only exist data at $298\,\text{K}$. As can be seen in Fig.~\ref{figure3}, predictions obtained using $\xi_{\text{ON}}=1.0$ overestimate the experimental values at all pressures. However, when $\xi_{\text{ON}}=1.15$ the agreement between simulation results and experimental data is improved, finding an excellent description of $\gamma$ at 1000 and 1500 bar. At 500 bar, the simulation result slightly underestimates the experimental data by $3\,\text{mN/m}$, approximately. Note also that the variation of the interfacial tension with pressure, at constant temperature, is softer when $\xi_{\text{ON}}=1.0$ than when $\xi_{\text{ON}}=1.15$. This results in a better description of $\gamma$ as a function of pressure at $298\,\text{K}$, as it has been already shown.

\subsection{H--L$_{\text{w}}$ equilibria: Solubility of N$_2$ in liquid water when in contact with N$_2$ hydrate.}

We have also determined the solubility of N$_2$ in the aqueous solution phase in contact with the N$_2$ hydrate phase via a planar interface at the same pressures considered in section A, i.e., $500$, $1000$, and $1500\,\text{bar}$. The initial simulation box is prepared in the same way for the three pressures. First, the N$_2$ hydrate phase is built up by replicating a unit cell of the crystallographic sII structure twice in each spatial direction, forming a $2\times 2\times 2$ unit cell of the solid structure. We also take explicitly into account that hydrates are proton-disordered structures. To obtain this arrangement, hydrogen atoms are placed using the algorithm proposed by Buch \emph{et al.}~\cite{Buch1998a} This allows to generate solid configurations satisfying the Bernal-Fowler rules,~\cite{Bernal1933a} with zero (or at least negligible) dipole moment. The unit cell of a sII hydrate has $136$ molecules of water and $24$ cages. Here we will assume that the hydrate exhibits single occupancy, i.e., only one N$_2$ molecule per cage. According to this, the total hydrate box contains $1088$ molecules of water and $192$ molecules of N$_2$. This hydrate box, with one N$_{2}$ molecule per cage, is equilibrated during 20 ns at each pressure in the $NPT$ ensemble, allowing the system to change independently the dimensions of the simulation box by using an anisotropic barostat. This is necessary in order to avoid any stress in the solid structure. When the N$_2$ hydrate is equilibrated, a pure water phase formed from $2176$ molecules of water (double the number of water molecules than in the hydrate structure) is added to the hydrate phase along the $z$ direction. The final simulation box has $3264$ molecules of water and $192$ molecules of N$_2$. The dimensions of this simulation box used to calculate the solubility of N$_{2}$ in the aqueous solution are $L_x=L_y\approx 3.5$ nm and $L_z\approx 9\,\text{nm}$. We recall that we use the same N$_{2}$ hydrate -- water initial simulation box for the three pressures. 

At this point, it is important to remark that the initial hydrate -- water simulation box could be prepared in a different way. Particularly, it is possible to have a hydrate phase in contact, via a planar interface, with an aqueous solution in which there are some guest molecules dissolved in that phase. This is how we prepared simulation boxes of methane and CO$_{2}$ hydrates in contact with an aqueous solution in our previous works.~\cite{Grabowska2022a,Algaba2023a} The aim of this approach was two-fold: (1) to start the simulations with an initial aqueous phase in which the composition of the guest is close to the equilibrium one when it is in contact with the hydrate phase; (2) to avoid that a considerable part of the hydrate melts to reach the equilibrium liberating guest molecules to the aqueous phase. However, in the current case (N$_2$ hydrate), the solubility of N$_{2}$ in the aqueous phase is low enough and it is possible to use a simulation box with pure water in contact with the N$_{2}$ hydrate phase. When some cages of the hydrate melt, the N$_2$ liberated to the aqueous phase rapidly saturates the system. According to this, we do not include N$_2$ molecules in the initial water phase. Since the initial aqueous phase is pure water, the hydrate must melt in order to release N$_2$ into the aqueous phase until reaches the equilibrium. Since the hydrate phase only can melt to reach the equilibrium, there is no growth process in which a double spontaneous occupancy could take place. Following this procedure, we can ensure that the solubility of N$_2$ in the aqueous phase is the exact solubility reached when a hydrate phase with single occupancy is in contact with an aqueous phase. In order to study the effect of the occupancy on the solubility would be necessary to use a multiple-occupied hydrate phase, but this is out of the scope of the current work.

The simulations are performed in the $NPT$ ensemble with an anisotropic barostat, at the corresponding pressure, to avoid any stress in the hydrate structure. The simulations run for $800\,\text{ns}$, the first $200\,\text{ns}$ for the equilibration of the interface. This first part is not taken into account for the analysis of the results. The density profiles are then obtained from the last $600\,\text{ns}$ following the same procedure as in Section A. Density profiles are obtained by dividing the simulation box into $200$ slabs perpendicular to the $z$ direction. The position of the center of mass of the molecules is assigned to each slab and the density profiles are determined from mass balance considerations.

\begin{figure}
\includegraphics[width=\columnwidth]{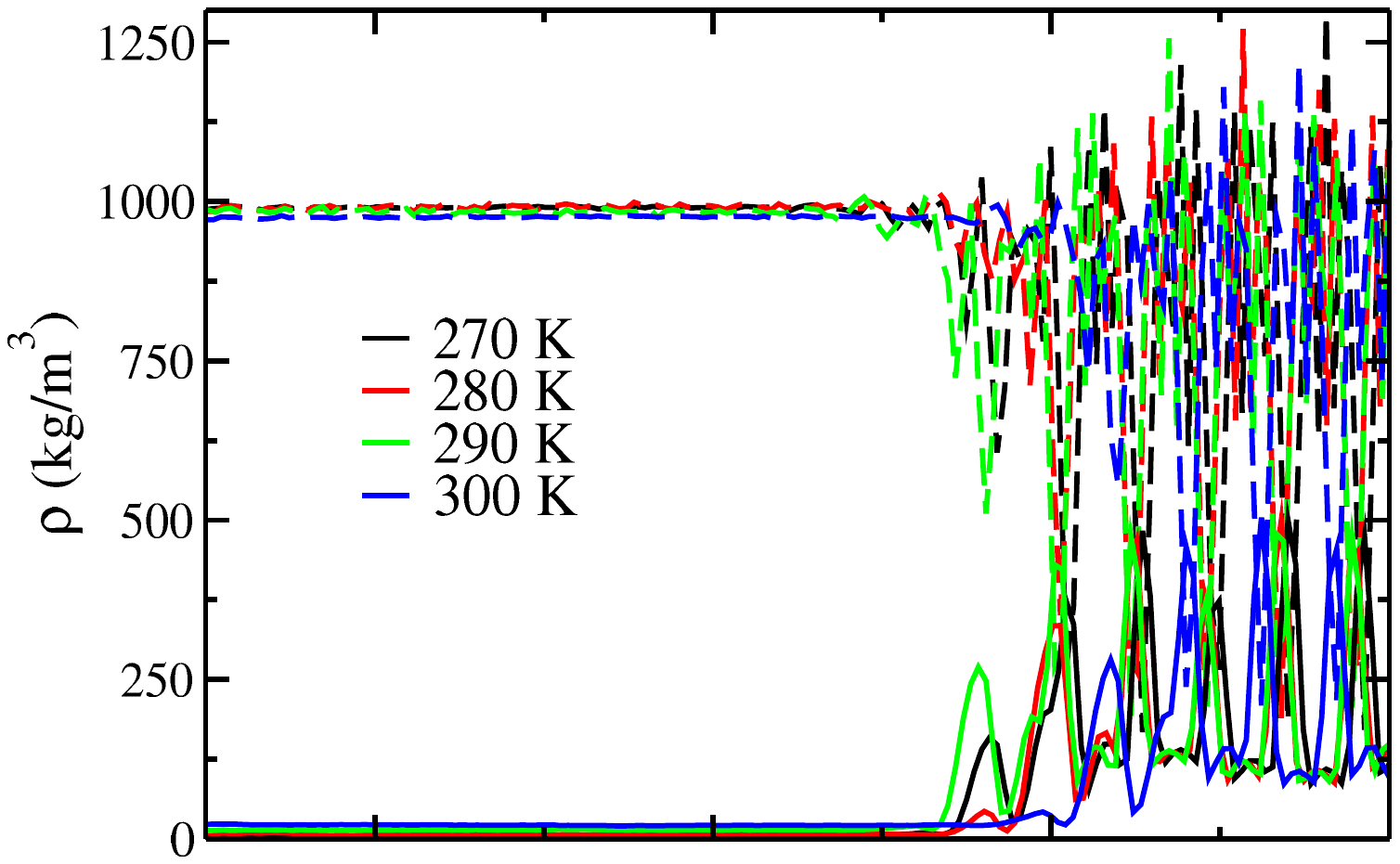}\\

\vspace{-0.9cm}
\includegraphics[width=\columnwidth]{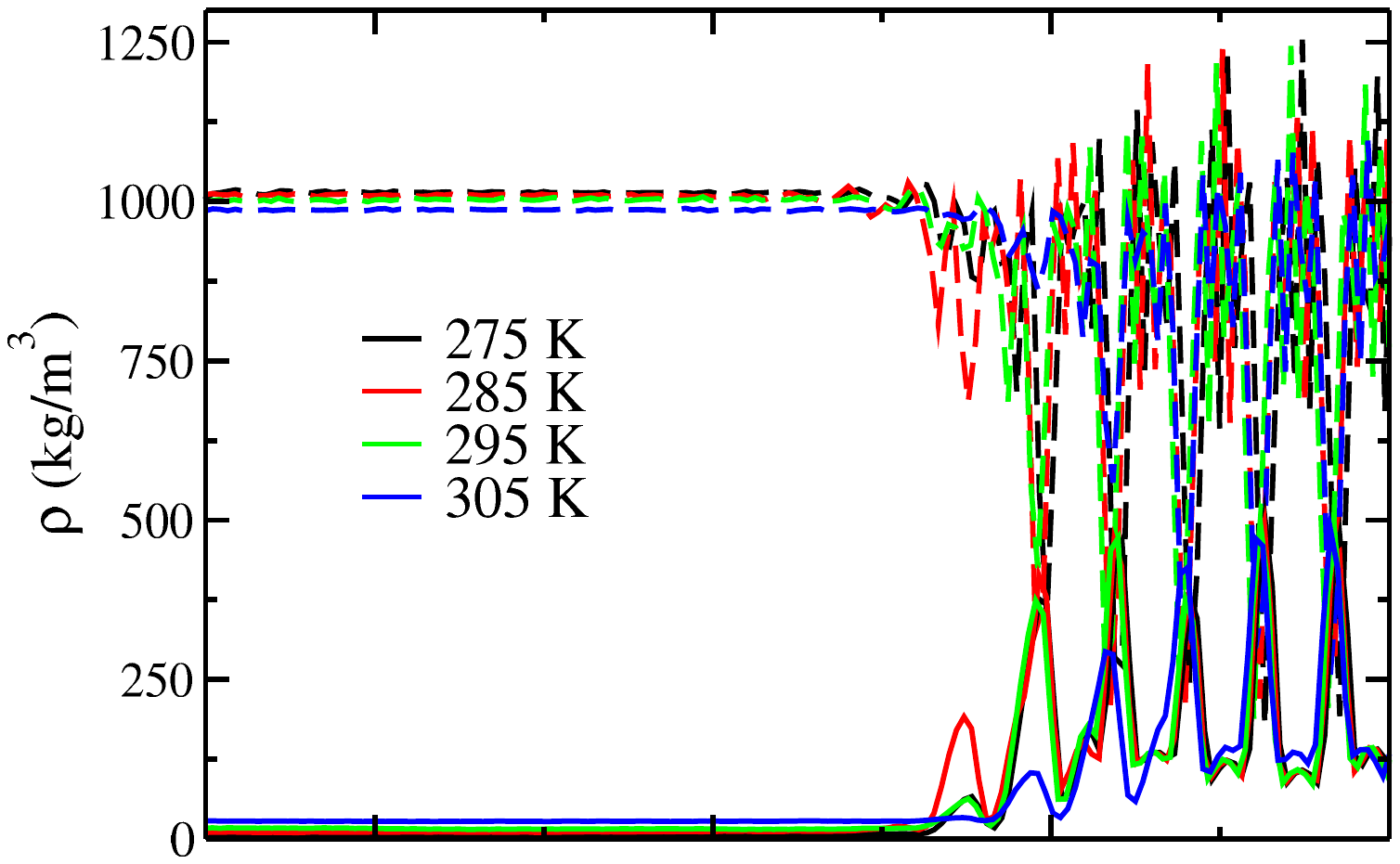}\\

\vspace{-0.9cm}
\includegraphics[width=\columnwidth]{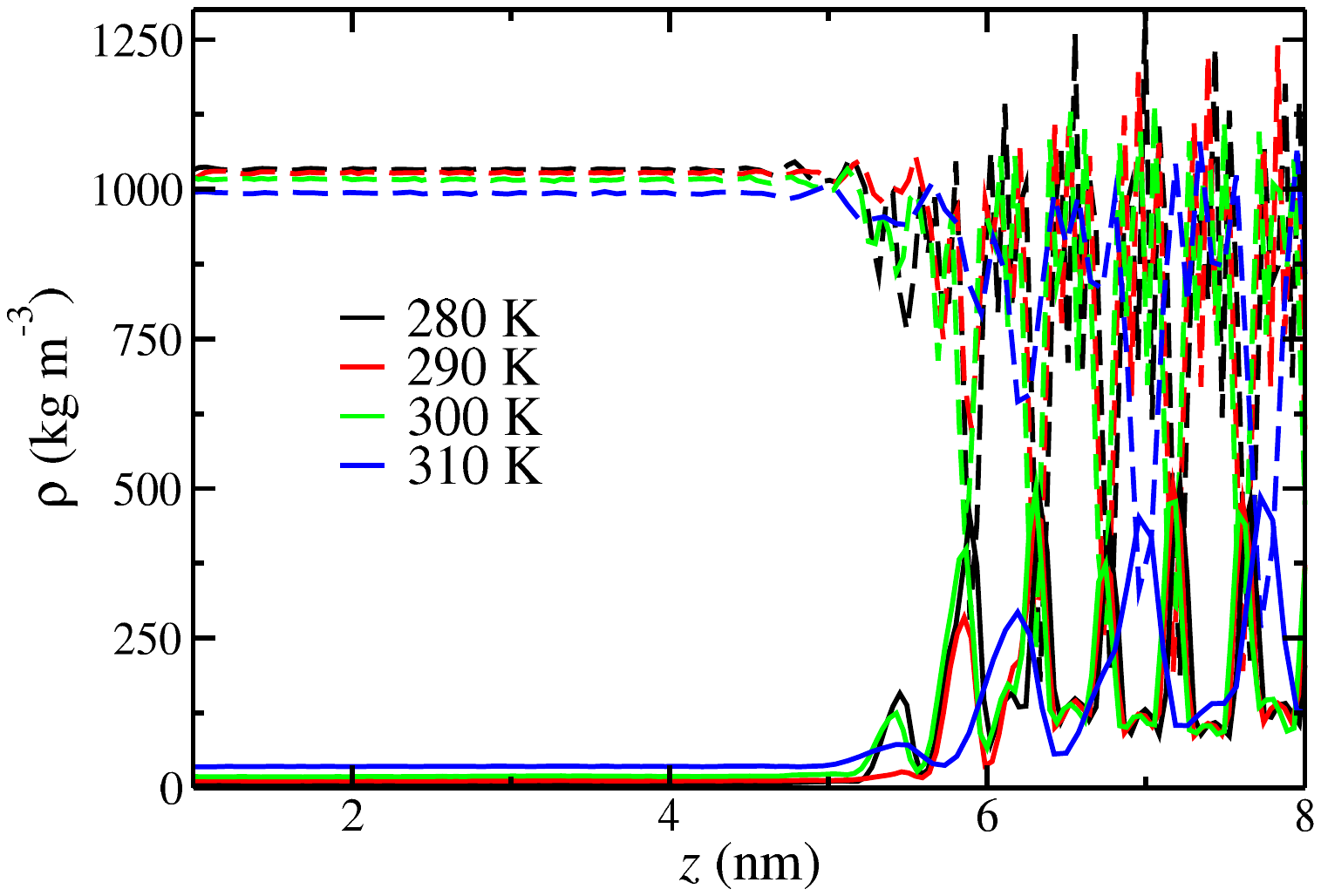}
\caption{Simulated equilibrium density profiles, $\rho(z)$, across the H--L$_{\text{w}}$ interface of N$_{2}$ (continuous curves) and water (dashed curves) as obtained from MD $NP_{z}\mathcal{A}T$ simulations, using $\xi_{\text{ON}}=1.15$, at $500$ (a), $1000$ (b), and $1500\,\text{bar}$ (c) and several temperatures (see legends).}
\label{figure4}
\end{figure}

Fig.~\ref{figure4} shows the equilibrium profiles of N$_{2}$ and water along the hydrate -- water planar interface. In order to avoid repetition, only one of the interfaces is presented. The behavior observed for the three studied pressures, $500$, $1000$, and $1500\,\text{bar}$, and for both values of $\xi_{\text{ON}}$, 1.0 and 1.15, is qualitatively the same. As happens with the L$_{\text{w}}$--L$_{\text{N}_{2}}$ density profiles, the density profiles at the three studied pressures exhibit the same qualitative behavior. We first focus on water density (dashed curves). The density of water in the aqueous phase (left side of the density profiles) decreases when the temperature is increased. The density profiles of water in the hydrate phase (right side of the profiles) show the usual oscillations expected for a solid phase in which the molecules exhibit long-range translational order. Particularly, the peaks observed for the density of water correspond to the equilibrium crystallographic positions occupied by the molecules in the hydrate structure. We also analyze of the density profiles of N$_{2}$ (continuous curves). The show the excepted behavior, in agreement with the density profiles of N$_{2}$. According to the results, the density of N$_{2}$ in the aqueous phase increases as the temperature is increased. 

\begin{figure}
\includegraphics[width=\columnwidth]{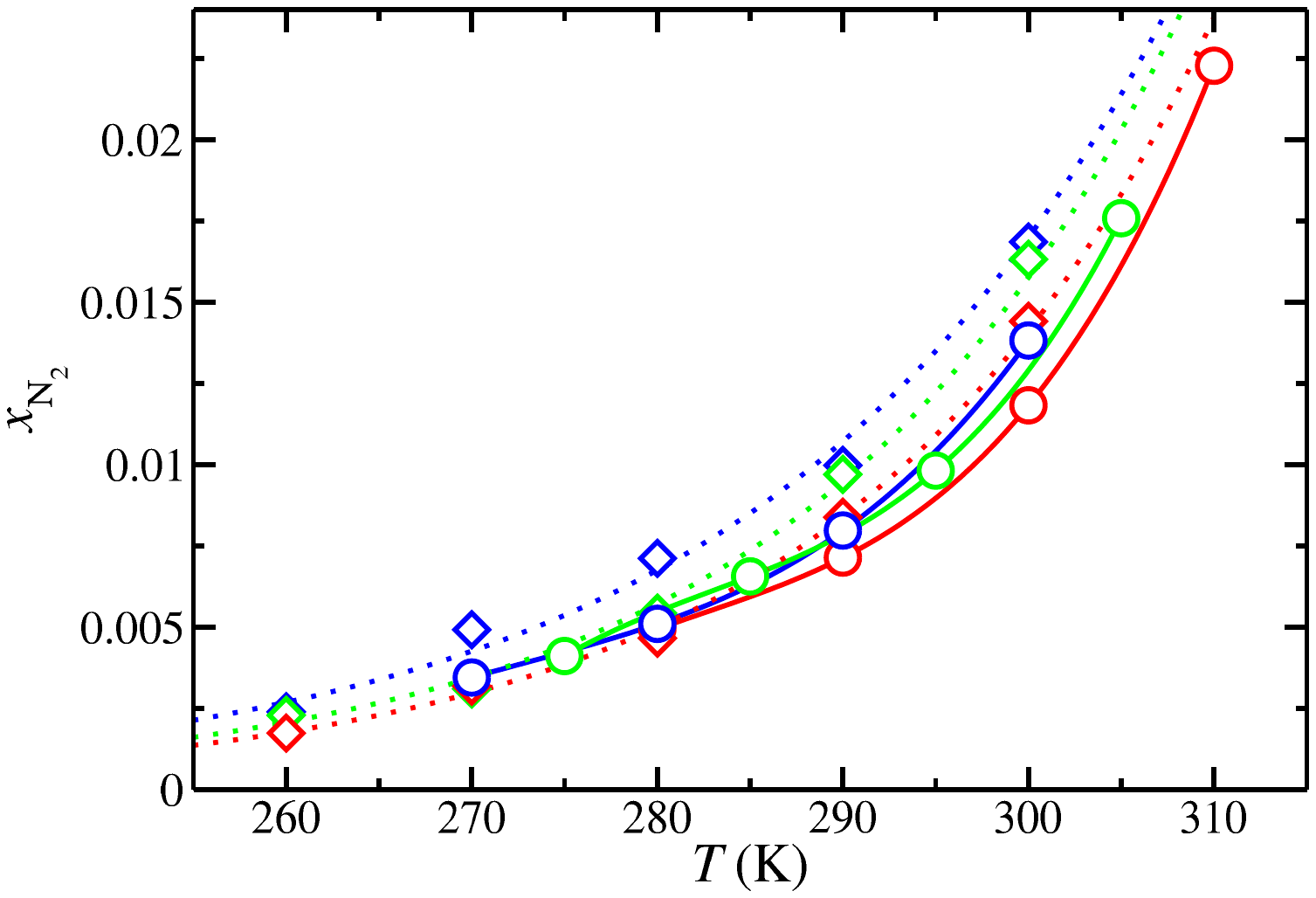}\\

\caption{Solubility of N$_2$ in the aqueous phase, as a function of temperature, at 
$500$ (blue), $1000$ (purple), and $1500\,\text{bar}$ (red) when the solution is in contact with the hydrate phase via a planar interface. The symbols correspond to solubility values obtained from MD $NP_{z}\mathcal{A}T$ simulations using $\xi_{\text{ON}}=1.0$ (open diamonds and dashed curves) and $\xi_{\text{ON}}=1.15$ (open circles and continuous curves). In all cases, the curves are included as guides to the eyes.}
\label{figure5}
\end{figure}

Similarly to the case of the L$_{\text{w}}$--L$_{\text{N}_{2}}$ density profiles, from the analysis of the H--L$_{\text{w}}$ density profiles it is possible to calculate the solubility of N$_2$ in the aqueous phase when it is in contact with the hydrate phase via a planar interface. As can be observed in Fig.~\ref{figure5}, the solubility of N$_2$ increases when the temperature is increased along the isobar solubility curve. Contrary, at constant temperature the solubility of N$_2$ decreases when the pressure is increased. The qualitative behavior of the N$_2$ solubility curves is the same for both values of $\xi_{\text{ON}}$, 1.0 and 1.15. However, as can be observed, an increment of the value of $\xi_{\text{ON}}$ provokes a reduction of the N$_2$ solubility value at all the studied conditions of temperature and pressure.

\subsection{Three phase coexistence line ($T_3$) determination from solubility method}

From the analysis of the solubilities of N$_2$ in the aqueous phase when it is in contact with the N$_2$-rich liquid phase (Section~III.A) and the N$_2$ hydrate phase  (Section~III.B), it is possible to calculate the three-phase coexistence line of this system. At this point, it is important to remark that the three-phase coexistence line, also commonly named the dissociation line of the hydrate, corresponds to the thermodynamic states at temperatures and pressures at which the N$_2$-rich liquid phase, the aqueous solution, and the hydrate phase coexist in equilibrium ($\text{H}-\text{L}_\text{w}-\text{L}_{\text{N}_\text{2}}$). This equilibrium of three phases can be also viewed as two simultaneous two-phase equilibria, hydrate -- aqueous solution equilibrium ($\text{H}-\text{L}_{\text{w}}$) and aqueous solution -- N$_{2}$-rich liquid equilibrium ($\text{L}_{\text{w}}-\text{L}_{\text{N}_\text{2}}$). As has been shown in Sections III.A and III.B, the solubility of N$_2$ in both equilibria depends on the temperature (at fixed pressure). At the $T_3$, the aqueous solution phase must be able to reach the same equilibrium conditions independently if this phase is in contact with the N$_2$-rich liquid phase, the N$_2$ hydrate phase, or with both phases at the same time. In other words, if the isobaric solubility curves of N$_2$ in the aqueous phase when it is in contact with the N$_2$-rich liquid phase and with the hydrate phase are plotted in the same graph, both curves must cross at $T_3$, the dissociation temperature of the hydrate at a given pressure. Note that this methodology has been successfully applied for the determination of the $T_3$ of the methane and carbon dioxide hydrates by some of us.~\cite{Grabowska2022a,Algaba2023a}

\begin{figure}[h!]
\includegraphics[width=\columnwidth]{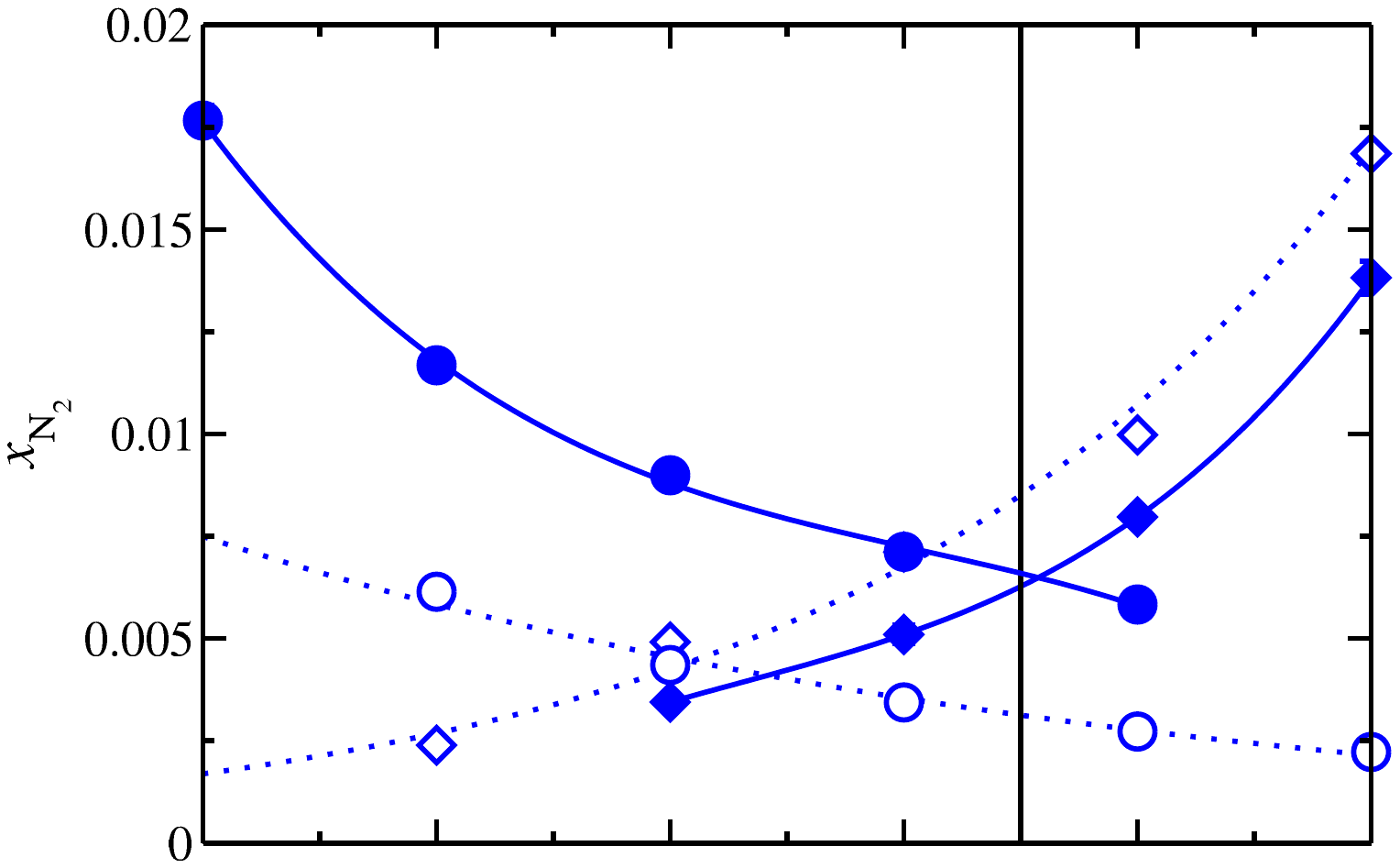}\\
\vspace{-0.9cm}

\includegraphics[width=\columnwidth]{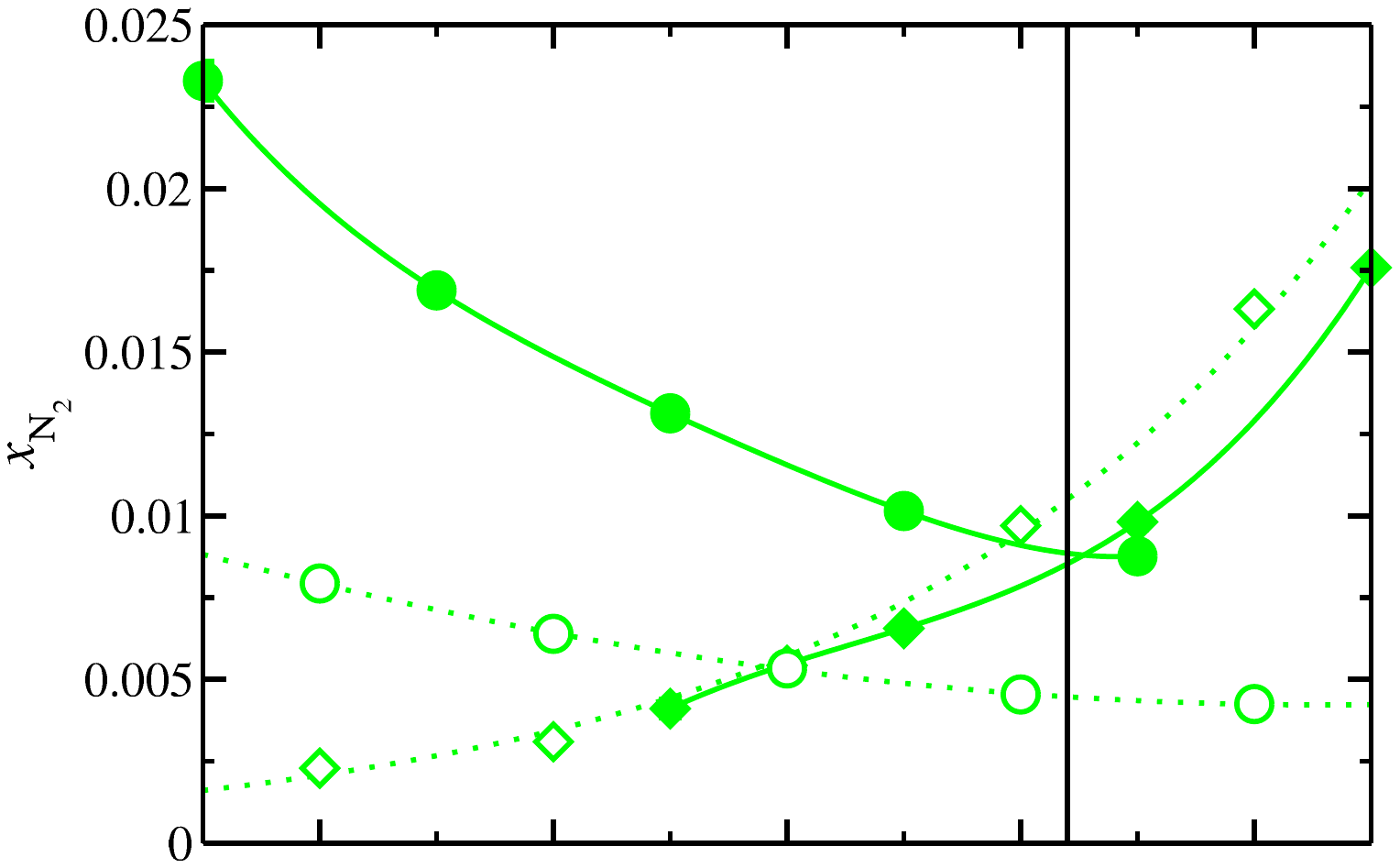}\\

\vspace{-0.9cm}
\includegraphics[width=\columnwidth]{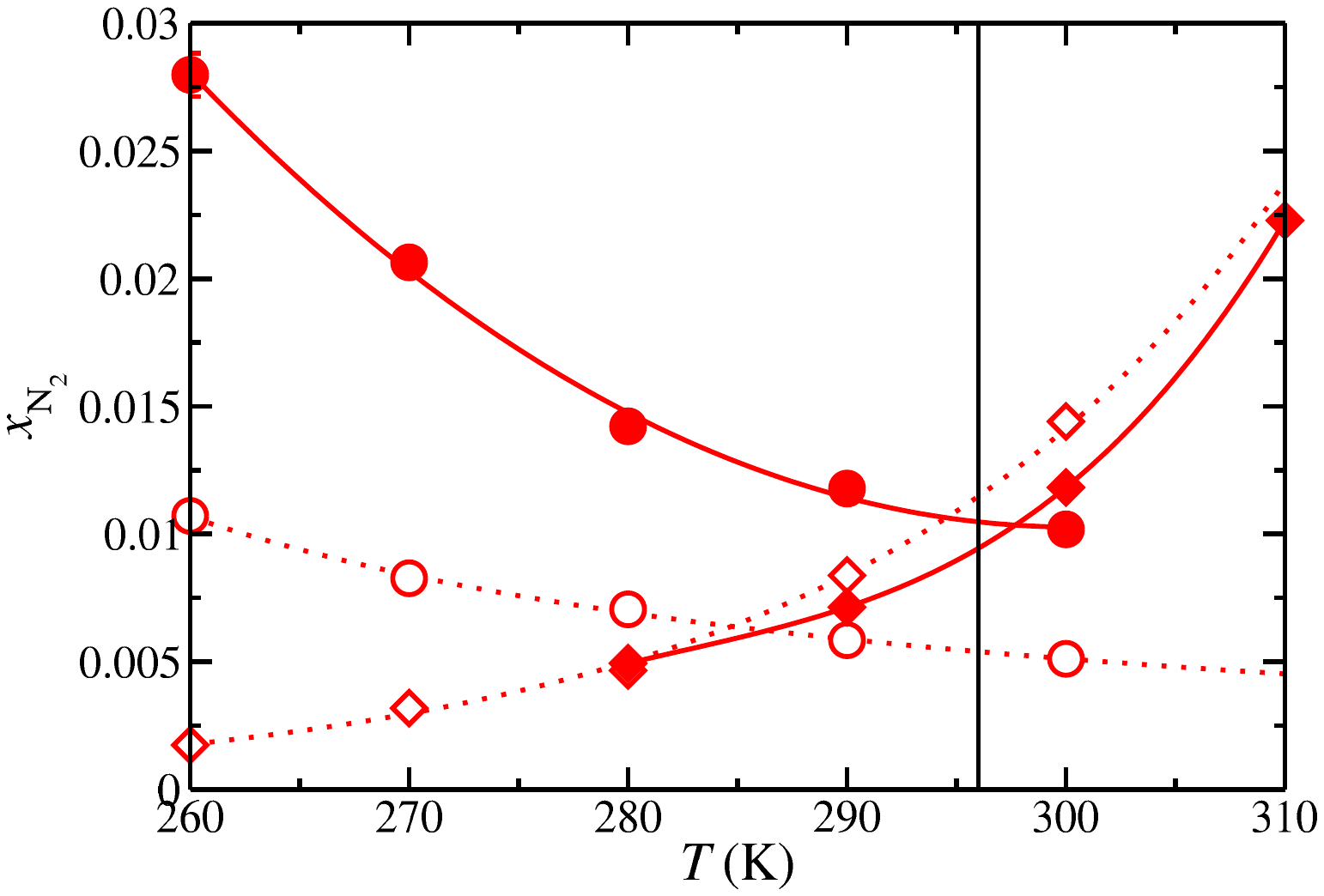}
\caption{Solubility of N$_2$ in the aqueous phase, as a function of temperature, at $500$ (blue), $1000$ (purple), and $1500\,\text{bar}$ (red) when the solution is in contact with the N$_{2}$ liquid and the hydrate via a planar interface. The symbols and colors are the same as those in Figs.~\ref{figure2} and \ref{figure5}. The crossing of the each pair of curves (continuous and dashed) determines the dissociation temperature of the hydrate, $T_3$, at the corresponding pressure. The vertical black lines correspond to the experimental $T_3$ values taken from the literature.\cite{Sugahara2002a,Marshall1964a} In all cases, the curves are included as guides to the eyes.}
\label{figure6}
\end{figure}

Figure \ref{figure6} shows the solubility of N$_2$ in the aqueous phase from the $\text{L}_{\text{w}}-\text{L}_{\text{N}_\text{2}}$ and $\text{H}-\text{L}_{\text{w}}$ equilibria at the three pressure studied in this work. The results obtained for $\xi_{\text{ON}}=1.0$ and 1.15 at each pressure have been represented together in order to analyze the effect of the modification of the Berthelot rule. As can be seen, in the three cases the solubility curves follow the same qualitative behavior: the solubility of N$_{2}$ in the aqueous phase when it is in contact with the N$_{2}$-rich liquid phase decreases with the temperature, as it happens with the solubility of most gases in water; contrary, the solubility of N$_{2}$ when it is contact with the hydrate phase increases with temperature. Note that this behavior with temperature is similar to that observed when a solid is dissolved in water.

We first concentrate on the results obtained at $500\,\text{bar}$. The experimental value of $T_3$ taken from the literature at this pressure is $284.5\,\text{K}$.~\cite{Sugahara2002a,Marshall1964a} The values obtained in this work using $\xi_{\text{ON}}=1.0$ and 1.15 are $271(2)$ and $286(2)\,\text{K}$, respectively. As can be seen, the value of $T_3$ obtained using $\xi_{\text{ON}}=1.0$ clearly underestimates the $T_3$ value by a $4.8\%$. However, it is important to remark that this value represents a pure prediction obtained from the parameters of the pure components. In other words, unlike interactions are pure predictions from the Bertherlot combining rule ($\xi_{\text{ON}}=1.0$). A difference of just $4.8\%$ is still a good result taking into account the complexity of the systems. The use of a larger value of $\xi_{\text{ON}}$, making the unlike interactions with N$_{2}$ and water more favorable, helps to predict the $T_{3}$ of the system better. Particularly, the agreement between the experimental value and the simulation result when $\xi_{\text{ON}}=1.15$ is excellent, since predicted values are inside the uncertainty ($286\pm 2\,\text{K}$). 

We now turn on the results obtained at $1000\,\text{bar}$ using $\xi_{\text{ON}}=1.0$ and 1.15. The $T_{3}$ values obtained for these values unlike dispersive interactions are $279(2)$ and $293(2)\,\text{K}$, respectively. The experimental value of the $T_3$ at $1000\,\text{bar}$ taken from the literature is $292\,\text{K}$.~\cite{Sugahara2002a,Marshall1964a} As it happens at $500\,\text{bar}$,
the computer simulation predictions obtained using $\xi_{\text{ON}}=1.15$ are in excellent agreement with the experimental value of $292\,\text{K}$ and inside the error bar. As it also occurs with the previous pressure, predictions using the original Berthelot rule ($\xi_{\text{ON}}=1.0$) slightly underestimates the experimental $T_3$ value (by a $4.5\%$).

Finally, the results obtained at $1500\,\text{bar}$ follow the expected behavior observed at $500$ and $1000\,\text{bar}$. Particularly, our simulation results indicate that the dissociation temperature of the N$_{2}$ hydrate at this pressure using $\xi_{\text{ON}}=1.15$ is $298(2)\,\text{K}$. Note that 
this value is also in excellent agreement with the experimental value of the dissociation temperature at this pressure, $296\,\text{K}$.~\cite{Sugahara2002a,Marshall1964a} Note that using the $\xi_{\text{ON}}=1.0$ value for unlike dispersive interactions we obtain the value $T_{3}=285(2)\,\text{K}$, that only underestimates the experimental value by a $3.8\%$. 

\begin{figure}
\includegraphics[width=\columnwidth]{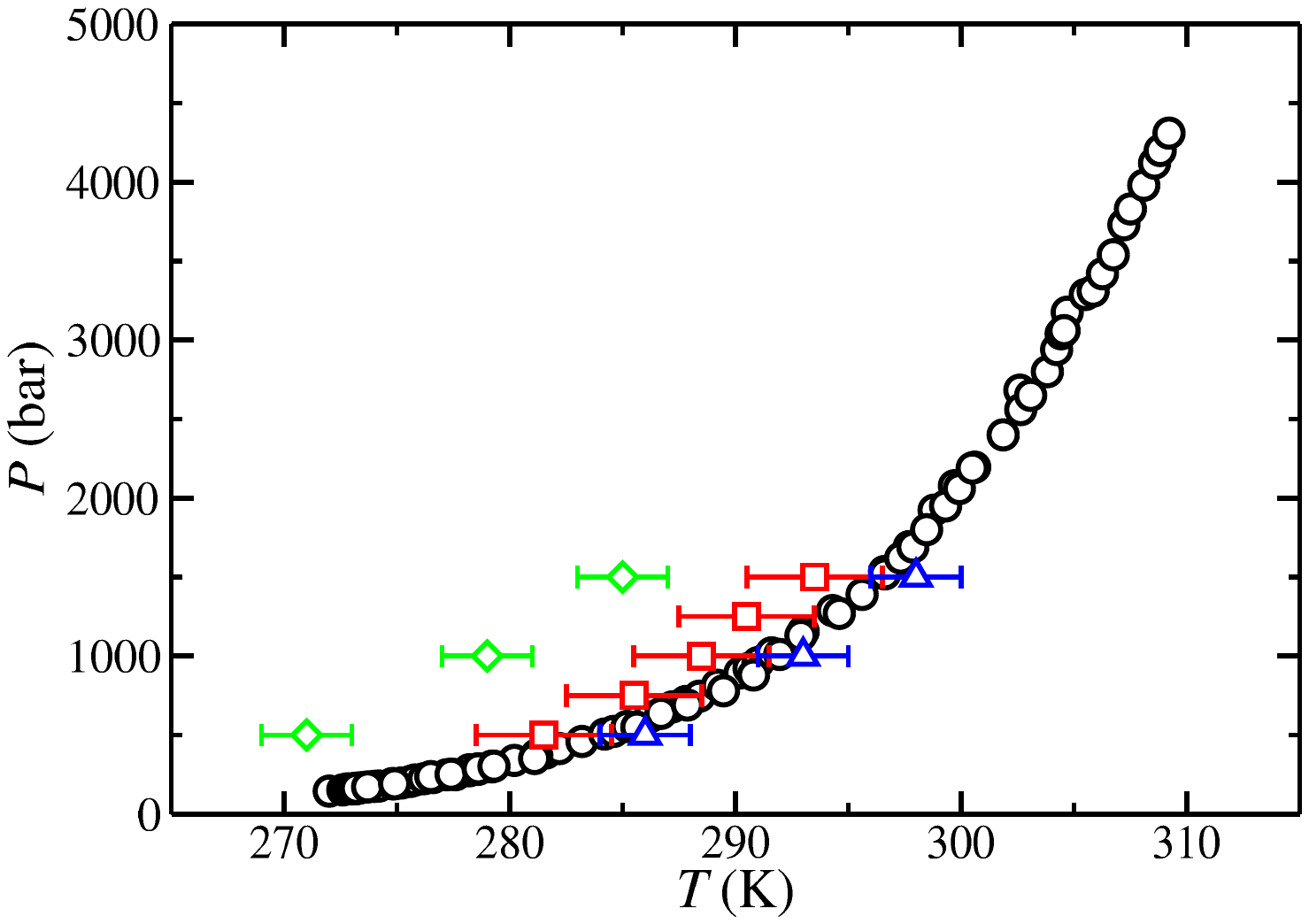}\\
\caption{Pressure-temperature or $PT$ projection of the dissociation line of the N$_{2}$ hydrate. purple diamonds and blue triangles up correspond to the results obtained in this work using the solubility method, the TIP4P/Ice model for water and the TraPPE model for N$_{2}$ with the original Berthelot rule $\xi_{\text{ON}}=1.0$ and the modified Bertherlot rule $\xi_{\text{ON}}=1.15$, respectively. Red squares are the results obtained by Yi \emph{et al.}~\cite{Yi2019a} using the direct coexistence technique, the TIP4P/2005 model for water, and the 2CLJQ model for N$_{2}$. Open black circles correspond to experimental data taken from the literature.~\cite{vanCleeff1960a,Marshall1964a,Jhaveri1965a,Sugahara2002a,Mohammadi2003a}}
\label{figure7}
\end{figure}

Figure~\ref{figure7} summarizes one of the most important results obtained in this work, the $PT$ projection of the dissociation line of the N$_{2}$ hydrate as obtained from molecular simulation. We also include experimental data taken from the literature~\cite{vanCleeff1960a,Marshall1964a,Jhaveri1965a,Sugahara2002a,Mohammadi2003a} to assess the accuracy of the simulation results. As can be seen, the results obtained using the original Berthelot rule $\xi_{\text{ON}}=1.0$ (purple diamonds) only provide a qualitative description of the three-phase line of the hydrate, systematically underestimating the dissociation line. A more accurate description of the dissociation line is obtained if the modified Bertherlot rule $\xi_{\text{ON}}=1.15$ is used (blue triangles up). It is important to take into account the previous work of Yi \emph{et al.},~\cite{Yi2019a} in which the authors have already studied the dissociation line of the N$_2$ hydrate combining molecular dynamics simulations and the direct coexistence technique (red squares). These authors use different models for water and N$_{2}$. Particularly, they use the TIP4P/2005 model for water molecules, more suitable to describe pure water and aqueous solutions in the absence of solid phases,~\cite{Abascal2005a} and a 2CLJQ model for N$_{2}$ molecules.~\cite{Vrabec2001a} As we have already mentioned in the Introduction, their results systematically underestimate the experimental data by $4\,\text{K}$ at pressures above $500\,\text{bar}$. Particularly, the simulation predictions of Yi and coworkers are slightly outside the error bars. However, as Yi \emph{et al.} claim,~\cite{Yi2019a}  the equilibrium curve obtained by them is found to be generally parallel to that obtained from experiments, suggesting that the models used by these authors are able to predict qualitatively the three-phase coexistence behavior of the system.

To recap, all the predictions obtained in this work at $500$, $1000$, and $1500\,\text{bar}$ using the Berthelot rule without the use of a modified unlike intermolecular dispersive energy, provide a good agreement with experimental data taken from the literature. Particularly, all deviations underestimate the experimental $T_{3}$ values by less than $4.8\%$. However, as we have already mentioned, it is possible to improve the simulation results using a modified Berthelot rule for the unlike dispersive interactions. This allows to predict, within the uncertainty of the simulations, the dissociation line of the N$_{2}$ hydrate in the whole range of pressures considered in this study.

As we have already mentioned, this is the first work in which the solubility method is used to obtain the $T_{3}$ of a hydrate that exhibits type sII crystallographic structure. According to this, we have only considered single occupancy and this is why we have not studied pressures above $1500\,\text{bar}$. Note that, according to the literature, multiple occupancy is expected at high pressures. It would be really interesting to consider the effect of multiple occupancy on the dissociation line of the hydrate. However, this is not a simple task. Multiple occupancy studies in hydrate systems are complicated to be undertaken by simulation. Recently, Michalis \emph{et al.} [\emph{J. Chem. Phys.} \textbf{157}, 154501, 2022] have determined the dissociation line of the H$_2$ hydrate at several pressures but only considering single occupancy. According to the authors, it is complicated to stabilize the hydrate structure when a multiple occupancy configuration is used. In fact, they couldn't stabilize the corresponding simulation boxes, which are similar to those used in this work since the H$_{2}$ hydrate also exhibits the type sII crystallographic structure. In a previous work of some of us,$^{49}$ the effect of the occupancy of the CO$_2$ hydrate was studied from the solubility method. According to these results, the effect of the occupancy on the solubility of CO$_{2}$ when in contact with the hydrate is negligible, and consequently, also on the $T_{3}$ of the CO$_{2}$ hydrate. However, the effect of multiple occupancy on the driving force for nucleation is noticeable. We think the same conclusions can be extrapolated to the N$_{2}$ but we also think more efforts are required in order to study the effect of multiple occupancies on the stability conditions of hydrates that exhibit type sII crystallographic structure. However, this is out of the scope of this work.

\subsection{Driving force for nucleation $\Delta\mu^{\text{EC}}_{N}$}

At the dissociation line of the hydrate, the N$_2$ liquid phase, the hydrate, and the aqueous solution are in equilibrium at a temperature $T_{3}$ (at a given pressure). Two possible scenarios can arise at this fixed pressure if temperature changes. If $T<T_3$, the hydrate is the most stable phase and it will grow until extinguishing one of the components (water or N$_2$) of the other two phases. The non-extinguished component will form a second phase in equilibrium with the hydrate. Contrary, if $T>T_3$, the liquid phases are more stable than the hydrate. As a consequence, the hydrate melts in two immiscible liquid phases in equilibrium (aqueous solution and N$_2$-rich liquid). However, as we have previously shown, it is possible to have two liquids in equilibrium via a planar interface, the aqueous solution and the N$_2$ liquid phase, at temperatures below the dissociation temperature of the hydrate at a given pressure ($T<T_3$). In this situation, the liquid phases are metastable compared with the hydrate phase. How the liquid phases can exist if they are thermodynamically less stable than the hydrate?

The explanation is that the formation of the hydrate from the aqueous solution of N$_{2}$ is an activated process. According to Classical Nucleation Theory (CNT),~\cite{Debenedetti1996a} the solid phase begins with the formation of a small cluster of ``solid'' molecules immersed in the liquid. The formation of this cluster is only possible if the system can break through a free energy barrier, which depends on the hydrate--aqueous solution interfacial free energy and the difference in chemical potential between the hydrate and the aqueous solution, usually known as the driving force for nucleation $\Delta\mu_{\text{N}}$. According to the CNT, the free energy barrier becomes smaller when the temperature is below the $T_{3}$ (supercooling conditions) and when the concentration of N$_2$ in the aqueous phase is increased (supersaturated conditions). In both cases, the driving force for nucleation increases (the difference in chemical potential becomes more negative). Obviously, this provokes that the liquid phases become less stable at the same time that the hydrate phase becomes more stable. This is also corroborated by the inspection of Fig.~\ref{figure2}: when $T$ is reduced, the solubility of N$_2$ in the aqueous phase is increased, reducing the stability of the liquid phases and increasing the stability of the hydrate phase. 

According to the literature,~\cite{Kashchiev2002a,Grabowska2022a,Algaba2023a} the formation of a hydrate in an aqueous solution can be viewed as a chemical reaction at constant $T$ and $P$,

\begin{equation}
\text{N}_{2} (\text{aq},x_{\text{N}_{2}}) +
5.67\,\text{H}_{2}\text{O} (\text{aq},x_{\text{N}_{2}}) 
\rightarrow [\text{N}_{2}(\text{H}_{2}\text{O})_{5.67}]_{\text{H}}
\label{reaction}
\end{equation}

\noindent Note that the compound $[\text{N}_{2}(\text{H}_{2}\text{O})_{5.67}]_{\text{H}}$ is the ``hydrate'' and we call one ``molecule'' of the hydrate in the solid the molecule $[\text{N}_{2}(\text{H}_{2}\text{O})_{5.67}]$. The 5.67 factor arises because of the stochiometric of the hydrate structure (the hydrate unit cell is built by 136 molecules of water and 24 molecules of N$_2$ or equivalently by $136/24\approx 5.67$ molecules of water per each molecule of N$_{2}$). Since we perform simulations at three different pressures, all quantities depend not only on temperature but also on pressure. Due to this, we keep in our notation the explicit dependency on $T$ and $P$. In this work, we assume that the occupancy is one molecule of N$_2$ in each cage of the hydrate, i.e., full single occupancy. Following the previous works of Grabowska~\emph{et~al.}~\cite{Grabowska2022b} and Algaba,~\emph{et~al.}~\cite{Algaba2023a} the chemical potential of a ``molecule'' of hydrate, as a function of $T$ and $P$, is given by,

\begin{equation}
\mu^{\text{H}}_{\text{H}}(P,T)=\mu_{\text{N}_{2}}^{\text{H}}(P,T)+5.67\,\mu_{\text{H}_{2}\text{O}}^{\text{H}}(P,T)
\label{chempot_hydrate}
\end{equation}

\noindent where $\mu_{\text{N}_{2}}^{\text{H}}(P,T)$ and $\mu_{\text{H}_{2}\text{O}}^{\text{H}}(P,T)$ are the chemical potentials of N$_2$ and water in the hydrate phase, respectively. Note that since we are assuming that the hydrate is fully occupied, the chemical potential of the hydrate does not depend on composition. Using Eq.~\eqref{chempot_hydrate}, the driving force for nucleation can be written as,~\cite{Kashchiev2002a,Grabowska2022a,Algaba2023a}

\begin{align}
\Delta\mu_{\text{N}}(P,T,x_{\text{N}_{2}})&=\mu^{\text{H}}_{\text{H}}(P,T) \nonumber\\
& -\mu^{\text{aq}}_{\text{N}_{2}}(P,T,x_{\text{N}_{2}})-5.67\,\mu^{\text{aq}}_{\text{H}_{2}\text{O}}(P,T,x_{\text{N}_{2}})
\label{driving_force}
\end{align}

\noindent $\mu^{\text{H}}_{\text{H}}(P,T)$ is the chemical potential of the hydrate defined in Eq.~\eqref{chempot_hydrate}, $\mu^{\text{aq}}_{\text{N}_{2}}(P,T,x_{\text{N}_{2}})$ is the chemical potential of N$_2$ in the aqueous phase and $\mu^{\text{aq}}_{\text{H}_{2}\text{O}}(P,T,x_{\text{N}_{2}})$ is the chemical potential of water in the aqueous phase. Note that we have expressed both chemical potentials in the aqueous solution as functions of $x_{\text{N}_{2}}$ since the composition of water is fully determined since we are dealing with a two-component system and $x_{\text{H}_{2}\text{O}}=1-x_{\text{N}_{2}}$.

At this point, it is important to remark that $\Delta\mu_{\text{N}}$ can be calculated at any value of $P$, $T$, and $x_{N_2}$. However, it is particularly interesting to investigate $\Delta\mu_{\text{N}}$ along the $\text{L}_{\text{w}}-\text{L}_{\text{N}_\text{2}}$ coexistence curve, conditions at which nucleation experiments are performed.~\cite{Grabowska2022a,Algaba2023a} Notice that the $\text{L}_{\text{w}}-\text{L}_{\text{N}_\text{2}}$ isobar coexistence curves at $500$, $1000$, and $1500\,\text{bar}$ have been previously obtained in this work. According to this, $x_{\text{N}_2}$ is a function of $P$ and $T$. Following the notation used by some of us in previous works,~\cite{Grabowska2022a,Algaba2023a} the driving force for nucleation at experimental conditions can be written as,

\begin{align}
\Delta\mu^{\text{EC}}_{\text{N}}(P,T)&=\mu^{\text{H}}_{\text{H}}(P,T) -\mu^{\text{aq}}_{\text{N}_{2}}(P,T,x_{\text{N}_{2}}^{\text{eq}}(P,T)) \nonumber\\
& -5.67\,\mu^{\text{aq}}_{\text{H}_{2}\text{O}}(P,T,x_{\text{N}_{2}}^{\text{eq}}(P,T))
\label{driving_force_EN}
\end{align}

\noindent Note that at $T_{3}$ all the chemical potentials are equal since all the phases are in thermodynamic equilibrium and $\Delta\mu^{\text{EC}}_{\text{N}}=0$. In addition to this, since we are interested in changes in chemical potentials and not in individual values, we can set to zero the chemical potential of N$_{2}$, water, and hydrate at $T_{3}$ at each pressure considered in this work.

In the next sections, we present two different routes for the calculation of $\Delta\mu^{\text{EC}}_{\text{N}}(P,T)$. These routes have been used by some of us in previous works to determine the driving force for nucleation of methane and carbon dioxide hydrates at experimental conditions.~\cite{Grabowska2022a,Algaba2023a} ~\cite{Debenedetti1996a} Also, it is important to point out, that all the driving force results obtained in this work have been obtained when $\xi_{\text{ON}}=1.15$.

\subsubsection{$\Delta\mu^{\text{EC}}_{\text{N}}$ via route 1}

The use of Route 1, originally proposed by Grabowska and collaborators,~\cite{Grabowska2022a} implies the calculation of the chemical potential of the ``hydrate molecule'' in the hydrate phase, as well as the calculation of the chemical potentials of water and N$_2$ in the aqueous phase at temperatures $T$ below the $T_3$. We first concentrate on the calculation of the chemical potential of N$_2$ in the aqueous phase. Along the $\text{L}_{\text{w}}-\text{L}_{\text{N}_\text{2}}$ solubility curve (at fixed pressure), the two liquid phases coexist in equilibrium, the aqueous solution phase and N$_2$-rich liquid phase. We have already checked that the solubility of water in the N$_2$-rich liquid phase is negligible (see Fig.~\ref{figure1}). According to this, we can assume that this phase is formed only from pure N$_2$. Since both phases are in equilibrium, the chemical potential of N$_2$ in the aqueous phase, $\mu_{\text{N}_{2}}^{\text{aq}}(P,T,x_{\text{N}_{2}}^{\text{aq}})$, and in the pure N$_2$ phase, $\mu_{\text{N}_{2}}(P,T)$, should be the same. Consequently, the chemical potential of N$_2$ in the aqueous phase can be expressed as,

\begin{equation}
\mu_{\text{N}_{2}}^{\text{aq}}(P,T,x_{\text{N}_{2}}^{\text{aq}})\approx
\mu_{\text{N}_{2}}^{\text{aq}}(P,T,x_{\text{N}_{2}}^{\text{aq}}\approx 1)\approx
\mu_{\text{N}_{2}}(P,T)
\end{equation}

\noindent
where $\mu_{\text{N}_{2}}(P,T)$ represents the chemical potential of pure N$_{2}$ at temperature $T$ and pressure $P$. It is possible to obtained the value of $\mu_{\text{N}_{2}}(P,T)$ along an isobar (constant pressure) from the knowledge of the enthalpy of pure N$_2$, $h_{\text{N}_{2}}(P,T)$, in the N$_2$ phase via the well-known Gibbs-Helmholtz thermodynamic relation for pure systems,

\begin{equation}
\Biggl(\dfrac{\partial(\mu_{\text{N}_{2}}(P,T)/T)}{\partial T}\Biggr)_{P,N_{\text{N}_{2}}}=-\dfrac{h_{\text{N}_{2}}(P,T)}{T^{2}}
\label{enthalpy}
\end{equation}

\noindent Note that the derivative of the chemical potential of pure N$_{2}$ is performed at constant pressure and number of molecules. Finally, it is possible to calculate the variation of $\mu_{\text{N}_{2}}(P,T)$ from its value at $T_{3}$ by integrating Eq.~\eqref{enthalpy} at constant pressure $P$,

\begin{equation}
\mu_{\text{N}_{2}}^{\text{aq}}(P,T,x_{\text{N}_{2}}^{\text{aq}})\approx
\dfrac{\mu_{\text{N}_{2}}(P,T)}{k_{B}T}=\dfrac{\mu_{\text{N}_{2}}(P,T_{3})}{k_{B}T_{3}}-\bigintsss_{T_{3}}^{T} \dfrac{h_{\text{N}_{2}}(P,T')}{k_{B}T'^{2}}\,dT'
\label{integral_N2}
\end{equation}

\noindent Since $\mu_{\text{N}_{2}}(P,T_{3})=0$, Eq.~\eqref{integral_N2} can be written as,

\begin{equation}
\mu_{\text{N}_{2}}^{\text{aq}}(P,T,x_{\text{N}_{2}}^{\text{aq}})\approx
\dfrac{\mu_{\text{N}_{2}}(P,T)}{k_{B}T}=
-\bigintsss_{T_{3}}^{T} \dfrac{h_{\text{N}_{2}}(P,T')}{k_{B}T'^{2}}\,dT'
\label{integral_N2_2}
\end{equation}

To evaluate $h_{\text{N}_{2}}(P,T)$, we use a simulation box with $1000$ molecules of N$_2$. We perform MD simulations at constant temperature and pressure in the isothermal-isobaric or $NPT$ ensemble. We use an isotropic Parrinello-Rahman barostat to fix the pressure that allows to modify the volume isotropically. At each pressure, $500$, $1000$, and $1500\,\text{bar}$, we consider several temperatures, from $250$ to $290\,\text{K}$ at $500\,\text{bar}$, from $255$ to $295\,\text{K}$ at $1000\,\text{bar}$, and from $260$ to $300\,\text{K}$ at $1500\,\text{bar}$. The simulation box size, $L_{x}=L_{y}=L_{z}=L$, varies from $L=4.85$ to $4.11\,\text{nm}$ depending on $T$ and $P$. The simulations are run during $100\,\text{ns}$. The first $20\,\text{ns}$ are taken as the equilibration time, and only the last $80\,\text{ns}$ are used to evaluate $h_{\text{N}_{2}}(P,T)$. 

As in the case of the calculation of the chemical potential of N$_{2}$, it is possible to calculate the chemical potential of the N$_{2}$ hydrate integrating the corresponding Gibbs-Helmholtz thermodynamic expression, at constant pressure, as, 

\begin{equation}
\dfrac{\mu_{\text{H}}^{\text{H}}(P,T)}{k_{B}T}=
-\bigintsss_{T_{3}}^{T} \dfrac{h_{\text{H}}^{\text{H}}(P,T')}{k_{B}T'^{2}}\,dT'
\label{integral_H}
\end{equation}

\noindent It is important to recall that integration of Eq.~\eqref{integral_H} is performed at constant pressure since we evaluate the chemical potential along a given isobar. Here $h_{\text{H}}^{\text{H}}(P,T')$ represents the enthalpy of the hydrate at pressure $P$ and temperature $T'$. Since the stoichiometry of the hydrate is fixed and we are assuming full occupancy, the hydrate can be considered a pure compound. According to this, the values of $h_{\text{H}}^{\text{H}}$ needed to calculate the chemical potential are obtained in the same way as those corresponding to the N$_{2}$ liquid phase: we perform MD simulations of N$_2$ hydrate in the isothermal-isobaric or $NPT$ ensemble. The simulation box is formed by replicating the unit cell of the N$_2$ hydrate twice in each space direction obtaining a $2\times 2\times 2$ configuration. This corresponds to a simulation box that contains $1088$ molecules of water and $192$ molecules of N$_2$. Similarly, the pressure is fixed using the isotropic Parrinello-Rahman barostat since the sII structure of the N$_{2}$ hydrate exhibits cubic symmetry. The simulations are performed at the same pressures and temperatures as those corresponding to pure N$_2$. In this case, the simulation box size is $L_x=L_y=L_z\approx3.46\,\text{nm}$. According to Eq.~\eqref{reaction}, $h_{\text{H}}^{\text{H}}(P,T')$ is obtained by dividing the total enthalpy of the simulation by the number of molecules of N$_2$, or the number of hydrate cages since we are assuming full but single occupancy per cage. The simulations are also run during $100\,\text{ns}$. The first $20\,\text{ns}$ are taken as the equilibration time, and only the last $80\,\text{ns}$ are taken as the production period.

The chemical potential of water in the aqueous phase, $\mu_{\text{H}_{\text{2}}\text{O}}^{\text{aq}}$, can not be determined with the same strategy used to calculate $\mu_{\text{N}_{\text{2}}}^{\text{aq}}$ and $\mu^{\text{H}}_{\text{H}}$. Unfortunately, an accurate prediction of the chemical potential of water critically depends on the solubility of N$_{2}$ and this effect should be taken into account.~\cite{Grabowska2022a,Algaba2023a} To overcome this issue, here we follow the same approach introduced by some of us for the calculation of driving forces for nucleation of methane and carbon dioxide hydrates.~\cite{Grabowska2022a,Algaba2023a}

According to that, $\mu_{\text{H}_{\text{2}}\text{O}}^{\text{aq}}$ is obtained in two steps. In the first step, we assume that the aqueous phase is formed from pure water since the solubility of N$_{2}$ is very small and calculate the change in the chemical potential of water in solution when the temperature varies from $T_{3}$ to $T$. This can be done using the same procedure previously used for the calculation of the chemical potential of N$_2$ in the aqueous solution and the chemical potential of the hydrate, i.e., calculating the molar enthalpy of pure water as a function of pressure and temperature, $h_{\text{H}_{\text{2}}\text{O}}(P,T)$. To this end, we perform MD simulations in the isothermal-isobaric or $NPT$ ensemble using an isotropic Parrinello-Rahman barostat. The simulation box is formed from $1000$ molecules of water. We perform simulations at the same thermodynamic conditions of $T$ and $P$ as in the case of pure liquid N$_2$ and N$_2$ hydrate. The simulation box size is $L_x=L_y=L_z\approx3.08\,\text{nm}$. As in the previous cases, the simulations run for $100\,\text{ns}$ taking the last $80\,\text{ns}$ as the production period. Note that the final change in the chemical potential of pure water when the temperature varies from $T_{3}$ and $T$ (at constant pressure) can be easily obtained by integrating the corresponding Gibbs-Helmholtz thermodynamic relation. 

In the second step, we take approximately into account that the aqueous phase is not really a pure water phase. This is necessary since there is a change in composition when the temperature varies from $T_{3}$ to $T$. Following the previous works of some of us,~\cite{Grabowska2022a,Algaba2023a} this second contribution to the chemical potential takes into account the change in the aqueous composition when the temperature varies from $T_3$, $x^{\text{eq}}_{\text{N}_{2}}(P,T_{3})$, to $T$, $x^{\text{eq}}_{\text{N}_{2}}(P,T)$, at constant pressure.

The chemical potential of water in the aqueous phase at $T$ and $P$ can be written as,

\begin{widetext}
\begin{equation}
\mu_{\text{H}_{2}\text{O}}^{\text{aq}}(P,T,x_{\text{N}^{\text{eq}}_{2}})\approx
-\bigintsss_{T_{3}}^{T} \dfrac{h_{\text{H}_{2}\text{O}}(P,T')}{k_{B}T'^{2}}\,dT'
 + \Big[k_{B}T\ln\{x^{\text{eq}}_{\text{H}_{2}\text{O}}(P,T)\}
-k_{B}T_{3}\ln\{x^{\text{eq}}_{\text{H}_{2}\text{O}}(P,T_{3})\}\Big]
\label{integral_water}
\end{equation}
\end{widetext}


\noindent
We recall here that $x^{\text{eq}}_{\text{H}_{2}\text{O}}$ is directly related to $x^{\text{eq}}_{\text{N}_{2}}$ since we are dealing with a two-component system, i.e, $x^{\text{eq}}_{\text{H}_{2}\text{O}}=1-x^{\text{eq}}_{\text{N}_{2}}$.

Finally, the driving force for nucleation at experimental conditions for the N$_{2}$ hydrate at $T$ and $P$ can be written combining Eqs.~\eqref{driving_force_EN}, and \eqref{integral_N2_2}--\eqref{integral_water} as,

\begin{widetext}
\begin{equation}
\dfrac{\Delta\mu^{\text{EC}}_{\text{N}}(P,T,x^{\text{eq}}_{\text{N}_{2}})}
{k_{B}T}=
-\bigintss_{T_{3}}^{T} 
\dfrac{h_{\text{H}}^{\text{H}}(P,T')- \Big\{h_{\text{N}_{2}}(P,T')+5.67\,
h_{\text{H}_{2}\text{O}}(P,T')\Big\}}{k_{B}T'^{2}}dT'
- \Big[k_{B}T\ln\{x^{\text{eq}}_{\text{H}_{2}\text{O}}(P,T)\}-k_{B}T_{3}\ln\{x^{\text{eq}}_{\text{H}_{2}\text{O}}(P,T_{3})\}\Big]
\label{driving_force_route1}
\end{equation}
\end{widetext}

\noindent Note that the term $h_{\text{H}}^{\text{H}}(P,T')- \Big\{h_{\text{N}_{2}}(P,T')+5.67\,
h_{\text{H}_{2}\text{O}}(P,T')\Big\}$ inside the integral is minus the dissociation enthalpy as a function of $P$ and $T'$, $h_{\text{H}}^{\text{diss}}(P,T')$.~\cite{Kashchiev2002a} Also, it is important to remark that the term outside the integral, $\ln\{x^{\text{eq}}_{\text{H}_{2}\text{O}}(T)\}-k_{B}T_{3}\ln\{x^{\text{eq}}_{\text{H}_{2}\text{O}}(T_{3})\}$, arises to take into account the change in the chemical potential of water in the aqueous phase because of the change in composition. For further details, we refer the reader to our previous works.~\cite{Grabowska2022a,Algaba2023a}

\subsubsection{$\Delta\mu^{\text{EC}}_{N}(T)$ via dissociation route}

According to our previous works,~\cite{Grabowska2022a,Algaba2023a} Eq.~\eqref{driving_force_route1} can be simplified following three approximations: (1) $h_{\text{H}}^{\text{diss}}$ does not change with the temperature; (2) its value is taken at $T_{3}$; and (3) $h_{\text{H}}^{\text{diss}}$ does not vary with the composition of N$_{2}$ in the aqueous phase. The first and second approximation allows to take out the $h_{\text{H}}^{\text{diss}}$ from the integral with a constant value, $h^{\text{diss}}_{\text{H}}(T_3)$. The third approximation allows to eliminate the last term of Eq.~\eqref{driving_force_route1}. Using the three approximations, it is possible to obtain a simplified version of Eq.~\eqref{driving_force_route1},

The chemical potential of water in the aqueous phase at $T$ and $P$ can be written as,

\begin{align}
\Delta\mu_{N}^{\text{EC}}(P,T)=&
k_{B}T \bigintsss_{T_{3}}^{T}\dfrac{h^{\text{diss}}_{\text{H}}(P,T_{3})}{k_{B}T'^{2}}\,dT'\approx  \nonumber\\
& -h^{\text{diss}}_{\text{H}}(P,T_{3}) \bigg(1-\dfrac{T}{T_{3}}\bigg)
\label{h_dissoc}
\end{align}

\noindent
To the best of our knowledge, Eq.~\eqref{h_dissoc} was presented for the first time by Kashchiev and A. Firoozabadi several years ago.~\cite{Kashchiev2002a}  Particularly, this equation corresponds to Eq.~(31) of the work of these authors. Since the final expression for $\Delta\mu_{N}^{\text{EC}}$ is given in terms of $h_{\text{H}}^{\text{diss}}$, we call this expression the dissociation route of the driving force for nucleation.~\cite{Grabowska2022a,Algaba2023a} As we will see in Section III.D.3, it can be considered a good approximation for the driving force for nucleation at small supercoolings.

\subsubsection{$\Delta\mu^{\text{EC}}_{N}(T)$ results}

\begin{figure}
\includegraphics[width=\columnwidth]{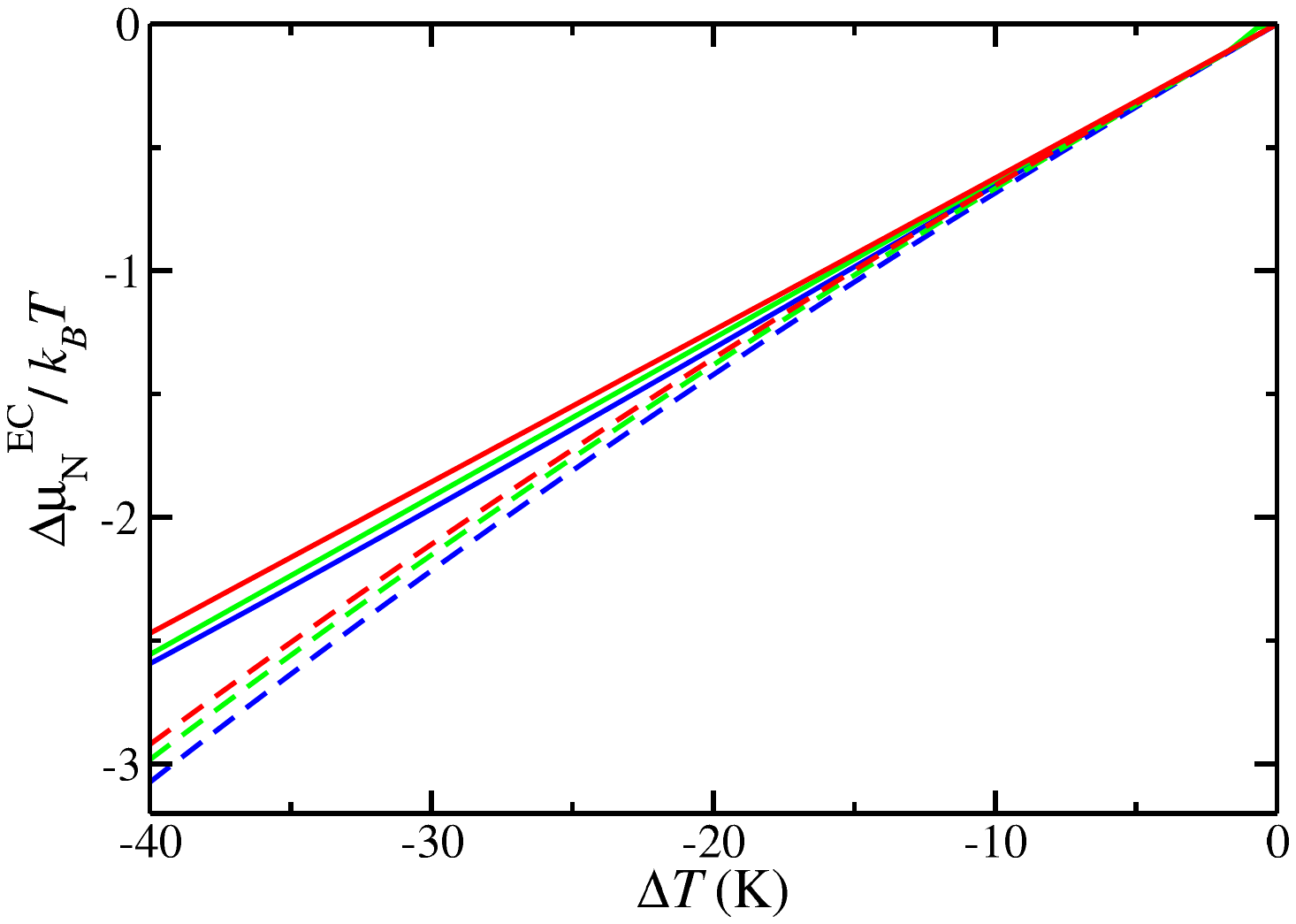}\\
\caption{Driving force for nucleation, $\Delta\mu^{\text{EC}}_{N}$, of N$_2$ hydrate at experimental conditions, as function of the supercooling, $\Delta T$, at $500$ (blue), $1000$ (purple), and $1500\,\text{bar}$ (red). The curves correspond to results obtained using route 1 (continuous curve) and dissociation route (dashed curves).}
\label{figure8}
\end{figure}

We have obtained the driving force for nucleation of the N$_{2}$ hydrate using route 1 and the dissociation route presented in Sections III D 1 and III D 2. Fig.~\ref{figure8} shows $\Delta\mu^{\text{EC}}_{N}$, as a function of the supercooling $\Delta T$, using both routes. We have included the predictions for the three pressures studied in this work. As can be observed, $\Delta\mu^{\text{EC}}_{N}$ obtained from both routes exhibits the same qualitative behavior in the whole range of temperatures. Particularly, both routes agree very well at low supercooled temperatures, from $|\Delta T|=0$ to $10-15\,\text{K}$, approximately. However, when $|\Delta T|$ increases deviations between both routes become larger. This is due to the approximations made in Eq.~\eqref{h_dissoc}. For $|\Delta T|\lesssim 15\,\text{K}$, the assumptions made do not have a significant effect on $\Delta\mu^{\text{EC}}_{N}(T)$ values. However, when the supercooling is increased, the variations of $h^{\text{diss}}_{\text{H}}$ and solubility of N$_2$ in the aqueous phase with temperature become important making the predictions of the dissociation route less accurate. Although the route given by Eq.~\eqref{h_dissoc} is an easy and fast way to estimate $\Delta\mu^{\text{EC}}_{N}$, we do not recommend it in general except for temperatures close to $T_{3}$. This conclusion is in agreement with our previous findings in the case of the CO$_{2}$ and CH$_{2}$ hydrates.~\cite{Grabowska2022a,Algaba2023a}

It is also interesting to analyze the effect of pressure on the driving force for nucleation of the N$_{2}$ hydrate. As can be seen, $\Delta\mu^{\text{EC}}_{N}$ values at $500$, $1000$, and $1500\,\text{bar}$ are very similar. The driving force for nucleation slightly becomes less negative as the pressure increases. This indicates that nucleation is favored when the pressure is decreased. However, the differences between the $\Delta\mu^{\text{EC}}_{N}$ values at the three pressures, at a given temperature, are really small. The maximum variation of the driving force for nucleation, $\Delta\mu_{\text{N}}^{\text{EC}}$, when the pressure is varied is achieved at the maximum supercooling, $\Delta T=-40\,\text{K}$. At these conditions,  $\Delta\mu_{\text{N}}^{\text{EC}}=-2.59$, $-2.55$, and $-2.47\,k_{B}T$ at $500$, $1000$, and $1500\,\text{bar}$, respectively. This means that the maximum variation of $\Delta\mu_{\text{N}}^{\text{EC}}$, at fixed $\Delta T$ and assuming single occupancy, is below $5\%$. From this perspective, the effect of pressure on the driving force for nucleation can be considered negligible. This result is relevant since it is the first time that the effect of pressure on the driving force for nucleation is studied for a hydrate with sII structure. According to results presented in Fig.~\ref{figure8}, we can assess that $\Delta\mu^{\text{EC}}_{N}$ does not depend on pressure in the range ($500-1500\,\text{bar}$) considered in this work and assuming single occupancy of the hydrate.

\section{Conclusions}

We have determined the solubility of N$_{2}$ when its aqueous solution is in contact with a N$_{2}$-rich liquid phase and with a N$_{2}$ hydrate phase via planar interfaces. Particularly, we have performed molecular dynamics simulations at several temperatures and at three different pressures, $500$, $1000$, and $1500\,\text{bar}$. We have also estimated the driving force for the nucleation of the N$_{2}$ hydrate using two different routes previously proposed by some of us to deal with CH$_{4}$ and CO$_{2}$ hydrates.~\cite{Grabowska2022a,Algaba2023a} Both, solubilities and driving forces for nucleation are key to understanding, from a thermodynamic point of view, the parameters that control nucleation of hydrates. As in our previous work, water is described using the TIP4P/Ice water model.~\cite{Grabowska2022a,Algaba2023a} In this work, we model N$_{2}$ using the TraPPE model.~\cite{Potoff2001a} Unlike dispersive interaction parameters between the oxygen atom of water and the nitrogen atoms of N$_{2}$, $\epsilon_{\text{ON}}=\xi_\text{{ON}}(\epsilon_{\text{OO}}\,\epsilon_\text{{NN}})^{1/2}$, are described using two different values for $\xi_\text{{ON}}$: (1) the standard value obtained from the Berthelot combining rule, $\xi_{\text{ON}}=1$; and (2) a modified value that correctly predicts the dissociation temperatures of the N$_{2}$ hydrate at all the pressures considered in this work, $\xi_{\text{ON}}=1.15$. We calculate all the solubilities using the direct coexistence technique between the two phases (aqueous solution -- N$_{2}$ and aqueous solution -- hydrate) at several pressures. We have also performed additional simulations of the pure systems, at several temperatures and pressures, to calculate the driving force for nucleation along the dissociation line of the hydrate.

We have analyzed the aqueous solution of N$_{2}$ when it is in contact with the liquid phase (pure N$_{2}$) and with the N$_{2}$ hydrate using the two values of $\xi_\text{{ON}}$ previously mentioned. From this information, we have obtained the solubility of N$_{2}$ in water when the solution is in contact with the N$_{2}$ liquid phase. The solubility of N$_{2}$ decreases with temperature, in a similar way to that of carbon dioxide and methane.~\cite{Grabowska2022a,Algaba2023a} Interestingly, the solubility of N$_{2}$ is similar to that of methane and one order of magnitude smaller than that of carbon dioxide. We have also studied the solubility of N$_{2}$ in the aqueous solution when it is in contact with the hydrate. This magnitude increases with the temperature, as it happens with the solubility of methane and carbon dioxide in water.~\cite{Grabowska2022a,Algaba2023a} 

Following our previous works,~\cite{Grabowska2022a,Algaba2023a}  we also determine the dissociation temperature of the N$_{2}$ hydrate ($T_{3}$), at three different pressures, $500$, $1000$, and $1500\,\text{bar}$, from the intersection of the solubility curves of N$_{2}$ when the aqueous phase it is in contact with the N$_{2}$-rich liquid phase and when it is in contact with the hydrate. This is possible since the formation of the hydrate phase below the dissociation line and the formation of the N$_{2}$ liquid phase above the dissociation line, at a given pressure, are activated processes. In other words, there exists metastability below and above the dissociation temperature of the hydrate. The temperature at which the intersection occurs is the $T_{3}$ of the hydrate at the corresponding pressure. We have also compared the predictions obtained from computer simulations with experimental data taken from the literature.~\cite{Sugahara2002a,Marshall1964a} Agreement between both results, within the error bars of the simulation values, is excellent for the three pressures studied in this work. This is the first time this technique is used and extended to obtain the dissociation line of a hydrate that exhibits the sII solid structure.

Finally, we also estimate the driving force for nucleation of the N$_{2}$ hydrate at three different pressures. Particularly, we have calculated $\Delta\mu^{\text{EC}}_{N}$ using two of the routes proposed in our previous paper (routes 1 and 3). Route 1 can be used with confidence since the solubility of N$_{2}$ in water is similar than that of methane. We also use the simple and approximated dissociation route to obtain a complementary estimation of $\Delta\mu^{\text{EC}}_{N}$. Although the dissociation route is a simple and fast way to estimate the driving force for nucleation, we do not recommend it in general except for temperatures close to $T_{3}$. Our results indicate that the effect of pressure on driving force for nucleation can be considered negligible, although $\Delta\mu^{\text{EC}}_{N}$ slightly decreases (in absolute value) when the pressure is increased. Note that this conclusion is valid for the range of pressures considered in this work and assuming single occupancy of the hydrate. To the best of our knowledge, this is the first time the driving force for nucleation of a hydrate that exhibits the crystallographic structure sII, along its dissociation line, is studied from computer simulation.

\section*{Conflicts of interest}

The authors have no conflicts to disclose.

\section*{Acknowledgements}
This work was financed by Ministerio de Ciencia e Innovaci\'on (Grant No.~PID2021-125081NB-I00), Junta de Andalucía (P20-00363), and Universidad de Huelva (P.O. FEDER UHU-1255522 and FEDER-UHU-202034), all four cofinanced by EU FEDER funds.  We acknowledge access to supercomputer time from RES from project FI-2023-2-0041. Additional  computational resources from Centro de Supercomputaci\'on de Galicia (CESGA, Santiago de Compostela, Spain) are also acknowledged.

\section*{Data availability}

The data that supports the findings of this study are available within the article.

\bibliography{bibfjblas}

\begin{thebibliography}{67}%
\makeatletter
\providecommand \@ifxundefined [1]{%
 \@ifx{#1\undefined}
}%
\providecommand \@ifnum [1]{%
 \ifnum #1\expandafter \@firstoftwo
 \else \expandafter \@secondoftwo
 \fi
}%
\providecommand \@ifx [1]{%
 \ifx #1\expandafter \@firstoftwo
 \else \expandafter \@secondoftwo
 \fi
}%
\providecommand \natexlab [1]{#1}%
\providecommand \enquote  [1]{``#1''}%
\providecommand \bibnamefont  [1]{#1}%
\providecommand \bibfnamefont [1]{#1}%
\providecommand \citenamefont [1]{#1}%
\providecommand \href@noop [0]{\@secondoftwo}%
\providecommand \href [0]{\begingroup \@sanitize@url \@href}%
\providecommand \@href[1]{\@@startlink{#1}\@@href}%
\providecommand \@@href[1]{\endgroup#1\@@endlink}%
\providecommand \@sanitize@url [0]{\catcode `\\12\catcode `\$12\catcode
  `\&12\catcode `\#12\catcode `\^12\catcode `\_12\catcode `\%12\relax}%
\providecommand \@@startlink[1]{}%
\providecommand \@@endlink[0]{}%
\providecommand \url  [0]{\begingroup\@sanitize@url \@url }%
\providecommand \@url [1]{\endgroup\@href {#1}{\urlprefix }}%
\providecommand \urlprefix  [0]{URL }%
\providecommand \Eprint [0]{\href }%
\providecommand \doibase [0]{http://dx.doi.org/}%
\providecommand \selectlanguage [0]{\@gobble}%
\providecommand \bibinfo  [0]{\@secondoftwo}%
\providecommand \bibfield  [0]{\@secondoftwo}%
\providecommand \translation [1]{[#1]}%
\providecommand \BibitemOpen [0]{}%
\providecommand \bibitemStop [0]{}%
\providecommand \bibitemNoStop [0]{.\EOS\space}%
\providecommand \EOS [0]{\spacefactor3000\relax}%
\providecommand \BibitemShut  [1]{\csname bibitem#1\endcsname}%
\let\auto@bib@innerbib\@empty
\bibitem [{\citenamefont {Sloan}\ and\ \citenamefont {Koh}(2008)}]{Sloan2008a}%
  \BibitemOpen
  \bibfield  {author} {\bibinfo {author} {\bibfnamefont {E.~D.}\ \bibnamefont
  {Sloan}}\ and\ \bibinfo {author} {\bibfnamefont {C.}~\bibnamefont {Koh}},\
  }\href@noop {} {\emph {\bibinfo {title} {{C}lathrate {H}ydrates of {N}atural
  {G}ases}}},\ \bibinfo {edition} {3rd}\ ed.\ (\bibinfo  {publisher} {CRC
  Press},\ \bibinfo {address} {New York},\ \bibinfo {year} {2008})\BibitemShut
  {NoStop}%
\bibitem [{\citenamefont {Ripmeester}\ and\ \citenamefont
  {Alavi}(2022)}]{Ripmeester2022a}%
  \BibitemOpen
  \bibfield  {author} {\bibinfo {author} {\bibfnamefont {J.~A.}\ \bibnamefont
  {Ripmeester}}\ and\ \bibinfo {author} {\bibfnamefont {S.}~\bibnamefont
  {Alavi}},\ }\href@noop {} {\emph {\bibinfo {title} {Clathrate Hydrates:
  Molecular Science and Characterization}}}\ (\bibinfo  {publisher} {Wiley-VCH:
  Weinheim, Germany},\ \bibinfo {year} {2022})\BibitemShut {NoStop}%
\bibitem [{\citenamefont {Kuhs}\ \emph {et~al.}(1997)\citenamefont {Kuhs},
  \citenamefont {Chazallon}, \citenamefont {Radaelli},\ and\ \citenamefont
  {Pauer}}]{Kuhs1997a}%
  \BibitemOpen
  \bibfield  {author} {\bibinfo {author} {\bibfnamefont {W.~F.}\ \bibnamefont
  {Kuhs}}, \bibinfo {author} {\bibfnamefont {B.}~\bibnamefont {Chazallon}},
  \bibinfo {author} {\bibfnamefont {P.~G.}\ \bibnamefont {Radaelli}}, \ and\
  \bibinfo {author} {\bibfnamefont {F.}~\bibnamefont {Pauer}},\ }\bibfield
  {title} {\enquote {\bibinfo {title} {Cage occupancy and compressibility of
  deuterated {N$_{2}$}--clathrate hydrate by neutron diffraction},}\
  }\href@noop {} {\bibfield  {journal} {\bibinfo  {journal} {J. Inclusion
  Phenom.}\ }\textbf {\bibinfo {volume} {29}},\ \bibinfo {pages} {65--77}
  (\bibinfo {year} {1997})}\BibitemShut {NoStop}%
\bibitem [{\citenamefont {van Klaveren}\ \emph
  {et~al.}(2001{\natexlab{a}})\citenamefont {van Klaveren}, \citenamefont
  {Michels}, \citenamefont {Schouten}, \citenamefont {Klug},\ and\
  \citenamefont {Tse}}]{vanKlaveren2001a}%
  \BibitemOpen
  \bibfield  {author} {\bibinfo {author} {\bibfnamefont {E.~P.}\ \bibnamefont
  {van Klaveren}}, \bibinfo {author} {\bibfnamefont {J.~P.~J.}\ \bibnamefont
  {Michels}}, \bibinfo {author} {\bibfnamefont {J.~A.}\ \bibnamefont
  {Schouten}}, \bibinfo {author} {\bibfnamefont {D.~D.}\ \bibnamefont {Klug}},
  \ and\ \bibinfo {author} {\bibfnamefont {J.~S.}\ \bibnamefont {Tse}},\
  }\bibfield  {title} {\enquote {\bibinfo {title} {Stability of doubly occupied
  {N$_{2}$} clathrate hydrates investigated by molecular dynamics
  simulations},}\ }\href@noop {} {\bibfield  {journal} {\bibinfo  {journal} {J.
  Chem. Phys.}\ }\textbf {\bibinfo {volume} {114}},\ \bibinfo {pages}
  {5745--5754} (\bibinfo {year} {2001}{\natexlab{a}})}\BibitemShut {NoStop}%
\bibitem [{\citenamefont {van Klaveren}\ \emph
  {et~al.}(2001{\natexlab{b}})\citenamefont {van Klaveren}, \citenamefont
  {Michels}, \citenamefont {Schouten}, \citenamefont {Klug},\ and\
  \citenamefont {Tse}}]{vanKlaveren2001b}%
  \BibitemOpen
  \bibfield  {author} {\bibinfo {author} {\bibfnamefont {E.~P.}\ \bibnamefont
  {van Klaveren}}, \bibinfo {author} {\bibfnamefont {J.~P.~J.}\ \bibnamefont
  {Michels}}, \bibinfo {author} {\bibfnamefont {J.~A.}\ \bibnamefont
  {Schouten}}, \bibinfo {author} {\bibfnamefont {D.~D.}\ \bibnamefont {Klug}},
  \ and\ \bibinfo {author} {\bibfnamefont {J.~S.}\ \bibnamefont {Tse}},\
  }\bibfield  {title} {\enquote {\bibinfo {title} {Molecular dynamics
  simulation study of the properties of doubly occupied {N$_{2}$} clathrate
  hydrates},}\ }\href@noop {} {\bibfield  {journal} {\bibinfo  {journal} {J.
  Chem. Phys.}\ }\textbf {\bibinfo {volume} {115}},\ \bibinfo {pages}
  {10500--10508} (\bibinfo {year} {2001}{\natexlab{b}})}\BibitemShut {NoStop}%
\bibitem [{\citenamefont {van Klaveren}\ \emph {et~al.}(2002)\citenamefont {van
  Klaveren}, \citenamefont {Michels}, \citenamefont {Schouten}, \citenamefont
  {Klug},\ and\ \citenamefont {Tse}}]{VanKlaveren2002a}%
  \BibitemOpen
  \bibfield  {author} {\bibinfo {author} {\bibfnamefont {E.~P.}\ \bibnamefont
  {van Klaveren}}, \bibinfo {author} {\bibfnamefont {J.~P.~J.}\ \bibnamefont
  {Michels}}, \bibinfo {author} {\bibfnamefont {J.~A.}\ \bibnamefont
  {Schouten}}, \bibinfo {author} {\bibfnamefont {D.~D.}\ \bibnamefont {Klug}},
  \ and\ \bibinfo {author} {\bibfnamefont {J.~S.}\ \bibnamefont {Tse}},\
  }\bibfield  {title} {\enquote {\bibinfo {title} {Computer simulations of the
  dynamics of doubly occupied {N$_{2}$} clathrate hydrates},}\ }\href@noop {}
  {\bibfield  {journal} {\bibinfo  {journal} {J. Chem. Phys.}\ }\textbf
  {\bibinfo {volume} {117}},\ \bibinfo {pages} {6637--6645} (\bibinfo {year}
  {2002})}\BibitemShut {NoStop}%
\bibitem [{\citenamefont {Chazallon}\ and\ \citenamefont
  {Kuhs}(2002)}]{Chazallon2002a}%
  \BibitemOpen
  \bibfield  {author} {\bibinfo {author} {\bibfnamefont {B.}~\bibnamefont
  {Chazallon}}\ and\ \bibinfo {author} {\bibfnamefont {W.~F.}\ \bibnamefont
  {Kuhs}},\ }\bibfield  {title} {\enquote {\bibinfo {title} {In situ structural
  properties of {N$_{2}$-}, {O$_{2}$-}, and air-clathrates by neutron
  diffraction},}\ }\href@noop {} {\bibfield  {journal} {\bibinfo  {journal} {J.
  Chem. Phys.}\ }\textbf {\bibinfo {volume} {117}},\ \bibinfo {pages}
  {308--320} (\bibinfo {year} {2002})}\BibitemShut {NoStop}%
\bibitem [{\citenamefont {Sasaki}\ \emph {et~al.}(2003)\citenamefont {Sasaki},
  \citenamefont {Hori}, \citenamefont {Kume},\ and\ \citenamefont
  {Shimizu}}]{Sasaki2003a}%
  \BibitemOpen
  \bibfield  {author} {\bibinfo {author} {\bibfnamefont {S.}~\bibnamefont
  {Sasaki}}, \bibinfo {author} {\bibfnamefont {S.}~\bibnamefont {Hori}},
  \bibinfo {author} {\bibfnamefont {T.}~\bibnamefont {Kume}}, \ and\ \bibinfo
  {author} {\bibfnamefont {H.}~\bibnamefont {Shimizu}},\ }\bibfield  {title}
  {\enquote {\bibinfo {title} {Microscopic observation and in situ raman
  scattering studies on high-pressure phase transformations of a synthetic
  nitrogen hydrate},}\ }\href@noop {} {\bibfield  {journal} {\bibinfo
  {journal} {J. Chem. Phys.}\ }\textbf {\bibinfo {volume} {118}},\ \bibinfo
  {pages} {7892--7897} (\bibinfo {year} {2003})}\BibitemShut {NoStop}%
\bibitem [{\citenamefont {Alavi}, \citenamefont {Ripmeester},\ and\
  \citenamefont {Klug}(2005)}]{Alavi2005a}%
  \BibitemOpen
  \bibfield  {author} {\bibinfo {author} {\bibfnamefont {S.}~\bibnamefont
  {Alavi}}, \bibinfo {author} {\bibfnamefont {J.~A.}\ \bibnamefont
  {Ripmeester}}, \ and\ \bibinfo {author} {\bibfnamefont {D.~D.}\ \bibnamefont
  {Klug}},\ }\bibfield  {title} {\enquote {\bibinfo {title} {Molecular-dynamics
  study of structure {II} hydrogen clathrates},}\ }\href@noop {} {\bibfield
  {journal} {\bibinfo  {journal} {J. Chem. Phys.}\ }\textbf {\bibinfo {volume}
  {123}},\ \bibinfo {pages} {024507} (\bibinfo {year} {2005})}\BibitemShut
  {NoStop}%
\bibitem [{\citenamefont {Alavi}, \citenamefont {Ripmeester},\ and\
  \citenamefont {Klug}(2006)}]{Alavi2006a}%
  \BibitemOpen
  \bibfield  {author} {\bibinfo {author} {\bibfnamefont {S.}~\bibnamefont
  {Alavi}}, \bibinfo {author} {\bibfnamefont {J.~A.}\ \bibnamefont
  {Ripmeester}}, \ and\ \bibinfo {author} {\bibfnamefont {D.~D.}\ \bibnamefont
  {Klug}},\ }\bibfield  {title} {\enquote {\bibinfo {title} {Molecular-dynamics
  simulations of binary structure {II} hydrogen and tetrahydrofurane
  clathrates},}\ }\href@noop {} {\bibfield  {journal} {\bibinfo  {journal} {J.
  Chem. Phys.}\ }\textbf {\bibinfo {volume} {124}},\ \bibinfo {pages} {014704}
  (\bibinfo {year} {2006})}\BibitemShut {NoStop}%
\bibitem [{\citenamefont {Barnes}\ and\ \citenamefont
  {Sum}(2013)}]{Barnes2013a}%
  \BibitemOpen
  \bibfield  {author} {\bibinfo {author} {\bibfnamefont {B.~C.}\ \bibnamefont
  {Barnes}}\ and\ \bibinfo {author} {\bibfnamefont {A.~K.}\ \bibnamefont
  {Sum}},\ }\bibfield  {title} {\enquote {\bibinfo {title} {Advances in
  molecular simulations of clathrate hydrates},}\ }\href@noop {} {\bibfield
  {journal} {\bibinfo  {journal} {Curr. Opin. Chem. Eng.}\ }\textbf {\bibinfo
  {volume} {105}},\ \bibinfo {pages} {184--190} (\bibinfo {year}
  {2013})}\BibitemShut {NoStop}%
\bibitem [{\citenamefont {Ripmeester}\ and\ \citenamefont
  {Alavi}(2016)}]{Ripmeester2016a}%
  \BibitemOpen
  \bibfield  {author} {\bibinfo {author} {\bibfnamefont {J.~A.}\ \bibnamefont
  {Ripmeester}}\ and\ \bibinfo {author} {\bibfnamefont {S.}~\bibnamefont
  {Alavi}},\ }\bibfield  {title} {\enquote {\bibinfo {title} {Some current
  challenges in clathrate hydrate science: Nucleation, decomposition and the
  memory effect},}\ }\href@noop {} {\bibfield  {journal} {\bibinfo  {journal}
  {Curr. Opin. Solid State Mater Sci.}\ }\textbf {\bibinfo {volume} {20}},\
  \bibinfo {pages} {344--351} (\bibinfo {year} {2016})}\BibitemShut {NoStop}%
\bibitem [{\citenamefont {Ratcliffe}(2022)}]{Ratcliffe2022a}%
  \BibitemOpen
  \bibfield  {author} {\bibinfo {author} {\bibfnamefont {C.~I.}\ \bibnamefont
  {Ratcliffe}},\ }\bibfield  {title} {\enquote {\bibinfo {title} {The
  development of clathrate hydrate science},}\ }\href@noop {} {\bibfield
  {journal} {\bibinfo  {journal} {Energy Fuels}\ }\textbf {\bibinfo {volume}
  {36}},\ \bibinfo {pages} {10412--10429} (\bibinfo {year} {2022})}\BibitemShut
  {NoStop}%
\bibitem [{\citenamefont {Tsimpanogiannis}\ and\ \citenamefont
  {Economou}(2017)}]{Tsimpanogiannis2017a}%
  \BibitemOpen
  \bibfield  {author} {\bibinfo {author} {\bibfnamefont {I.~N.}\ \bibnamefont
  {Tsimpanogiannis}}\ and\ \bibinfo {author} {\bibfnamefont {I.~G.}\
  \bibnamefont {Economou}},\ }\bibfield  {title} {\enquote {\bibinfo {title}
  {Monte carlo simulation studies of clathrate hydrates: A review},}\
  }\href@noop {} {\bibfield  {journal} {\bibinfo  {journal} {J. Supercrit.
  Fluids}\ }\textbf {\bibinfo {volume} {134}},\ \bibinfo {pages} {51--60}
  (\bibinfo {year} {2017})}\BibitemShut {NoStop}%
\bibitem [{\citenamefont {Brumby}\ \emph {et~al.}(2019)\citenamefont {Brumby},
  \citenamefont {Yuhara}, \citenamefont {Hasegawa}, \citenamefont {Wu},
  \citenamefont {Sum},\ and\ \citenamefont {Yasuoka}}]{Brumby2019a}%
  \BibitemOpen
  \bibfield  {author} {\bibinfo {author} {\bibfnamefont {P.~E.}\ \bibnamefont
  {Brumby}}, \bibinfo {author} {\bibfnamefont {D.}~\bibnamefont {Yuhara}},
  \bibinfo {author} {\bibfnamefont {T.}~\bibnamefont {Hasegawa}}, \bibinfo
  {author} {\bibfnamefont {D.~T.}\ \bibnamefont {Wu}}, \bibinfo {author}
  {\bibfnamefont {A.~K.}\ \bibnamefont {Sum}}, \ and\ \bibinfo {author}
  {\bibfnamefont {K.}~\bibnamefont {Yasuoka}},\ }\bibfield  {title} {\enquote
  {\bibinfo {title} {Cage occupancies, lattice constants, and guest chemical
  potentials for structure ii hydrogen clathrate hydrate from gibbs ensemble
  monte carlo simulations},}\ }\href@noop {} {\bibfield  {journal} {\bibinfo
  {journal} {J. Chem. Phys.}\ }\textbf {\bibinfo {volume} {150}},\ \bibinfo
  {pages} {134503} (\bibinfo {year} {2019})}\BibitemShut {NoStop}%
\bibitem [{\citenamefont {Yi}\ \emph {et~al.}(2019)\citenamefont {Yi},
  \citenamefont {Zhou}, \citenamefont {He}, \citenamefont {Cai}, \citenamefont
  {Zhao}, \citenamefont {Zhang},\ and\ \citenamefont {Shao}}]{Yi2019a}%
  \BibitemOpen
  \bibfield  {author} {\bibinfo {author} {\bibfnamefont {L.}~\bibnamefont
  {Yi}}, \bibinfo {author} {\bibfnamefont {X.}~\bibnamefont {Zhou}}, \bibinfo
  {author} {\bibfnamefont {Y.}~\bibnamefont {He}}, \bibinfo {author}
  {\bibfnamefont {Z.}~\bibnamefont {Cai}}, \bibinfo {author} {\bibfnamefont
  {L.}~\bibnamefont {Zhao}}, \bibinfo {author} {\bibfnamefont {W.}~\bibnamefont
  {Zhang}}, \ and\ \bibinfo {author} {\bibfnamefont {Y.}~\bibnamefont {Shao}},\
  }\bibfield  {title} {\enquote {\bibinfo {title} {Molecular dynamics
  simulation study on the growth of structure ii nitrogen hydrate},}\
  }\href@noop {} {\bibfield  {journal} {\bibinfo  {journal} {The Journal of
  Physical Chemistry B}\ }\textbf {\bibinfo {volume} {123}},\ \bibinfo {pages}
  {9180--9186} (\bibinfo {year} {2019})}\BibitemShut {NoStop}%
\bibitem [{\citenamefont {Michalis}\ \emph {et~al.}(2022)\citenamefont
  {Michalis}, \citenamefont {Economou}, \citenamefont {Stubos},\ and\
  \citenamefont {Tsimpanogiannis}}]{Michalis2022a}%
  \BibitemOpen
  \bibfield  {author} {\bibinfo {author} {\bibfnamefont {V.~K.}\ \bibnamefont
  {Michalis}}, \bibinfo {author} {\bibfnamefont {I.~G.}\ \bibnamefont
  {Economou}}, \bibinfo {author} {\bibfnamefont {A.~K.}\ \bibnamefont
  {Stubos}}, \ and\ \bibinfo {author} {\bibfnamefont {I.~N.}\ \bibnamefont
  {Tsimpanogiannis}},\ }\bibfield  {title} {\enquote {\bibinfo {title} {Phase
  equilibria molecular simulations of hydrogen hydrates via the direct phase
  coexistence approach},}\ }\href@noop {} {\bibfield  {journal} {\bibinfo
  {journal} {J. Chem. Phys.}\ }\textbf {\bibinfo {volume} {157}},\ \bibinfo
  {pages} {154501} (\bibinfo {year} {2022})}\BibitemShut {NoStop}%
\bibitem [{\citenamefont {Kashchiev}(2000)}]{Kashchiev2000a}%
  \BibitemOpen
  \bibfield  {author} {\bibinfo {author} {\bibfnamefont {D.}~\bibnamefont
  {Kashchiev}},\ }\href@noop {} {\emph {\bibinfo {title} {Nucleation}}}\
  (\bibinfo  {publisher} {Butterworth-Heinemann: Oxford, UK},\ \bibinfo {year}
  {2000})\BibitemShut {NoStop}%
\bibitem [{\citenamefont {Kashchiev}\ and\ \citenamefont
  {Firoozabadi}(2002{\natexlab{a}})}]{Kashchiev2002a}%
  \BibitemOpen
  \bibfield  {author} {\bibinfo {author} {\bibfnamefont {D.}~\bibnamefont
  {Kashchiev}}\ and\ \bibinfo {author} {\bibfnamefont {A.}~\bibnamefont
  {Firoozabadi}},\ }\bibfield  {title} {\enquote {\bibinfo {title} {Driving
  force for crystallization of gas hydrates},}\ }\href@noop {} {\bibfield
  {journal} {\bibinfo  {journal} {J. Cryst. Growth}\ }\textbf {\bibinfo
  {volume} {241}},\ \bibinfo {pages} {220--230} (\bibinfo {year}
  {2002}{\natexlab{a}})}\BibitemShut {NoStop}%
\bibitem [{\citenamefont {Kashchiev}\ and\ \citenamefont
  {Firoozabadi}(2002{\natexlab{b}})}]{Kashchiev2002b}%
  \BibitemOpen
  \bibfield  {author} {\bibinfo {author} {\bibfnamefont {D.}~\bibnamefont
  {Kashchiev}}\ and\ \bibinfo {author} {\bibfnamefont {A.}~\bibnamefont
  {Firoozabadi}},\ }\bibfield  {title} {\enquote {\bibinfo {title} {Nucleation
  of gas hydrates},}\ }\href@noop {} {\bibfield  {journal} {\bibinfo  {journal}
  {J. Cryst. Growth}\ }\textbf {\bibinfo {volume} {243}},\ \bibinfo {pages}
  {476--489} (\bibinfo {year} {2002}{\natexlab{b}})}\BibitemShut {NoStop}%
\bibitem [{\citenamefont {Kashchiev}\ and\ \citenamefont {van
  Rosmalen}(2003)}]{Kashchiev2003a}%
  \BibitemOpen
  \bibfield  {author} {\bibinfo {author} {\bibfnamefont {D.}~\bibnamefont
  {Kashchiev}}\ and\ \bibinfo {author} {\bibfnamefont {G.~M.}\ \bibnamefont
  {van Rosmalen}},\ }\bibfield  {title} {\enquote {\bibinfo {title} {Review:
  Nucleation in solutions revisited},}\ }\href {\doibase
  https://doi.org/10.1002/crat.200310070} {\bibfield  {journal} {\bibinfo
  {journal} {Cryst. Res. Technol.}\ }\textbf {\bibinfo {volume} {38}},\
  \bibinfo {pages} {555--574} (\bibinfo {year} {2003})}\BibitemShut {NoStop}%
\bibitem [{\citenamefont {van Cleeff}\ and\ \citenamefont
  {Diepen}(1960)}]{vanCleeff1960a}%
  \BibitemOpen
  \bibfield  {author} {\bibinfo {author} {\bibfnamefont {A.}~\bibnamefont {van
  Cleeff}}\ and\ \bibinfo {author} {\bibfnamefont {G.~A.~M.}\ \bibnamefont
  {Diepen}},\ }\bibfield  {title} {\enquote {\bibinfo {title} {Gas hydrates of
  nitrogen and oxygen},}\ }\href@noop {} {\bibfield  {journal} {\bibinfo
  {journal} {Rec. Trav. Chim.}\ }\textbf {\bibinfo {volume} {79}},\ \bibinfo
  {pages} {582--586} (\bibinfo {year} {1960})}\BibitemShut {NoStop}%
\bibitem [{\citenamefont {Marshall}, \citenamefont {Saito},\ and\ \citenamefont
  {Kobayashi}(1964)}]{Marshall1964a}%
  \BibitemOpen
  \bibfield  {author} {\bibinfo {author} {\bibfnamefont {D.~R.}\ \bibnamefont
  {Marshall}}, \bibinfo {author} {\bibfnamefont {S.}~\bibnamefont {Saito}}, \
  and\ \bibinfo {author} {\bibfnamefont {R.}~\bibnamefont {Kobayashi}},\
  }\bibfield  {title} {\enquote {\bibinfo {title} {Hydrates at high pressures:
  Part i. methane-water, argon-water, and nitrogen-water systems},}\
  }\href@noop {} {\bibfield  {journal} {\bibinfo  {journal} {AIChE Journal}\
  }\textbf {\bibinfo {volume} {10}},\ \bibinfo {pages} {202--205} (\bibinfo
  {year} {1964})}\BibitemShut {NoStop}%
\bibitem [{\citenamefont {Jhaveri}\ and\ \citenamefont
  {Robinson}(1965)}]{Jhaveri1965a}%
  \BibitemOpen
  \bibfield  {author} {\bibinfo {author} {\bibfnamefont {J.}~\bibnamefont
  {Jhaveri}}\ and\ \bibinfo {author} {\bibfnamefont {D.~B.}\ \bibnamefont
  {Robinson}},\ }\bibfield  {title} {\enquote {\bibinfo {title} {Hydrates in
  the methane-nitrogen system},}\ }\href@noop {} {\bibfield  {journal}
  {\bibinfo  {journal} {Can. J. Chem Eng.}\ }\textbf {\bibinfo {volume} {43}},\
  \bibinfo {pages} {75--78} (\bibinfo {year} {1965})}\BibitemShut {NoStop}%
\bibitem [{\citenamefont {Sugahara}\ \emph {et~al.}(2002)\citenamefont
  {Sugahara}, \citenamefont {Tanaka}, \citenamefont {Sugahara},\ and\
  \citenamefont {Ohgaki}}]{Sugahara2002a}%
  \BibitemOpen
  \bibfield  {author} {\bibinfo {author} {\bibfnamefont {K.}~\bibnamefont
  {Sugahara}}, \bibinfo {author} {\bibfnamefont {Y.}~\bibnamefont {Tanaka}},
  \bibinfo {author} {\bibfnamefont {T.}~\bibnamefont {Sugahara}}, \ and\
  \bibinfo {author} {\bibfnamefont {K.}~\bibnamefont {Ohgaki}},\ }\bibfield
  {title} {\enquote {\bibinfo {title} {Thermodynamic stability and structure of
  nitrogen hydrate crystal},}\ }\href@noop {} {\bibfield  {journal} {\bibinfo
  {journal} {Journal of Supramolecular Chemistry}\ }\textbf {\bibinfo {volume}
  {2}},\ \bibinfo {pages} {365--368} (\bibinfo {year} {2002})}\BibitemShut
  {NoStop}%
\bibitem [{\citenamefont {Mohammadi}, \citenamefont {Tohidi},\ and\
  \citenamefont {Burgass}(2003)}]{Mohammadi2003a}%
  \BibitemOpen
  \bibfield  {author} {\bibinfo {author} {\bibfnamefont {A.~H.}\ \bibnamefont
  {Mohammadi}}, \bibinfo {author} {\bibfnamefont {B.}~\bibnamefont {Tohidi}}, \
  and\ \bibinfo {author} {\bibfnamefont {R.~W.}\ \bibnamefont {Burgass}},\
  }\bibfield  {title} {\enquote {\bibinfo {title} {Equilibrium data and
  thermodynamic modeling of nitrogen, oxygen, and air clathrate hydrates},}\
  }\href@noop {} {\bibfield  {journal} {\bibinfo  {journal} {J. Chem. Eng.
  Data}\ }\textbf {\bibinfo {volume} {48}},\ \bibinfo {pages} {612--616}
  (\bibinfo {year} {2003})}\BibitemShut {NoStop}%
\bibitem [{\citenamefont {Waals}\ and\ \citenamefont
  {Hermans}(1950)}]{vanderWaals1950a}%
  \BibitemOpen
  \bibfield  {author} {\bibinfo {author} {\bibfnamefont {J.~H. V.~D.}\
  \bibnamefont {Waals}}\ and\ \bibinfo {author} {\bibfnamefont {J.~J.}\
  \bibnamefont {Hermans}},\ }\bibfield  {title} {\enquote {\bibinfo {title}
  {Thermodynamic properties of mixtures of alkanes differing in chain length.
  i. heats of mixing},}\ }\href@noop {} {\bibfield  {journal} {\bibinfo
  {journal} {Rec. Tran. Chim.}\ }\textbf {\bibinfo {volume} {69}},\ \bibinfo
  {pages} {949--970} (\bibinfo {year} {1950})}\BibitemShut {NoStop}%
\bibitem [{\citenamefont {Waals}(1951)}]{vanderWaals1951a}%
  \BibitemOpen
  \bibfield  {author} {\bibinfo {author} {\bibfnamefont {J.~H. V.~D.}\
  \bibnamefont {Waals}},\ }\bibfield  {title} {\enquote {\bibinfo {title}
  {Thermodynamic properties of mixtures of alkanes differing in chain length,
  iii: The system hexane-dodecane},}\ }\href@noop {} {\bibfield  {journal}
  {\bibinfo  {journal} {Rec. Tran. Chim.}\ }\textbf {\bibinfo {volume} {70}},\
  \bibinfo {pages} {101--104} (\bibinfo {year} {1951})}\BibitemShut {NoStop}%
\bibitem [{\citenamefont {S.~Dufal}\ and\ \citenamefont
  {Haslam}(2012)}]{Dufal2012a}%
  \BibitemOpen
  \bibfield  {author} {\bibinfo {author} {\bibfnamefont {G.~J.}\ \bibnamefont
  {S.~Dufal}, \bibfnamefont {A.~Galindo}}\ and\ \bibinfo {author}
  {\bibfnamefont {A.~J.}\ \bibnamefont {Haslam}},\ }\bibfield  {title}
  {\enquote {\bibinfo {title} {Modelling the effect of methanol, glycol
  inhibitors and electrolytes on the equilibrium stability of hydrates with the
  saft-vr approach},}\ }\href {\doibase
  https://doi.org/10.1080/00268976.2012.664662} {\bibfield  {journal} {\bibinfo
   {journal} {Molecular Physics}\ }\textbf {\bibinfo {volume} {110}},\ \bibinfo
  {pages} {1223--1240} (\bibinfo {year} {2012})}\BibitemShut {NoStop}%
\bibitem [{\citenamefont {Tanaka}\ and\ \citenamefont
  {Kiyohara}(1993{\natexlab{a}})}]{Tanaka1993a}%
  \BibitemOpen
  \bibfield  {author} {\bibinfo {author} {\bibfnamefont {H.}~\bibnamefont
  {Tanaka}}\ and\ \bibinfo {author} {\bibfnamefont {K.}~\bibnamefont
  {Kiyohara}},\ }\bibfield  {title} {\enquote {\bibinfo {title} {On the
  thermodynamic stability of clathrate hydrate. {I}},}\ }\href@noop {}
  {\bibfield  {journal} {\bibinfo  {journal} {J. Chem. Phys.}\ }\textbf
  {\bibinfo {volume} {98}},\ \bibinfo {pages} {4098--4109} (\bibinfo {year}
  {1993}{\natexlab{a}})}\BibitemShut {NoStop}%
\bibitem [{\citenamefont {Tanaka}\ and\ \citenamefont
  {Kiyohara}(1993{\natexlab{b}})}]{Tanaka1993b}%
  \BibitemOpen
  \bibfield  {author} {\bibinfo {author} {\bibfnamefont {H.}~\bibnamefont
  {Tanaka}}\ and\ \bibinfo {author} {\bibfnamefont {K.}~\bibnamefont
  {Kiyohara}},\ }\bibfield  {title} {\enquote {\bibinfo {title} {The
  thermodynamic stability of clathrate hydrate. {II}. simultaneous occupation
  of larger and smaller cages},}\ }\href@noop {} {\bibfield  {journal}
  {\bibinfo  {journal} {J. Chem. Phys.}\ }\textbf {\bibinfo {volume} {98}},\
  \bibinfo {pages} {8110--8118} (\bibinfo {year}
  {1993}{\natexlab{b}})}\BibitemShut {NoStop}%
\bibitem [{\citenamefont {Tanaka}(1994)}]{Tanaka1994a}%
  \BibitemOpen
  \bibfield  {author} {\bibinfo {author} {\bibfnamefont {H.}~\bibnamefont
  {Tanaka}},\ }\bibfield  {title} {\enquote {\bibinfo {title} {The
  thermodynamic stability of clathrate hydrate. {III}. accommodation of
  nonspherical propane and ethane molecules},}\ }\href@noop {} {\bibfield
  {journal} {\bibinfo  {journal} {J. Chem. Phys.}\ }\textbf {\bibinfo {volume}
  {101}},\ \bibinfo {pages} {10833--10842} (\bibinfo {year}
  {1994})}\BibitemShut {NoStop}%
\bibitem [{\citenamefont {Tanaka}, \citenamefont {Nakatsuka},\ and\
  \citenamefont {Koga}(2004)}]{Tanaka2004a}%
  \BibitemOpen
  \bibfield  {author} {\bibinfo {author} {\bibfnamefont {H.}~\bibnamefont
  {Tanaka}}, \bibinfo {author} {\bibfnamefont {T.}~\bibnamefont {Nakatsuka}}, \
  and\ \bibinfo {author} {\bibfnamefont {K.}~\bibnamefont {Koga}},\ }\bibfield
  {title} {\enquote {\bibinfo {title} {On the thermodynamic stability of
  clathrate hydrates {IV}: Double occupancy of cages},}\ }\href@noop {}
  {\bibfield  {journal} {\bibinfo  {journal} {J. Chem. Phys.}\ }\textbf
  {\bibinfo {volume} {121}},\ \bibinfo {pages} {5488--5493} (\bibinfo {year}
  {2004})}\BibitemShut {NoStop}%
\bibitem [{\citenamefont {Tsimpanogiannis}, \citenamefont {Papadimitriou},\
  and\ \citenamefont {Stubos}(2012)}]{Tsimpanogiannis2012a}%
  \BibitemOpen
  \bibfield  {author} {\bibinfo {author} {\bibfnamefont {I.~N.}\ \bibnamefont
  {Tsimpanogiannis}}, \bibinfo {author} {\bibfnamefont {N.~I.}\ \bibnamefont
  {Papadimitriou}}, \ and\ \bibinfo {author} {\bibfnamefont {A.~K.}\
  \bibnamefont {Stubos}},\ }\bibfield  {title} {\enquote {\bibinfo {title} {On
  the limitation of the van der {Waals-Platteeuw}-based thermodynamic models
  for hydrates with multiple occupancy of cavities},}\ }\href@noop {}
  {\bibfield  {journal} {\bibinfo  {journal} {Mol. Phys.}\ }\textbf {\bibinfo
  {volume} {110}},\ \bibinfo {pages} {1213--1221} (\bibinfo {year}
  {2012})}\BibitemShut {NoStop}%
\bibitem [{\citenamefont {Abascal}\ and\ \citenamefont
  {Vega}(2005)}]{Abascal2005a}%
  \BibitemOpen
  \bibfield  {author} {\bibinfo {author} {\bibfnamefont {J.~L.}\ \bibnamefont
  {Abascal}}\ and\ \bibinfo {author} {\bibfnamefont {C.}~\bibnamefont {Vega}},\
  }\bibfield  {title} {\enquote {\bibinfo {title} {A general purpose model for
  the condensed phases of water: {TIP4P/2005}},}\ }\href@noop {} {\bibfield
  {journal} {\bibinfo  {journal} {J. Chem. Phys.}\ }\textbf {\bibinfo {volume}
  {123}},\ \bibinfo {pages} {234505/1--12} (\bibinfo {year}
  {2005})}\BibitemShut {NoStop}%
\bibitem [{\citenamefont {Vrabec}, \citenamefont {Stoll},\ and\ \citenamefont
  {Hasse}(2001)}]{Vrabec2001a}%
  \BibitemOpen
  \bibfield  {author} {\bibinfo {author} {\bibfnamefont {J.}~\bibnamefont
  {Vrabec}}, \bibinfo {author} {\bibfnamefont {J.}~\bibnamefont {Stoll}}, \
  and\ \bibinfo {author} {\bibfnamefont {H.}~\bibnamefont {Hasse}},\ }\bibfield
   {title} {\enquote {\bibinfo {title} {A set of molecular models for symmetric
  quadrupolar fluids},}\ }\href@noop {} {\bibfield  {journal} {\bibinfo
  {journal} {The Journal of Physical Chemistry B}\ }\textbf {\bibinfo {volume}
  {105}},\ \bibinfo {pages} {12126--12133} (\bibinfo {year}
  {2001})}\BibitemShut {NoStop}%
\bibitem [{\citenamefont {Abascal}\ \emph {et~al.}(2005)\citenamefont
  {Abascal}, \citenamefont {Sanz}, \citenamefont {Fern\'andez},\ and\
  \citenamefont {Vega}}]{Abascal2005b}%
  \BibitemOpen
  \bibfield  {author} {\bibinfo {author} {\bibfnamefont {J.~L.~F.}\
  \bibnamefont {Abascal}}, \bibinfo {author} {\bibfnamefont {E.}~\bibnamefont
  {Sanz}}, \bibinfo {author} {\bibfnamefont {R.~G.}\ \bibnamefont
  {Fern\'andez}}, \ and\ \bibinfo {author} {\bibfnamefont {C.}~\bibnamefont
  {Vega}},\ }\bibfield  {title} {\enquote {\bibinfo {title} {A potential model
  for the study of ices and amorphous water: {TIP4P/Ice}},}\ }\href@noop {}
  {\bibfield  {journal} {\bibinfo  {journal} {J. Chem. Phys.}\ }\textbf
  {\bibinfo {volume} {122}},\ \bibinfo {pages} {234511--1--234511--9} (\bibinfo
  {year} {2005})}\BibitemShut {NoStop}%
\bibitem [{\citenamefont {Conde}\ and\ \citenamefont
  {Vega}(2013)}]{Conde2013a}%
  \BibitemOpen
  \bibfield  {author} {\bibinfo {author} {\bibfnamefont {M.~M.}\ \bibnamefont
  {Conde}}\ and\ \bibinfo {author} {\bibfnamefont {C.}~\bibnamefont {Vega}},\
  }\bibfield  {title} {\enquote {\bibinfo {title} {Note: A simple correlation
  to locate the three phase coexistence line in methane-hydrate simulations},}\
  }\href {\doibase https://doi.org/10.1063/1.4790647} {\bibfield  {journal}
  {\bibinfo  {journal} {J. Chem. Phys.}\ }\textbf {\bibinfo {volume} {138}},\
  \bibinfo {pages} {056101} (\bibinfo {year} {2013})}\BibitemShut {NoStop}%
\bibitem [{\citenamefont {Ladd}\ and\ \citenamefont
  {Woodcock}(1977)}]{Ladd1977a}%
  \BibitemOpen
  \bibfield  {author} {\bibinfo {author} {\bibfnamefont {J.~C.}\ \bibnamefont
  {Ladd}}\ and\ \bibinfo {author} {\bibfnamefont {L.~V.}\ \bibnamefont
  {Woodcock}},\ }\bibfield  {title} {\enquote {\bibinfo {title} {Triple-point
  coexistence properties of the lennard-jones system},}\ }\href@noop {}
  {\bibfield  {journal} {\bibinfo  {journal} {Chem. Phys. Lett.}\ }\textbf
  {\bibinfo {volume} {51}},\ \bibinfo {pages} {159155} (\bibinfo {year}
  {1977})}\BibitemShut {NoStop}%
\bibitem [{\citenamefont {Conde}\ and\ \citenamefont
  {Vega}(2010)}]{Conde2010a}%
  \BibitemOpen
  \bibfield  {author} {\bibinfo {author} {\bibfnamefont {M.~M.}\ \bibnamefont
  {Conde}}\ and\ \bibinfo {author} {\bibfnamefont {C.}~\bibnamefont {Vega}},\
  }\bibfield  {title} {\enquote {\bibinfo {title} {Determining the three-phase
  coexistence line in methane hydrates using computer simulations},}\ }\href
  {\doibase https://doi.org/10.1063/1.3466751} {\bibfield  {journal} {\bibinfo
  {journal} {J. Chem. Phys.}\ }\textbf {\bibinfo {volume} {133}},\ \bibinfo
  {pages} {064507} (\bibinfo {year} {2010})}\BibitemShut {NoStop}%
\bibitem [{\citenamefont {M{\'\i}guez}\ \emph {et~al.}(2015)\citenamefont
  {M{\'\i}guez}, \citenamefont {Conde}, \citenamefont {Torr{\'e}},
  \citenamefont {Blas}, \citenamefont {Pi{\~n}eiro},\ and\ \citenamefont
  {Vega}}]{Miguez2015a}%
  \BibitemOpen
  \bibfield  {author} {\bibinfo {author} {\bibfnamefont {J.~M.}\ \bibnamefont
  {M{\'\i}guez}}, \bibinfo {author} {\bibfnamefont {M.~M.}\ \bibnamefont
  {Conde}}, \bibinfo {author} {\bibfnamefont {J.-P.}\ \bibnamefont
  {Torr{\'e}}}, \bibinfo {author} {\bibfnamefont {F.~J.}\ \bibnamefont {Blas}},
  \bibinfo {author} {\bibfnamefont {M.~M.}\ \bibnamefont {Pi{\~n}eiro}}, \ and\
  \bibinfo {author} {\bibfnamefont {C.}~\bibnamefont {Vega}},\ }\bibfield
  {title} {\enquote {\bibinfo {title} {Molecular dynamics simulation of
  {CO$_2$} hydrates: Prediction of three phase coexistence line},}\ }\href
  {\doibase https://doi.org/10.1063/1.4916119} {\bibfield  {journal} {\bibinfo
  {journal} {J. Chem. Phys.}\ }\textbf {\bibinfo {volume} {142}},\ \bibinfo
  {pages} {124505} (\bibinfo {year} {2015})}\BibitemShut {NoStop}%
\bibitem [{\citenamefont {P{\'e}rez-Rodr{\'\i}guez}\ \emph
  {et~al.}(2017)\citenamefont {P{\'e}rez-Rodr{\'\i}guez}, \citenamefont
  {Vidal-Vidal}, \citenamefont {M{\'\i}guez}, \citenamefont {Blas},
  \citenamefont {Torr{\'e}},\ and\ \citenamefont
  {Pi{\~n}eiro}}]{Perez-Rodriguez2017a}%
  \BibitemOpen
  \bibfield  {author} {\bibinfo {author} {\bibfnamefont {M.}~\bibnamefont
  {P{\'e}rez-Rodr{\'\i}guez}}, \bibinfo {author} {\bibfnamefont
  {A.}~\bibnamefont {Vidal-Vidal}}, \bibinfo {author} {\bibfnamefont
  {J.}~\bibnamefont {M{\'\i}guez}}, \bibinfo {author} {\bibfnamefont {F.~J.}\
  \bibnamefont {Blas}}, \bibinfo {author} {\bibfnamefont {J.-P.}\ \bibnamefont
  {Torr{\'e}}}, \ and\ \bibinfo {author} {\bibfnamefont {M.~M.}\ \bibnamefont
  {Pi{\~n}eiro}},\ }\bibfield  {title} {\enquote {\bibinfo {title}
  {Computational study of the interplay between intermolecular interactions and
  {CO$_{2}$} orientations in type {I} hydrates},}\ }\href {\doibase
  https://doi.org/10.1039/C6CP07097C} {\bibfield  {journal} {\bibinfo
  {journal} {Phys. Chem. Chem. Phys.}\ }\textbf {\bibinfo {volume} {19}},\
  \bibinfo {pages} {3384--3393} (\bibinfo {year} {2017})}\BibitemShut {NoStop}%
\bibitem [{\citenamefont {Michalis}\ \emph {et~al.}(2015)\citenamefont
  {Michalis}, \citenamefont {Costandy}, \citenamefont {Tsimpanogiannis},
  \citenamefont {Stubos},\ and\ \citenamefont {Economou}}]{Michalis2015a}%
  \BibitemOpen
  \bibfield  {author} {\bibinfo {author} {\bibfnamefont {V.~K.}\ \bibnamefont
  {Michalis}}, \bibinfo {author} {\bibfnamefont {J.}~\bibnamefont {Costandy}},
  \bibinfo {author} {\bibfnamefont {I.~N.}\ \bibnamefont {Tsimpanogiannis}},
  \bibinfo {author} {\bibfnamefont {A.~K.}\ \bibnamefont {Stubos}}, \ and\
  \bibinfo {author} {\bibfnamefont {I.~G.}\ \bibnamefont {Economou}},\
  }\bibfield  {title} {\enquote {\bibinfo {title} {Prediction of the phase
  equilibria of methane hydrates using the direct phase coexistence
  methodology},}\ }\href {\doibase https://doi.org/10.1063/1.4905572}
  {\bibfield  {journal} {\bibinfo  {journal} {J. Chem. Phys.}\ }\textbf
  {\bibinfo {volume} {142}},\ \bibinfo {pages} {044501} (\bibinfo {year}
  {2015})}\BibitemShut {NoStop}%
\bibitem [{\citenamefont {Costandy}\ \emph {et~al.}(2015)\citenamefont
  {Costandy}, \citenamefont {Michalisa}, \citenamefont {Tsimpanogiannis},
  \citenamefont {Stubos},\ and\ \citenamefont {Economou}}]{Costandy2015a}%
  \BibitemOpen
  \bibfield  {author} {\bibinfo {author} {\bibfnamefont {J.}~\bibnamefont
  {Costandy}}, \bibinfo {author} {\bibfnamefont {V.~K.}\ \bibnamefont
  {Michalisa}}, \bibinfo {author} {\bibfnamefont {I.~N.}\ \bibnamefont
  {Tsimpanogiannis}}, \bibinfo {author} {\bibfnamefont {A.~K.}\ \bibnamefont
  {Stubos}}, \ and\ \bibinfo {author} {\bibfnamefont {I.~G.}\ \bibnamefont
  {Economou}},\ }\bibfield  {title} {\enquote {\bibinfo {title} {The role of
  intermolecular interactions in the prediction of the phase equilibria of
  carbon dioxide hydrates},}\ }\href {\doibase
  https://doi.org/10.1063/1.4929805} {\bibfield  {journal} {\bibinfo  {journal}
  {J. Chem. Phys.}\ }\textbf {\bibinfo {volume} {143}},\ \bibinfo {pages}
  {094506} (\bibinfo {year} {2015})}\BibitemShut {NoStop}%
\bibitem [{\citenamefont {Fern{\'a}ndez-Fern{\'a}ndez}\ \emph
  {et~al.}(2019)\citenamefont {Fern{\'a}ndez-Fern{\'a}ndez}, \citenamefont
  {P{\'e}rez-Rodr{\'\i}guez}, \citenamefont {Comesa{\~n}a},\ and\ \citenamefont
  {Pi{\~n}eiro}}]{Fernandez-Fernandez2019a}%
  \BibitemOpen
  \bibfield  {author} {\bibinfo {author} {\bibfnamefont {A.~M.}\ \bibnamefont
  {Fern{\'a}ndez-Fern{\'a}ndez}}, \bibinfo {author} {\bibfnamefont
  {M.}~\bibnamefont {P{\'e}rez-Rodr{\'\i}guez}}, \bibinfo {author}
  {\bibfnamefont {A.}~\bibnamefont {Comesa{\~n}a}}, \ and\ \bibinfo {author}
  {\bibfnamefont {M.~M.}\ \bibnamefont {Pi{\~n}eiro}},\ }\bibfield  {title}
  {\enquote {\bibinfo {title} {Three-phase equilibrium curve shift for methane
  hydrate in oceanic conditions calculated from molecular dynamics
  simulations},}\ }\href {\doibase
  https://doi.org/10.1016/j.molliq.2018.10.146} {\bibfield  {journal} {\bibinfo
   {journal} {J. Mol. Liq.}\ }\textbf {\bibinfo {volume} {274}},\ \bibinfo
  {pages} {426--433} (\bibinfo {year} {2019})}\BibitemShut {NoStop}%
\bibitem [{\citenamefont {Fern{\'a}ndez-Fern{\'a}ndez}, \citenamefont
  {Pi{\~n}eiro},\ and\ \citenamefont
  {P{\'e}rez-Rodr{\'{\i}}guez}(2021)}]{Fernandez-Fernandez2021a}%
  \BibitemOpen
  \bibfield  {author} {\bibinfo {author} {\bibfnamefont {A.~M.}\ \bibnamefont
  {Fern{\'a}ndez-Fern{\'a}ndez}}, \bibinfo {author} {\bibfnamefont {M.~M.}\
  \bibnamefont {Pi{\~n}eiro}}, \ and\ \bibinfo {author} {\bibfnamefont
  {M.}~\bibnamefont {P{\'e}rez-Rodr{\'{\i}}guez}},\ }\bibfield  {title}
  {\enquote {\bibinfo {title} {Molecular dynamics of fluoromethane type {I}
  hydrates},}\ }\href@noop {} {\bibfield  {journal} {\bibinfo  {journal} {J.
  Mol. Liq.}\ }\textbf {\bibinfo {volume} {339}},\ \bibinfo {pages} {116720}
  (\bibinfo {year} {2021})}\BibitemShut {NoStop}%
\bibitem [{\citenamefont {Blazquez}, \citenamefont {Vega},\ and\ \citenamefont
  {Conde}(2023)}]{Blazquez2023b}%
  \BibitemOpen
  \bibfield  {author} {\bibinfo {author} {\bibfnamefont {S.}~\bibnamefont
  {Blazquez}}, \bibinfo {author} {\bibfnamefont {C.}~\bibnamefont {Vega}}, \
  and\ \bibinfo {author} {\bibfnamefont {M.~M.}\ \bibnamefont {Conde}},\
  }\bibfield  {title} {\enquote {\bibinfo {title} {Three phase equilibria of
  the methane hydrate in nacl solutions: A simulation study},}\ }\href@noop {}
  {\bibfield  {journal} {\bibinfo  {journal} {J. Mol. Liq.}\ }\textbf {\bibinfo
  {volume} {383}},\ \bibinfo {pages} {122031} (\bibinfo {year}
  {2023})}\BibitemShut {NoStop}%
\bibitem [{\citenamefont {Grabowska}\ \emph {et~al.}(2022)\citenamefont
  {Grabowska}, \citenamefont {Bl{\'a}zquez}, \citenamefont {Sanz},
  \citenamefont {Zer{\'o}n}, \citenamefont {Algaba}, \citenamefont
  {M{\'{\i}}guez}, \citenamefont {Blas},\ and\ \citenamefont
  {Vega}}]{Grabowska2022a}%
  \BibitemOpen
  \bibfield  {author} {\bibinfo {author} {\bibfnamefont {J.}~\bibnamefont
  {Grabowska}}, \bibinfo {author} {\bibfnamefont {S.}~\bibnamefont
  {Bl{\'a}zquez}}, \bibinfo {author} {\bibfnamefont {E.}~\bibnamefont {Sanz}},
  \bibinfo {author} {\bibfnamefont {I.~M.}\ \bibnamefont {Zer{\'o}n}}, \bibinfo
  {author} {\bibfnamefont {J.}~\bibnamefont {Algaba}}, \bibinfo {author}
  {\bibfnamefont {J.~M.}\ \bibnamefont {M{\'{\i}}guez}}, \bibinfo {author}
  {\bibfnamefont {F.~J.}\ \bibnamefont {Blas}}, \ and\ \bibinfo {author}
  {\bibfnamefont {C.}~\bibnamefont {Vega}},\ }\bibfield  {title} {\enquote
  {\bibinfo {title} {Solubility of methane in water: some useful results for
  hydrate nucleation},}\ }\href@noop {} {\bibfield  {journal} {\bibinfo
  {journal} {J. Phys. Chem. B}\ }\textbf {\bibinfo {volume} {126}},\ \bibinfo
  {pages} {8553--8570} (\bibinfo {year} {2022})}\BibitemShut {NoStop}%
\bibitem [{\citenamefont {Algaba}\ \emph {et~al.}(2023)\citenamefont {Algaba},
  \citenamefont {Zer{\'o}n}, \citenamefont {M{\'\i}guez}, \citenamefont
  {Grabowska}, \citenamefont {Blazquez}, \citenamefont {Sanz}, \citenamefont
  {Vega},\ and\ \citenamefont {Blas}}]{Algaba2023a}%
  \BibitemOpen
  \bibfield  {author} {\bibinfo {author} {\bibfnamefont {J.}~\bibnamefont
  {Algaba}}, \bibinfo {author} {\bibfnamefont {I.~M.}\ \bibnamefont
  {Zer{\'o}n}}, \bibinfo {author} {\bibfnamefont {J.~M.}\ \bibnamefont
  {M{\'\i}guez}}, \bibinfo {author} {\bibfnamefont {J.}~\bibnamefont
  {Grabowska}}, \bibinfo {author} {\bibfnamefont {S.}~\bibnamefont {Blazquez}},
  \bibinfo {author} {\bibfnamefont {E.}~\bibnamefont {Sanz}}, \bibinfo {author}
  {\bibfnamefont {C.}~\bibnamefont {Vega}}, \ and\ \bibinfo {author}
  {\bibfnamefont {F.~J.}\ \bibnamefont {Blas}},\ }\bibfield  {title} {\enquote
  {\bibinfo {title} {Solubility of carbon dioxide in water: Some useful results
  for hydrate nucleation},}\ }\href@noop {} {\bibfield  {journal} {\bibinfo
  {journal} {The Journal of Chemical Physics}\ }\textbf {\bibinfo {volume}
  {158}} (\bibinfo {year} {2023})}\BibitemShut {NoStop}%
\bibitem [{\citenamefont {{van der Spoel}}\ \emph {et~al.}(2005)\citenamefont
  {{van der Spoel}}, \citenamefont {Lindahl}, \citenamefont {Hess},
  \citenamefont {Groenhof}, \citenamefont {Mark},\ and\ \citenamefont
  {Berendsen}}]{VanDerSpoel2005a}%
  \BibitemOpen
  \bibfield  {author} {\bibinfo {author} {\bibfnamefont {D.}~\bibnamefont {{van
  der Spoel}}}, \bibinfo {author} {\bibfnamefont {E.}~\bibnamefont {Lindahl}},
  \bibinfo {author} {\bibfnamefont {B.}~\bibnamefont {Hess}}, \bibinfo {author}
  {\bibfnamefont {G.}~\bibnamefont {Groenhof}}, \bibinfo {author}
  {\bibfnamefont {A.~E.}\ \bibnamefont {Mark}}, \ and\ \bibinfo {author}
  {\bibfnamefont {H.~J.}\ \bibnamefont {Berendsen}},\ }\bibfield  {title}
  {\enquote {\bibinfo {title} {Gromacs: Fast, flexible, and free},}\
  }\href@noop {} {\bibfield  {journal} {\bibinfo  {journal} {J. Comput. Chem.}\
  }\textbf {\bibinfo {volume} {26}},\ \bibinfo {pages} {1701--1718} (\bibinfo
  {year} {2005})}\BibitemShut {NoStop}%
\bibitem [{\citenamefont {Potoff}\ and\ \citenamefont
  {Siepmann}(2001)}]{Potoff2001a}%
  \BibitemOpen
  \bibfield  {author} {\bibinfo {author} {\bibfnamefont {J.~J.}\ \bibnamefont
  {Potoff}}\ and\ \bibinfo {author} {\bibfnamefont {J.~I.}\ \bibnamefont
  {Siepmann}},\ }\bibfield  {title} {\enquote {\bibinfo {title} {Vapor-liquid
  equilibria of mixtures containing alkanes, carbon dioxide, and nitrogen},}\
  }\href@noop {} {\bibfield  {journal} {\bibinfo  {journal} {AIChE Journal.}\
  }\textbf {\bibinfo {volume} {47}},\ \bibinfo {pages} {1676--1682} (\bibinfo
  {year} {2001})}\BibitemShut {NoStop}%
\bibitem [{\citenamefont {Cuendet}\ and\ \citenamefont
  {Gunsteren}(2007)}]{Cuendet2007a}%
  \BibitemOpen
  \bibfield  {author} {\bibinfo {author} {\bibfnamefont {M.~A.}\ \bibnamefont
  {Cuendet}}\ and\ \bibinfo {author} {\bibfnamefont {W.~F.~V.}\ \bibnamefont
  {Gunsteren}},\ }\bibfield  {title} {\enquote {\bibinfo {title} {On the
  calculation of velocity-dependent properties in molecular dynamics
  simulations using the leapfrog integration algorithm},}\ }\href@noop {}
  {\bibfield  {journal} {\bibinfo  {journal} {J. Chem. Phys.}\ }\textbf
  {\bibinfo {volume} {127}},\ \bibinfo {pages} {184102/1--9} (\bibinfo {year}
  {2007})}\BibitemShut {NoStop}%
\bibitem [{\citenamefont {Nos{\'e}}(1984)}]{Nose1984a}%
  \BibitemOpen
  \bibfield  {author} {\bibinfo {author} {\bibfnamefont {S.}~\bibnamefont
  {Nos{\'e}}},\ }\bibfield  {title} {\enquote {\bibinfo {title} {A molecular
  dynamics method for simulations in the canonical ensemble},}\ }\href@noop {}
  {\bibfield  {journal} {\bibinfo  {journal} {Mol. Phys.}\ }\textbf {\bibinfo
  {volume} {52}},\ \bibinfo {pages} {255--268} (\bibinfo {year}
  {1984})}\BibitemShut {NoStop}%
\bibitem [{\citenamefont {Parrinello}\ and\ \citenamefont
  {Rahman}(1981)}]{Parrinello1981a}%
  \BibitemOpen
  \bibfield  {author} {\bibinfo {author} {\bibfnamefont {M.}~\bibnamefont
  {Parrinello}}\ and\ \bibinfo {author} {\bibfnamefont {A.}~\bibnamefont
  {Rahman}},\ }\bibfield  {title} {\enquote {\bibinfo {title} {Polymorphic
  transitions in single crystals: A new molecular dynamics method},}\
  }\href@noop {} {\bibfield  {journal} {\bibinfo  {journal} {J. Appl. Phys.}\
  }\textbf {\bibinfo {volume} {52}},\ \bibinfo {pages} {7182--7190} (\bibinfo
  {year} {1981})}\BibitemShut {NoStop}%
\bibitem [{\citenamefont {Essmann}\ \emph {et~al.}(1995)\citenamefont
  {Essmann}, \citenamefont {Perera}, \citenamefont {Berkowitz}, \citenamefont
  {Darden}, \citenamefont {Lee},\ and\ \citenamefont
  {Pedersen}}]{Essmann1995a}%
  \BibitemOpen
  \bibfield  {author} {\bibinfo {author} {\bibfnamefont {U.}~\bibnamefont
  {Essmann}}, \bibinfo {author} {\bibfnamefont {L.}~\bibnamefont {Perera}},
  \bibinfo {author} {\bibfnamefont {M.~L.}\ \bibnamefont {Berkowitz}}, \bibinfo
  {author} {\bibfnamefont {T.}~\bibnamefont {Darden}}, \bibinfo {author}
  {\bibfnamefont {H.}~\bibnamefont {Lee}}, \ and\ \bibinfo {author}
  {\bibfnamefont {L.~G.}\ \bibnamefont {Pedersen}},\ }\bibfield  {title}
  {\enquote {\bibinfo {title} {A smooth particle mesh {Ewald} method},}\
  }\href@noop {} {\bibfield  {journal} {\bibinfo  {journal} {J. Chem. Phys.}\
  }\textbf {\bibinfo {volume} {103}},\ \bibinfo {pages} {8577--8593} (\bibinfo
  {year} {1995})}\BibitemShut {NoStop}%
\bibitem [{\citenamefont {Wiebe}, \citenamefont {Gaddy},\ and\ \citenamefont
  {Heins~Jr}(1933)}]{Wiebe1933}%
  \BibitemOpen
  \bibfield  {author} {\bibinfo {author} {\bibfnamefont {R.}~\bibnamefont
  {Wiebe}}, \bibinfo {author} {\bibfnamefont {V.}~\bibnamefont {Gaddy}}, \ and\
  \bibinfo {author} {\bibfnamefont {C.}~\bibnamefont {Heins~Jr}},\ }\bibfield
  {title} {\enquote {\bibinfo {title} {The solubility of nitrogen in water at
  50, 75 and 100 from 25 to 1000 atmospheres},}\ }\href@noop {} {\bibfield
  {journal} {\bibinfo  {journal} {Journal of the American Chemical Society}\
  }\textbf {\bibinfo {volume} {55}},\ \bibinfo {pages} {947--953} (\bibinfo
  {year} {1933})}\BibitemShut {NoStop}%
\bibitem [{\citenamefont {Frenkel}\ and\ \citenamefont
  {Smit}(2002)}]{Frenkel2002a}%
  \BibitemOpen
  \bibfield  {author} {\bibinfo {author} {\bibfnamefont {D.}~\bibnamefont
  {Frenkel}}\ and\ \bibinfo {author} {\bibfnamefont {B.}~\bibnamefont {Smit}},\
  }\href@noop {} {\emph {\bibinfo {title} {Understanding Molecular
  Simulations}}}\ (\bibinfo  {publisher} {2nd Ed. Academic, San Diego},\
  \bibinfo {year} {2002})\BibitemShut {NoStop}%
\bibitem [{\citenamefont {Allen}\ and\ \citenamefont
  {Tildesley}(2017)}]{Allen2017a}%
  \BibitemOpen
  \bibfield  {author} {\bibinfo {author} {\bibfnamefont {M.~P.}\ \bibnamefont
  {Allen}}\ and\ \bibinfo {author} {\bibfnamefont {D.~J.}\ \bibnamefont
  {Tildesley}},\ }\href@noop {} {\emph {\bibinfo {title} {Computer Simulation
  of Liquids, 2 Ed.}}}\ (\bibinfo  {publisher} {Oxford University
  PressClaredon, Oxford},\ \bibinfo {year} {2017})\BibitemShut {NoStop}%
\bibitem [{\citenamefont {Rowlinson}\ and\ \citenamefont
  {Widom}(1982)}]{Rowlinson1982b}%
  \BibitemOpen
  \bibfield  {author} {\bibinfo {author} {\bibfnamefont {J.~S.}\ \bibnamefont
  {Rowlinson}}\ and\ \bibinfo {author} {\bibfnamefont {B.}~\bibnamefont
  {Widom}},\ }\href@noop {} {\emph {\bibinfo {title} {Molecular Theory of
  Capillarity}}}\ (\bibinfo  {publisher} {Claredon Press},\ \bibinfo {year}
  {1982})\BibitemShut {NoStop}%
\bibitem [{\citenamefont {Miguel}, \citenamefont {Blas},\ and\ \citenamefont
  {R\'{\i}o}(2006)}]{deMiguel2006a}%
  \BibitemOpen
  \bibfield  {author} {\bibinfo {author} {\bibfnamefont {E.~D.}\ \bibnamefont
  {Miguel}}, \bibinfo {author} {\bibfnamefont {F.~J.}\ \bibnamefont {Blas}}, \
  and\ \bibinfo {author} {\bibfnamefont {E.~M.~D.}\ \bibnamefont {R\'{\i}o}},\
  }\bibfield  {title} {\enquote {\bibinfo {title} {Molecular simulation of
  model liquid crystals in a strong aligning field},}\ }\href@noop {}
  {\bibfield  {journal} {\bibinfo  {journal} {Mol. Phys.}\ }\textbf {\bibinfo
  {volume} {104}},\ \bibinfo {pages} {2919--2927} (\bibinfo {year}
  {2006})}\BibitemShut {NoStop}%
\bibitem [{\citenamefont {Miguel}\ and\ \citenamefont
  {Jackson}(2006)}]{deMiguel2006b}%
  \BibitemOpen
  \bibfield  {author} {\bibinfo {author} {\bibfnamefont {E.~D.}\ \bibnamefont
  {Miguel}}\ and\ \bibinfo {author} {\bibfnamefont {G.}~\bibnamefont
  {Jackson}},\ }\bibfield  {title} {\enquote {\bibinfo {title} {The nature of
  the calculation of the pressure in molecular simulations of continuous models
  from volume perturbations},}\ }\href@noop {} {\bibfield  {journal} {\bibinfo
  {journal} {J. Chem. Phys.}\ }\textbf {\bibinfo {volume} {125}},\ \bibinfo
  {pages} {164109/1--12} (\bibinfo {year} {2006})}\BibitemShut {NoStop}%
\bibitem [{\citenamefont {Flyvbjerg}\ and\ \citenamefont
  {Petersen}(1989)}]{Flyvbjerg1989a}%
  \BibitemOpen
  \bibfield  {author} {\bibinfo {author} {\bibfnamefont {H.}~\bibnamefont
  {Flyvbjerg}}\ and\ \bibinfo {author} {\bibfnamefont {H.}~\bibnamefont
  {Petersen}},\ }\bibfield  {title} {\enquote {\bibinfo {title} {Error
  estimates on averages of correlated data},}\ }\href@noop {} {\bibfield
  {journal} {\bibinfo  {journal} {J. Chem. Phys.}\ }\textbf {\bibinfo {volume}
  {91}},\ \bibinfo {pages} {461--466} (\bibinfo {year} {1989})}\BibitemShut
  {NoStop}%
\bibitem [{\citenamefont {Wiegand}\ and\ \citenamefont
  {Franck}(1994)}]{Wiegand1994}%
  \BibitemOpen
  \bibfield  {author} {\bibinfo {author} {\bibfnamefont {G.}~\bibnamefont
  {Wiegand}}\ and\ \bibinfo {author} {\bibfnamefont {E.}~\bibnamefont
  {Franck}},\ }\bibfield  {title} {\enquote {\bibinfo {title} {Interfacial
  tension between water and non-polar fluids up to 473 k and 2800 bar},}\
  }\href@noop {} {\bibfield  {journal} {\bibinfo  {journal} {Berichte der
  Bunsengesellschaft f{\"u}r physikalische Chemie}\ }\textbf {\bibinfo {volume}
  {98}},\ \bibinfo {pages} {809--817} (\bibinfo {year} {1994})}\BibitemShut
  {NoStop}%
\bibitem [{\citenamefont {Buch}, \citenamefont {Sandler},\ and\ \citenamefont
  {Sadlej}(1998)}]{Buch1998a}%
  \BibitemOpen
  \bibfield  {author} {\bibinfo {author} {\bibfnamefont {V.}~\bibnamefont
  {Buch}}, \bibinfo {author} {\bibfnamefont {P.}~\bibnamefont {Sandler}}, \
  and\ \bibinfo {author} {\bibfnamefont {J.}~\bibnamefont {Sadlej}},\
  }\bibfield  {title} {\enquote {\bibinfo {title} {Simulations of h2o solid,
  liquid, and clusters, with an emphasis on ferroelectric ordering transition
  in hexagonal ice},}\ }\href@noop {} {\bibfield  {journal} {\bibinfo
  {journal} {J. Phys. Chem. B}\ }\textbf {\bibinfo {volume} {102}},\ \bibinfo
  {pages} {8641--8653} (\bibinfo {year} {1998})}\BibitemShut {NoStop}%
\bibitem [{\citenamefont {Bernal}\ and\ \citenamefont
  {Fowler}(1933)}]{Bernal1933a}%
  \BibitemOpen
  \bibfield  {author} {\bibinfo {author} {\bibfnamefont {J.~D.}\ \bibnamefont
  {Bernal}}\ and\ \bibinfo {author} {\bibfnamefont {R.~H.}\ \bibnamefont
  {Fowler}},\ }\bibfield  {title} {\enquote {\bibinfo {title} {Simulations of
  h2o solid, liquid, and clusters, with an emphasis on ferroelectric ordering
  transition in hexagonal ice},}\ }\href@noop {} {\bibfield  {journal}
  {\bibinfo  {journal} {J. Chem. Phys.}\ }\textbf {\bibinfo {volume} {1}},\
  \bibinfo {pages} {515--548} (\bibinfo {year} {1933})}\BibitemShut {NoStop}%
\bibitem [{\citenamefont {Debenedetti}(1997)}]{Debenedetti1996a}%
  \BibitemOpen
  \bibfield  {author} {\bibinfo {author} {\bibfnamefont {P.~G.}\ \bibnamefont
  {Debenedetti}},\ }\href@noop {} {\emph {\bibinfo {title} {Metastable Liquids:
  Concepts and Principles}}}\ (\bibinfo  {publisher} {Princeton University
  Press},\ \bibinfo {year} {1997})\BibitemShut {NoStop}%
\bibitem [{\citenamefont {Grabowska}\ \emph {et~al.}(2023)\citenamefont
  {Grabowska}, \citenamefont {Bl{\'a}zquez}, \citenamefont {Sanz},
  \citenamefont {Noya}, \citenamefont {Zer{\'o}n}, \citenamefont {Algaba},
  \citenamefont {M{\'{\i}}guez}, \citenamefont {Blas},\ and\ \citenamefont
  {Vega}}]{Grabowska2022b}%
  \BibitemOpen
  \bibfield  {author} {\bibinfo {author} {\bibfnamefont {J.}~\bibnamefont
  {Grabowska}}, \bibinfo {author} {\bibfnamefont {S.}~\bibnamefont
  {Bl{\'a}zquez}}, \bibinfo {author} {\bibfnamefont {E.}~\bibnamefont {Sanz}},
  \bibinfo {author} {\bibfnamefont {E.~G.}\ \bibnamefont {Noya}}, \bibinfo
  {author} {\bibfnamefont {I.~M.}\ \bibnamefont {Zer{\'o}n}}, \bibinfo {author}
  {\bibfnamefont {J.}~\bibnamefont {Algaba}}, \bibinfo {author} {\bibfnamefont
  {J.~M.}\ \bibnamefont {M{\'{\i}}guez}}, \bibinfo {author} {\bibfnamefont
  {F.~J.}\ \bibnamefont {Blas}}, \ and\ \bibinfo {author} {\bibfnamefont
  {C.}~\bibnamefont {Vega}},\ }\bibfield  {title} {\enquote {\bibinfo {title}
  {Homogeneous nucleation rate of methane hydrate formation under experimental
  conditions from seeding simulations},}\ }\href@noop {} {\bibfield  {journal}
  {\bibinfo  {journal} {J. Chem. Phys.}\ }\textbf {\bibinfo {volume} {158}}
  (\bibinfo {year} {2023})}\BibitemShut {NoStop}%
\end{thebibliography}%

\end{document}